\documentclass[aps,reprint,twocolumn,superscriptaddress]{revtex4-1}
\usepackage[colorlinks,bookmarks=true,citecolor=blue,linkcolor=red,urlcolor=blue]{hyperref}
\usepackage{amsmath,amsfonts,amssymb,mathtools}
\usepackage{mathrsfs}
\usepackage{graphicx,float}
\usepackage{multirow}
\usepackage{makecell}
\usepackage[ruled,vlined]{algorithm2e}
\usepackage{algorithmic}
\usepackage{color}
\usepackage{dcolumn}
\usepackage{bm}
\usepackage{subfiles}
\usepackage{verbatim} 
\usepackage{tikz}
\usepackage{longtable}
\usepackage{rotating} 
\usepackage{float}

\usepackage[T1]{fontenc}
\usepackage[utf8]{inputenc}
\usepackage{lmodern} 

\usepackage{url}
\usepackage[title]{appendix}

\usepackage{braket}

\begin{document}

\title{Wire Construction of  Topological Crystalline Superconductors}
\author{Bingrui Peng}
\email{pengbingrui@iphy.ac.cn}
\affiliation{Beijing National Laboratory for Condensed Matter Physics and Institute of Physics, Chinese Academy of Sciences, Beijing 100190, China}
\affiliation{University of Chinese Academy of Sciences, Beijing 100049, China}
\author{Hongming Weng}
\affiliation{Beijing National Laboratory for Condensed Matter Physics and Institute of Physics, Chinese Academy of Sciences, Beijing 100190, China}
\affiliation{University of Chinese Academy of Sciences, Beijing 100049, China}
\affiliation{Songshan Lake Materials Laboratory, Dongguan, Guangdong 523808, China}
\author{Chen Fang}
\affiliation{Beijing National Laboratory for Condensed Matter Physics and Institute of Physics, Chinese Academy of Sciences, Beijing 100190, China}
\affiliation{University of Chinese Academy of Sciences, Beijing 100049, China}
\affiliation{Songshan Lake Materials Laboratory, Dongguan, Guangdong 523808, China}
\affiliation{Kavli Institute for Theoretical Sciences, Chinese Academy of Sciences, Beijing 100190, China}

\begin{abstract}
We present a unified framework to construct and classify topological crystalline superconductors (TCSCs).
The building blocks are one-dimensional topological superconductors (TSCs) protected solely by onsite symmetries, which are arranged and glued by crystalline symmetries in real space.
We call this real-space scheme ``wire construction'', and we show its procedure can be formulated mathematically.
For illustration, we treat TCSCs of Altland-Zirnbauer (AZ) class\cite{altland1997nonstandard} DIII protected by wallpaper group as well as layer group symmetries,
with the resulting states by wire construction being TCSCs in two-dimension. 
We also discuss how these real-space TCSCs by wire construction are characterized by  anomalous boundary states.
Our method provides a real-space picture for TCSCs versus momentum-space pictures.

\end{abstract}
\maketitle

\section{Introduction}

Topological quantum matters have received wide concerns among the condensed matter physics community in recent years\cite{kane2005quantum,kane2005z,bernevig2006quantum,konig2007quantum,kitaev2003fault,nayak2008non,fu2007topological,hasan2010colloquium,qi2011topological,zhang2009topological,wen2017colloquium,chen2013symmetry,senthil2015symmetry,bansil2016colloquium,fu2011topological,hsieh2012topological,wang2012dirac,wang2013three,liu2014discovery,weng2015weyl,xu2015discovery,lv2015experimental,armitage2018weyl,chiu2016classification,ando2015topological,sato2017topological}.
Beyond Landau's theory of symmetry breaking which classifies matters according to symmetries, the studies of topological quantum matters reveal many different phases of matter with the same symmetries, which entail explanations from the perspective of topology. 
One significant problem in the studies of topological matters is how many topologically inequivalent classes of states exist for a specific kind of physical system, i.e., topological classification.
This issue has been resolved, completely or incompletely, for many physical systems, especially for weakly interacting systems, which have single-particle band descriptions.
For free fermion systems protected by only onsite symmetries, the classification problems have been solved by K-theory-based method, which is also known as the ``tenfold way'' \cite{kitaev2009periodic,schnyder2008classification,ryu2010topological}.
The classification of topological insulators protected by spacial symmetries, i.e., topological crystalline insulators (TCIs), has been completed by different methods\cite{kruthoff2017topological,khalaf2018symmetry,song2018quantitative,song2019topological,shiozaki2017topological,elcoro2021magnetic,peng2021topological}.
By contrast, the study of classification problems of topological crystalline superconductors (TCSCs) are still ongoing with many efforts \cite{chiu2013classification,shiozaki2014topology,fang2014new,wang2016topological,shiozaki2016topology,fang2017topological,langbehn2017reflection,geier2018second,khalaf2018higher,trifunovic2019higher,cornfeld2019classification,zhang2020kitaev,cornfeld2021tenfold,huang2021faithful,zou2021new,khalaf2018higher,ahn2021unconventional,zhang2021classification,vu2020time,ono2021symmetry,sumita2022topological,shiozaki2022classification}.

Historically, topological classifications always began with momentum-space frameworks, where one can mathematically define topological invariants in terms of Bloch wave functions.
Besides, the diagnosis of topological states sometimes can be easily accessible in momentum space, where one can acquire (usually partial) information of the topological invariants by using only the symmetry representations of Bloch wave functions at high symmetry points in Brillouin zone (BZ).
The diagnosis of band topology through symmetries traced back to the Fu-Kane formula\cite{fu2007topological} and was further promoted into the theory of symmetry-based indicators\cite{po2017symmetry,bradlyn2017topological,watanabe2018structure,ono2018unified,ono2019symmetry,po2020symmetry,ono2020refined,geier2020symmetry,shiozaki2019variants,skurativska2020atomic,ono2021z}.
However, momentum-space schemes for topological classification often involve sophisticated mathematical tools, such as K-theory, and the boundary anomalies characterized by the topological states could be obscure.
A different type of methods developed later, which provide another perspective to crystalline topological states, are more physically transparent than momentum-space-based methods.
These methods can be summarised as real-space construction\cite{song2017topological,huang2017building,shiozaki2018generalized,else2019crystalline,rasmussen2020classification,song2019topological,song2020real,zhang2020construction,zhang2020real,peng2021topological,song2020bosonic}, where topological states protected adjointly by onsite symmetries and spacial symmetries are constructed by arranging topological states protected by onsite symmetries in the ways that maintain spacial symmetries.
One of the simplest examples is that, 3D weak topological insulators (TIs) can be constructed by stacking an array of 2D TIs in a specific direction.
From this example, it is not hard for one to see crystalline topological states could be understood intuitively from real-space perspective.

Despite success in the classification of topological crystalline insulators (TCIs)\cite{song2018quantitative,song2019topological,elcoro2021magnetic,peng2021topological}, the extension of real-space construction to TCSCs is not straightforward, as many detailed techniques become more sophisticated. 
In real-space construction of TCIs, only 2D building blocks, which are actually (mirror) Chern insulators and 2D TIs, are used, while for TCSC, due to the existence of particle-hole symmetry, the building blocks can be diverse.
In addition to 2D building blocks, 1D building blocks also have to be taken into consideration, for TCSCs of several AZ classes.
Compared with 2D building blocks which either coincide with mirror planes or not, 1D building blocks can have more complicated symmetries such as $C_n$-rotations or $C_{nv}$, which provide more ingredients for topological classifications. 
Additionally, while classifying TCIs involve group representation theory, classifying TCSCs involve projective representations, due to nontrivial representations of crystalline symmetries acting on the superconducting gap function.

In this letter, we focus on real-space construction by 1D building blocks, i.e., ``wire construction'',
which we claim can give complete classification of TCSCs in 2D, including weak TSCs (protected by translation symmetries) and second-order TSCs.
We deal with 2D TCSCs in AZ class DIII, which have always received special concerns.  
It is worth noting that, 2D strong TSCs with helical  Majorana edge states are beyond the scope of wire construction.
Therefore we remark our method does not give the full classification of TSCs in 2D.
In addition, one should be careful when discussing 2D spacial symmetries for spinful electron systems.
For instance, even though $p2$ and $p\bar{1}$, which are generated by twofold rotation $C_2$ and spatial inversion $\mathcal{I}$, respectively, plus lattice translations, share the same lattice in 2D, their symmetry representations on spin-$\frac{1}{2}$ particles are different, i.e., $C_2^2 = -1$ while $\mathcal{I}^2 = 1$.
This difference can significantly influence the TCSC classification, as will be shown in the following context.
For a comprehensive discussion, in this work, we treat both wallpaper groups and layer groups.

Before going further, we briefly outline the contents.
Sec.\ref{Section formulation} is devoted to the formulation of wire construction, where we begin with the geometric structure in real space to decorate with ``wires'', i.e., 1D TSCs protected solely by onsite symmetries, and then show the condition for the wire-like building blocks to be glued together without ``opened boundaries'' (see Sec.\ref{subsection gluing condition}), as well as the condition for them to be stable under real-space trivialization process (Sec.\ref{subsection bubble equivalence}).
In Sec.\ref{Section example P4}, we give an example of wallpaper group $p 4$ to show the above procedure, and more examples can be found in the appendix.
In Sec.\ref{section wallpaper groups vs layer groups} we compare wallpaper groups and layer groups in terms of TCSC classification.
In Sec.\ref{Section classification results} we summarise the results of TCSC classification by wire construction, with detailed tables in Appendix.\ref{Appendix classification tables}.
In Sec.\ref{Section boundary states}, we discuss the topological invariants and anomalous boundary states which characterize the TCSCs by wire construction.
Lastly, in Sec.\ref{Section discussion} we summarise this work and discuss further directions.

\section{Formulation\label{Section formulation}}

In this section, we explicitly present the complete procedure of wire construction.

\subsection{Prerequisite: symmetries in superconductors}

Here we formulate the symmetries of BdG Hamiltonian following Ref.\cite{fang2017topological,ono2020refined}, which
is the basis for further discussions.

Consider a generic superconducting Hamiltonian 
\begin{align}
    \mathcal{H} = \epsilon ( \hat{c}^{\dagger} \hat{c} + h.c. ) + \Delta^{\dagger} \hat{c}^{\dagger} \hat{c}^{\dagger} + \Delta \hat{c} \hat{c} 
\end{align}
where $\hat{c}$ is the shorthand for $ ( \hat{c}_1, \hat{c}_2, \cdots, \hat{c}_n )$, $\epsilon$ is the normal state spectrum, and $\Delta$ is the superconducting gap function.

A crystalline symmetry $g$ of normal state Hamiltonian could no longer be a symmetry for the superconducting Hamiltonian, as it may be broken by the superconducting gap function $\Delta$, i.e.,
\begin{align}
    \hat{U}_g ( \Delta \hat{c} \hat{c} ) \hat{U}_g^{-1} = e^{i \theta} \Delta \hat{c} \hat{c}
\end{align}
where $e^{i \theta}$ is a $U(1)$ phase factor.
Nevertheless, we can redefine the crystalline symmetry as 
\begin{align}
    \hat{\mathcal{U}}_g  = \hat{U}_g \hat{V}_{\theta} = \hat{U}_g e^{i \frac{\theta}{2} \hat{Q} }
\end{align}
where $\hat{Q}$ is the charge operator.
Then we have
\begin{align}
    \hat{\mathcal{U}}_g' ( \Delta \hat{c} \hat{c} ) \hat{\mathcal{U}}^{-1}_g = \Delta \hat{c} \hat{c}
\end{align}
So we get to restore the crystalline symmetry of the Hamiltonian.
Now for $\hat{\mathcal{U}}_g$, we have
\begin{align}
    \{ \hat{\mathcal{U}}_g, \hat{\mathcal{T}} \} = 0, \;
    [ \hat{\mathcal{U}}_g, \hat{\mathcal{P}} ] = 0
\end{align}
where $\hat{\mathcal{T}}$ is time-reversal symmetry and $\hat{\mathcal{P}}$ is particle-hole symmetry, by noting
\begin{align}
    [ \hat{U}_g, \hat{\mathcal{T}} ] = 0, \;
    [ \hat{U}_g, \hat{\mathcal{P}} ] = 0
\end{align}
and 
\begin{align}
    \{ \hat{V}_{\theta}, \hat{\mathcal{T}} \} = 0, \;
    [ \hat{V}_{\theta}, \hat{\mathcal{P}} ] = 0
\end{align}
In the first-quantization language, we can write
\begin{align}
    \mathcal{U}_g = \left(
        \begin{array}{cc}
          U_g e^{i\frac{\theta}{2}}  & 0    \\
          0  &  U^{*}_g e^{-i\frac{\theta}{2}}   
         \end{array}
         \right) 
\end{align}
where $U_g$ is the matrix representation of $g$ for normal state.

However, we move on for a new convention as we hope to make  $\hat{\mathcal{U}}_g$ commutes with $\hat{\mathcal{T}}$ as it does for the normal state. 
We do this by
\begin{align}
     \mathcal{U}_g \rightarrow \mathcal{U}_g e^{-i\frac{\theta}{2}} = \left(
        \begin{array}{cc}
          U_g  & 0    \\
          0  &  U^{*}_g e^{-i\theta}
         \end{array}
         \right)
\end{align}
This actually amounts to changing $\hat{c}$ and $\hat{c}^{\dagger}$ as
\begin{align}
    \hat{c} \rightarrow \hat{c} e^{-i\frac{\theta}{2}}, \;\;
    \hat{c}^{\dagger} \rightarrow \hat{c}^{\dagger} e^{i\frac{\theta}{2}}
\end{align}
In such a convention, we have
\begin{align}
    [ \hat{\mathcal{U}}_g, \hat{\mathcal{T}} ] = 0, \;\;
     \hat{\mathcal{U}}_g \hat{\mathcal{P}} =
     e^{i\theta}
     \hat{\mathcal{P}}
     \hat{\mathcal{U}}_g 
\end{align}
In the first-quantization language, the particle-hole symmetry is given by
\begin{align}
    \mathcal{P} =  \left(
        \begin{array}{cc}
          0  & \xi \bm{1_n}    \\
          \bm{1_n}  & 0    
         \end{array}
         \right) \mathcal{K}
\end{align}
where $\mathcal{K}$ is the complex conjugation operator, $\xi = \pm 1$ depends on physical realization.
In presence of $\hat{\mathcal{T}}$, $\theta$ is restricted to be $0$ or $\theta$, thus we have
\begin{align}
    \mathcal{U}_g = \left(
        \begin{array}{cc}
          U_g  & 0    \\
          0  &  \pm U^{*}_g 
         \end{array}
         \right)    
\end{align}
which leads to 
\begin{align}
    \mathcal{P} \mathcal{U}_g = \pm \mathcal{U}_g \mathcal{P}
\end{align}
and further
\begin{align}
    \mathcal{S} \mathcal{U}_g = \pm \mathcal{U}_g \mathcal{S}    
\end{align}
where $\mathcal{S} = \mathcal{T} \mathcal{P}$ is the Chiral symmetry.

To sum up, due to the nontrivial representation of superconducting gap function under crystalline symmetry (2), the redefined crystalline symmetry $\mathcal{U}_g$, in presence of time-reversal,  could either commute or anti-commute with particle-hole symmetry $\mathcal{P}$ (or Chiral symmetry $\mathcal{S}$), i.e., 
\begin{align}
    \mathcal{P} \mathcal{U}_g = \pm \mathcal{U}_g \mathcal{P} \Longleftrightarrow \mathcal{S} \mathcal{U}_g = \pm \mathcal{U}_g \mathcal{S}   
\end{align}
As a result, the whole symmetry group generated by crystalline symmetry group $G_c$, time-reversal symmetry $\mathcal{T}$ and particle-hole symmetry $\mathcal{P}$ (or Chiral symmetry $\mathcal{S}$), is projectively represented, with the factor system given by
\begin{align}
    z_{g \mathcal{S}, g\mathcal{S}} \; U_{g \mathcal{S} g \mathcal{S}} = U_{g \mathcal{S} } U_{g \mathcal{S}}
\end{align}
where $z_{g \mathcal{S}, g\mathcal{S}} = \chi_g = \pm 1 $ depending on whether $U_g$ commutes or anti-commutes with $\mathcal{S}$.

\subsection{Cell decomposition}

The first step of wire construction (as well as for other real-space construction schemes) is to decompose the $n$D space into small units by applying crystalline symmetries, which we call $d$-cells, where $d=0,1,2,\cdots,n$.
The resulting geometric structure composed of all $d$-cells is called a cell complex, which is a CW-complex in mathematics.
Each $d$-cell has no two points inside it related by crystalline symmetry operations.
In our case, we have $n=2$. 
The 2-cells are small patches that cover the whole 2D space without overlaps, and a single 2-cell is also called an asymmetric unit (AU).
1-cells are edges of 2-cells, and 0-cells are ends of 1-cells. 
Each $(d-1)$-cell is shared by more than one $d$-cells. 

It is worth noting that there could be different ways to construct the cell complex for a specific crystalline symmetry group.
However, these seemingly different cell complexes are essentially equivalent and can be deformed into each other while preserving symmetries.

\begin{figure}
	\centering
	\includegraphics[width=0.4\textwidth]{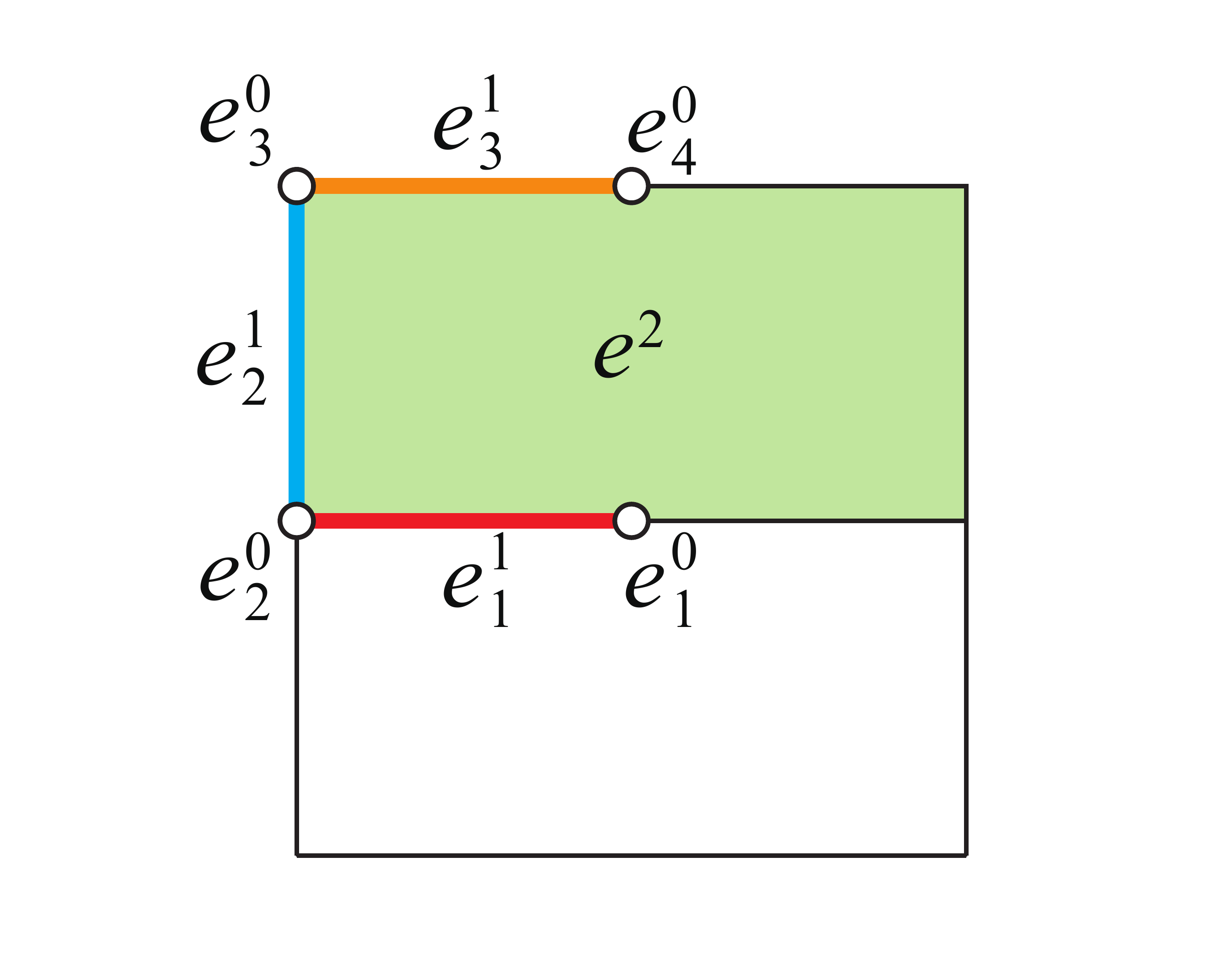}
	\caption{\label{cell complex of P2} The cell complex of wallpaper group $p2$. The big square is the unit cell, which is partitioned by $C_2$ rotation into two 2-cells, and we denote one of them as $e^2_1$. There are three symmetry-independent 1-cells, colored by three different colors, and denoted as $e^1_1$, $e^1_2$ and $e^1_3$. 
	There are four symmetry-independent 0-cells, which we denote as $e^0_1$, $e^0_2$, $e^0_3$ and $e^0_4$.
	Note we only have to label the symmetry-independent cells, since other ones are related to them by either inversion or translations. }
\end{figure}

Here we give an example, i.e., the cell complex of 2D space group $p2$, which is shown in Figure.\ref{cell complex of P2}.

\subsection{Building blocks}

Upon having the cell complex, one has to decide what topological states can be decorated on the 1-cells as building blocks for wire construction.
On each 1-cell, no crystalline symmetry operation can relate two distinct points, therefore the topological states that can be placed on it are those protected solely by onsite symmetries. 
Note that apart from $\mathcal{T}$, $\mathcal{P}$ and $\mathcal{S}$, the onsite symmetries could include crystalline operations that keep every point on the 1-cell invariant, which form a subgroup $G_c^1$ of the entire onsite symmetry group $G^1$.
We remark that, although $G_c^1$ generically could contain lattice translations, here we ignore them and always regard $G_c^1$ as a point group.

For wallpaper groups, $G^1_c$ is always simple, which depends on whether the 1-cell coincides with some mirror line or not and turns out to be either $\bm{1}$ or $\bm{M}$,
where $\bm{1}$ stands for the point group with identity operation only, and $\bm{M}$ for the point group having mirror reflection besides identity.
Different types of 1-cells host different topological states protected by onsite symmetries, depending on specific $G_c^1$ thus $G^1$.
And it further depends on the projective representations of $G^1$, as in presence of nontrivial symmetry representations of the gap function, the representations of $G^1$ are projective.

Actually, the projective irreducible representations of $G^1$ can be obtained from the irreducible representations (irreps) of $G_c^1$.
Given the irreps of $G_c^1$, the inclusion of $\mathcal{T}$, $\mathcal{P}$ and $\mathcal{S}$ can have effects on these irreps, making them either invariant or paired, and the resulting representations, together with those of $\mathcal{T}$, $\mathcal{P}$ and $\mathcal{S}$, form  projective irreps of $G^1$.
For each sector labeled by an irrep of $G_c^1$, we determine its effective Altland-Zirnbaue (AZ) class under $\mathcal{T}$, $\mathcal{P}$, and $\mathcal{S}$, so that the topological classification of this sector can be directly known from the ``tenfold way'' results\cite{ryu2010topological}.
Finally, the total classification on the 1-cell is obtained by combining the classification of each sector.
As two different sectors could be paired under $\mathcal{T}$, $\mathcal{P}$ and $\mathcal{S}$, the independent sectors which are labeled by projective irreps of $G^1$ could be less than the sectors labeled by irreps of $G^1_c$, as two paired sectors always have the same or opposite topological numbers.
Therefore we can define an independent topological index for each new sector labeled by projective irrep of $G^1$, see following contents.  

To figure out the effective AZ classes and the behaviors of the irreps of $G^1_c$ under $\mathcal{T}$, $\mathcal{P}$ and $\mathcal{S}$, one can employ a powerful tool, the generalized Wigner test\cite{shiozaki2018atiyah,bradley2009mathematical,wigner2012group,herring1937effect}, which can handle not only time-reversal but also particle-hole and chiral symmetries.
It states that for a group $G$ which can be decomposed into left cosets as 
\begin{align}
    G = a G_0 \oplus b G_0 \oplus c G_0 \oplus ab G_0
\end{align}
where $G_0$ is a unitary group, $a$ is a $\mathcal{T}$-like operator, $b$ is a $\mathcal{P}$-like operator and $ab$ is $\mathcal{S}$-like operator, there are three indices to compute in order to determine the effects of $\mathcal{T}$, $\mathcal{P}$ and $\mathcal{S}$ on a irrep $\alpha$ of $G_0$: 
\begin{align}
    & W^{\alpha}(\mathcal{T}) 
    = \frac{1}{|G_0|} \sum_{g \in G_0 } z_{a g, a g} \; \widetilde{\chi}_{\alpha}((a g)^2) 
    \\
    & W^{\alpha}(\mathcal{P}) = \frac{1}{|G_0|} \sum_{g \in G_0 } z_{b g, b g} \; \widetilde{\chi}_{\alpha}((b g)^2) 
    \\
    & W^{\alpha}(\mathcal{S}) = \frac{1}{|G_0|}
    \sum_{g \in G_0} \frac{z_{g, a b }}{z_{a b, (a b)^{\text{-}1} g (a b)}} \widetilde{\chi}_{\alpha}(g) \widetilde{\chi}_{\alpha}^{*}((ab) g (ab)^{\text{-}1}) 
\end{align}
with $W^{\alpha}(\mathcal{T}) \in \{ \pm 1, 0 \}$, $W^{\alpha}(\mathcal{P}) \in \{ \pm1, 0 \}$, and $W^{\alpha}(\mathcal{S}) \in \{1, 0 \}$.
Here $|G_0|$ is the number of elements in $G_0$, $z_{g_1, g_2} U_{g_1 g_2} = U_{g_1} U_{g_2}$ specifies the factor system, and $\widetilde{\chi}_{\alpha}$ denotes the character of the irrep $\alpha$. 
Specifically, 
\begin{align}
    W^{\alpha}(U =\mathcal{T}, \mathcal{P}, \mathcal{S}) = 1 
\end{align}
indicates $\alpha$ is invariant under $U$, 
\begin{align}
    W^{\alpha}(U =\mathcal{T}, \mathcal{P}, \mathcal{S}) = -1 
\end{align}
indicates $\alpha$ is paired with itself, and 
\begin{align}
    W^{\alpha}(U =\mathcal{T}, \mathcal{P}, \mathcal{S}) = 0 
\end{align}
indicates $\alpha$ is paired with another irrep $\alpha'$. 
The effective AZ class for $\alpha$ is determined as 
\begin{align}
    [ W^{\alpha}(\mathcal{T}), \;  W^{\alpha}(\mathcal{P}), \; W^{\alpha}(\mathcal{S}) ] = [ \mathcal{T}^2, \mathcal{P}^2, \mathcal{S}^2 ]
\end{align}
For instance, 
\begin{align}
    [ W^{\alpha}(\mathcal{T}), \;  W^{\alpha}(\mathcal{P}), \; W^{\alpha}(\mathcal{S}) ] = [ -1, 0, 0 ]
\end{align}
dictates $\alpha$ has effective AZ class AII.

\begin{figure}
	\centering
	\includegraphics[width=0.3\textwidth]{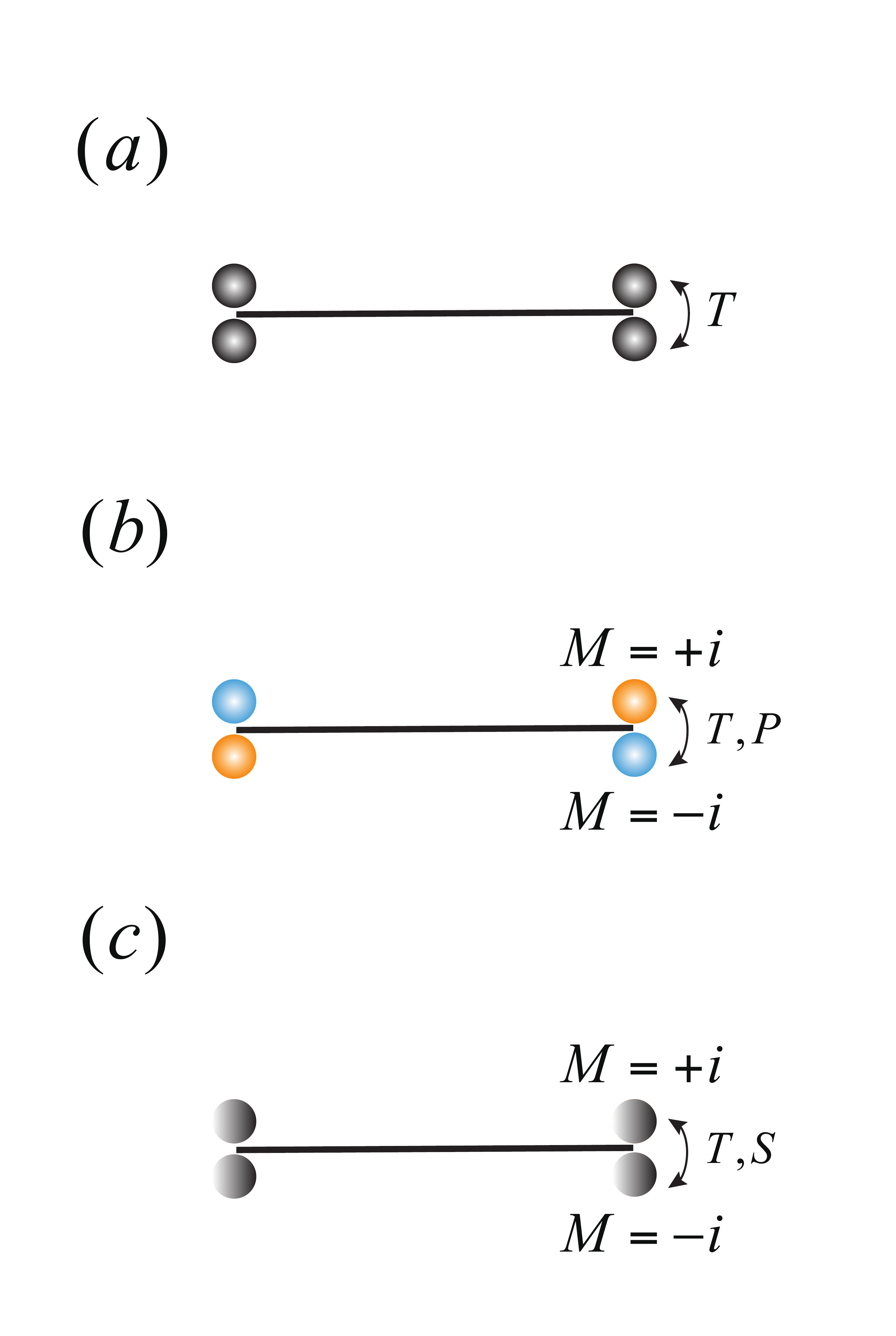}
	\caption{\label{building blocks of WGs} 
	Building blocks for wallpaper groups.
	(a) The decoration on 1-cells with trivial $G^1_c$, which is a 1D TSC in AZ class DIII with $\mathbb{Z}_2$ topological index.
	At its 0D boundary, there is a Kramer's pair of Majorana zero modes. 
	(b) The decoration on 1-cells protected by onsite mirror reflection $M$ with $\chi_M = 1$.
	The two sectors with $M$ eigenvalues $\pm i$ are related by $\mathcal{T}$ and $\mathcal{P}$, and each one is invariant under $\mathcal{S}$ thus belongs to AZ class AIII with $\mathbb{Z}$ topological index in 1D, i.e., the 1D winding number.
	At its 0D boundary, there are $|w_M| \in \mathbb{Z}$ pairs of Majorana zero modes, where $w_M = \frac{1}{2} ( w^{+} - w^{-} )$ is the mirror winding number.
	(c) The decoration on 1-cells protected by mirror reflection $M$ with $\chi_M = -1$.
	The two sectors with $M$ eigenvalues $\pm i$ are related by $\mathcal{T}$ and $\mathcal{S}$, and each one is invariant under $\mathcal{P}$ thus belongs to AZ class D with $\mathbb{Z}_2$ topological index in 1D.
	At each end, there is a pair of Majorana zero modes as boundary state.}
\end{figure}

In fact, it is straightforward to figure out all the building blocks for wallpaper groups.
As the irreps of point group $\bm{1}$ and $\bm{M}$ are all 1D, it is each to see the behaviors of each sector under $\mathcal{T}$, $\mathcal{P}$ and $\mathcal{S}$, i.e., whether it is unchanged or get paired with either itself or another irrep.
The results are illustrated in Fig.\ref{building blocks of WGs}.
By contrast, 1-cells with higher symmetries exist in layer groups, the building blocks on which are more complicated and not straightforward to determine.

\subsubsection{building blocks for wallpapar groups}{\label{subsubsection building blocks for WGs}}

\begin{enumerate}

\item 1-cells with $C^1_c = \bm{1}$

These 1-cells do not coincide with any mirror lines, and their $G_c$ is trivial.
Therefore the onsite symmetry group $G^1$ is generated by $\mathcal{T}$ and $\mathcal{P}$ (or$\mathcal{S}$).
Apparently, the effective AZ class is DIII with $\mathbb{Z}_2$ classification in 1D, and the corresponding topological invariant $\nu^{\mathcal{T}}$ is defined as\cite{sato2017topological}
\begin{align}
    (-1)^{\nu^{\mathcal{T}}} = e^{i\gamma^{\text{I}(\text{II})}}, \; \;
    \gamma^{\text{I}(\text{II})} = \int^{\pi}_{-\pi} dk \mathcal{A}^{\text{I}(\text{II})}_{(-)} (k) 
\end{align}
where $\mathcal{A}^{\text{I}}_{(-)}$ and $\mathcal{A}^{\text{II}}_{(-)}$ represent the Berry phase of the Kramer's pair formed by the occupied states.
The boundary state of such a building block with $\mathbb{Z}_2 = 1$ is a Kramer's pair of Majorana zero modes.
Actually such a pair of zero modes can be eigenstates of chiral symmetry $\mathcal{S}$, with opposite eigenvalues of $\mathcal{S}$, i.e., $\pm 1$.
One can write out the symmetry representation taken by the zero modes as
\begin{align}
    & \mathcal{T} = i\sigma_2 \mathcal{K}
    \\
    & \mathcal{S} = \sigma_3
\end{align}
which satisfies $\mathcal{T}^2= -1$, $\mathcal{S}^2 = 1$ and $\mathcal{P}^2 = ( \mathcal{T} \mathcal{S} )^2 = 1$.

\item 1-cells with $G_c = \bm{M}$

This type of 1-cells lie on mirror lines.
The Hilbert space is divided into two sectors with mirror eigenvalues $\pm i$ since $M^2 = -1$. 
One has to discuss whether the gap function is even or odd under $M$, i.e., $\chi_M = 1$ or $\chi_M = -1$.

\begin{enumerate}
    
    \item $\chi_M = 1$

    Since $[ M, \mathcal{T} ] = 0$ and $[ M, \mathcal{S} ] = 0$, it is easy to see each of the two sectors is closed under $\mathcal{S}$, and they are interchanged under $\mathcal{T}$ and $\mathcal{P}$.
    Therefore each sector belongs to AZ class AIII with $\mathbb{Z}$ classification in 1D.
    The corresponding topological invariant is the 1D winding number
    \begin{align}
        w^{\pm} = \frac{i}{4\pi} \int^{\pi}_{-\pi} dk \; \text{tr} [ \mathcal{S} \mathcal{H}_{\pm}^{-1}(k) \partial_{k} \mathcal{H}_{\pm}(k) ]
    \end{align}
    where $\mathcal{H}_{\pm}$ stands for the Hamiltonian of the $ M = \pm i$ sector.
    Since the two sectors are not independent but related by $\mathcal{T}$ and $\mathcal{P}$, there is only one independent $\mathbb{Z}$ index and actually $w^{+} = - w^{-} $.
    We can define the independent index as $w_M = \frac{1}{2} (w^{+} - w^{-}) $.
    
    For a $|w_M| = n$ state, at its 0D boundary there are $n$ Majorana zero modes with $M = +i$ and $\mathcal{S} = \pm 1$, as well as $n$ Majorana zero modes with $M = -i$ and $\mathcal{S} = \mp 1$, forming $n$ Majorana Kramer's pairs.
    In fact, one can write out the symmetry representation
    carried by the zero modes as
    \begin{align}
        & \mathcal{T} = i \sigma_2 \mathcal{K}
        \\
        & \mathcal{S} = \sigma_3
        \\
        & M = i \sigma_3
    \end{align}
    which satisfies $\mathcal{T}^2 = -1$, $\mathcal{S}^2 = 1$, $ \mathcal{P}^2 =  ( \mathcal{T} \mathcal{S} )^2 = 1$, $[ \mathcal{T}, M ] = 0$, and $[ \mathcal{S}, M ] = 0$.

    \item $\chi_M = -1$
    
    Like the above case, it is easy to find each of the two mirror sectors is closed under $\mathcal{P}$ but change into the other one under $\mathcal{T}$ and $\mathcal{S}$.
    Therefore each sector belongs to AZ class D with $\mathbb{Z}_2$ classification in 1D. 
    The topological index $\nu$ is given by\cite{sato2017topological}
    \begin{align}
        (-1)^{\nu} = e^{i\gamma}, \;\;
        \gamma = \int^{\pi}_{-\pi} dk \mathcal{A}_{(-)} (k)
    \end{align}
    where $\mathcal{A}_{(-)}$ stands for Berry connection of occupied states.
    Since each of the two sectors is not independent and they are related by $\mathcal{T}$ and $\mathcal{S}$,
    they have the same $\nu$ and 
    there is only one independent $\mathbb{Z}_2$ index.
    The Majorana zero modes at its 0D boundary are described by the symmetries
    \begin{align}
        & \mathcal{T} = i \sigma_2 \mathcal{K}
        \\
        & \mathcal{S} = \sigma_1 \; 
        \\
        & M = i \sigma_3
    \end{align}
    which satisfy $\mathcal{T}^2 = -1$, $\mathcal{S}^2 = 1$, $ \mathcal{P}^2 =  ( \mathcal{T} \mathcal{S} )^2 = 1$, $[ \mathcal{T}, M ] = 0$, and $\{ \mathcal{S}, M \} = 0$.

    \end{enumerate}
    
\end{enumerate}

\subsubsection{building blocks for layer groups} 
    
Firstly note that all the building blocks for wallpaper groups discussed above are also for layer groups, as wallpaper groups can be viewed as subset of layer groups.
Besides, for layer groups there can be mirror reflection vertical to the lattice plane, which we denote as $M_z$.
When combined with $M_x$ ($M_y$), the generated point group is $\bm{C_{2v}}$.
As a consequence, the 1-cells coinciding with such $C_{2v}$ axis have $C^1_c = \bm{C_{2v}}$.
The determination of building blocks on such 1-cells is not straightforward and one need turn to the Wigner test.

\begin{figure}
	\centering
	\includegraphics[width=0.27\textwidth]{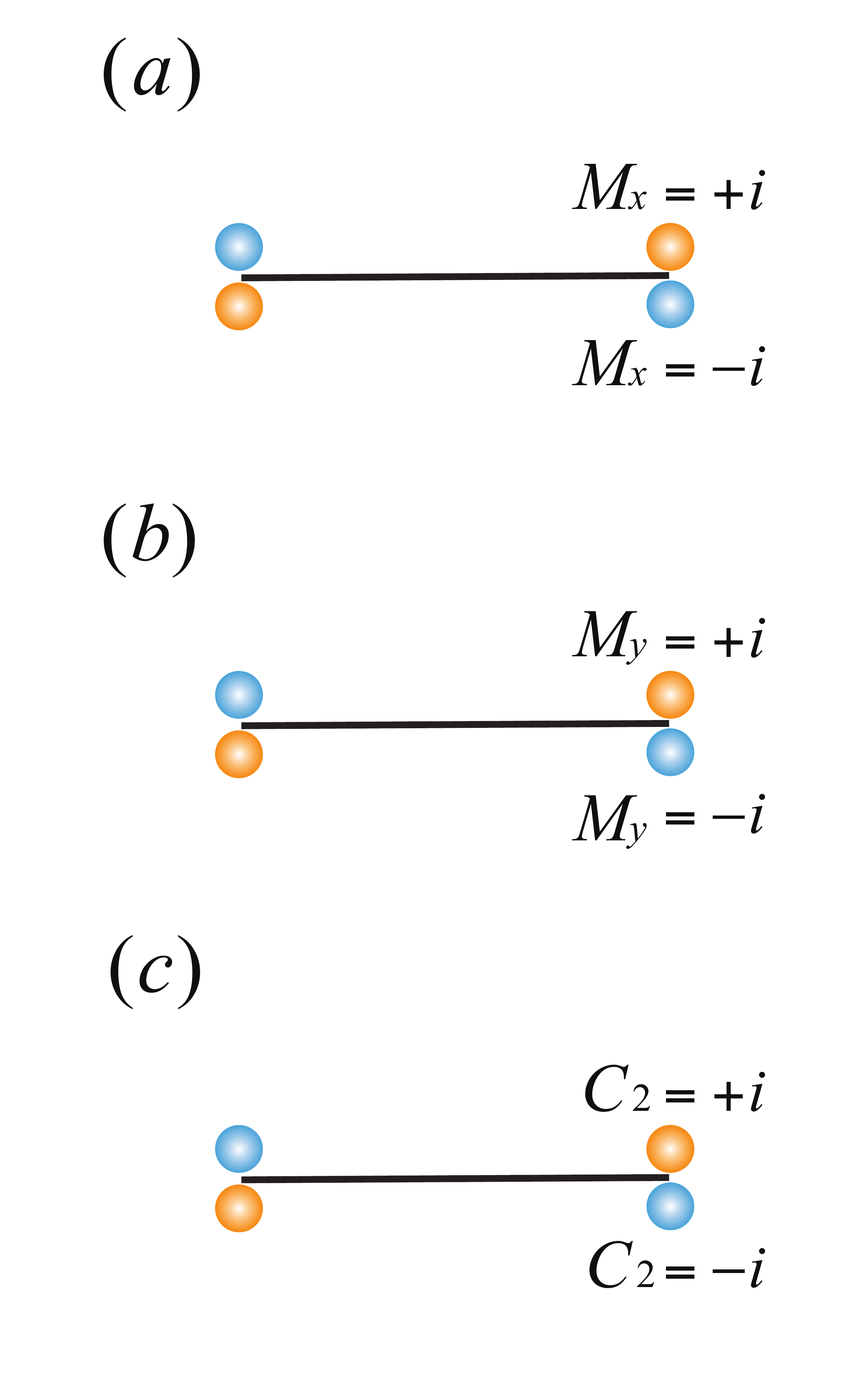}
	\caption{\label{building blocks of LGs} 
	Nontrivial Building blocks on $C_{2v}$-axis.
	The three types of building blocks are for $A_2$, $B_1$ and $B_2$, respectively.
	The trivial building block for $A_1$ is not plotted. 
	(a) Building block for $A_2$.
	The topological index is mirror winding number $w_{M_x} \in \mathbb{Z}$. 
	The generator with $w_{M_x} = 1 $ has a pair of Majorana zero modes at its 0D boundary (blue and orange dots), with $M_x$ eigenvalues $\pm i$ and $\mathcal{S}$ eigenvalues $\pm 1$.
	(b) Building block for $B_1$.
	The topological index is mirror winding number $w_{M_y} \in \mathbb{Z}$. 
	The generator with $w_{M_y} = 1$ has a pair of Majorana zero modes at its boundary (blue and orange dots), with $M_y$ eigenvalues $\pm i$ and $\mathcal{S}$ eigenvalues $\pm 1$.
	(c) Building block for $B_2$.
	The topological index is $C_2$ winding number $w_{C_2} \in \mathbb{Z}$. 
	The generator with $w_{C_2} = 1$ has a pair of Majorana zero modes at its boundary (blue and orange dots), with $C_2$ eigenvalues $\pm i$ and $\mathcal{S}$ eigenvalues $\pm 1$.}
\end{figure}

For $\bm{C_{2v}}$, there are four representations of the gap function, $A_1$, $B_1$, $A_2$ and $B_2$.
Meanwhile, there is only one double-valued irrep, $\bar{\Gamma}_5$.
By careful analysis, we find that for $A_1$ representation, the classification on the 1-cell is trivial, while for the remaining three representations, i.e., $A_2$, $B_1$ and $B_2$, the classifications are all $\mathbb{Z}$, with the corresponding topological index being the 1D winding numbers defined regarding either $M$ or $C_2$ sectors.
Here we summarise the three nontrivial building blocks in Fig.\ref{building blocks of LGs}, and readers interested in details can refer to Appendix.\ref{Appendix building blocks for layer groups}.

\subsection{Gluing condition\label{subsection  gluing condition}}

By decorating 1-cells with the above building blocks, we have a resulting state in 2D, which we call a decoration.
However, not all those decorations are what we need.
As the building blocks are 1D TSCs, they naturally have topologically protected Majorana zero modes at their 0D boundaries, which coincide with 0-cells.  
To make a decoration a TCSC fully gapped in the bulk, one has to ensure all the 0D zero modes at the 0-cells are gapped out, i.e., there are not ``opened boundaries'' in the interior of the decoration, which we call gluing condition at 0-cells.

\begin{figure}
	\centering
	\includegraphics[width=0.45\textwidth]{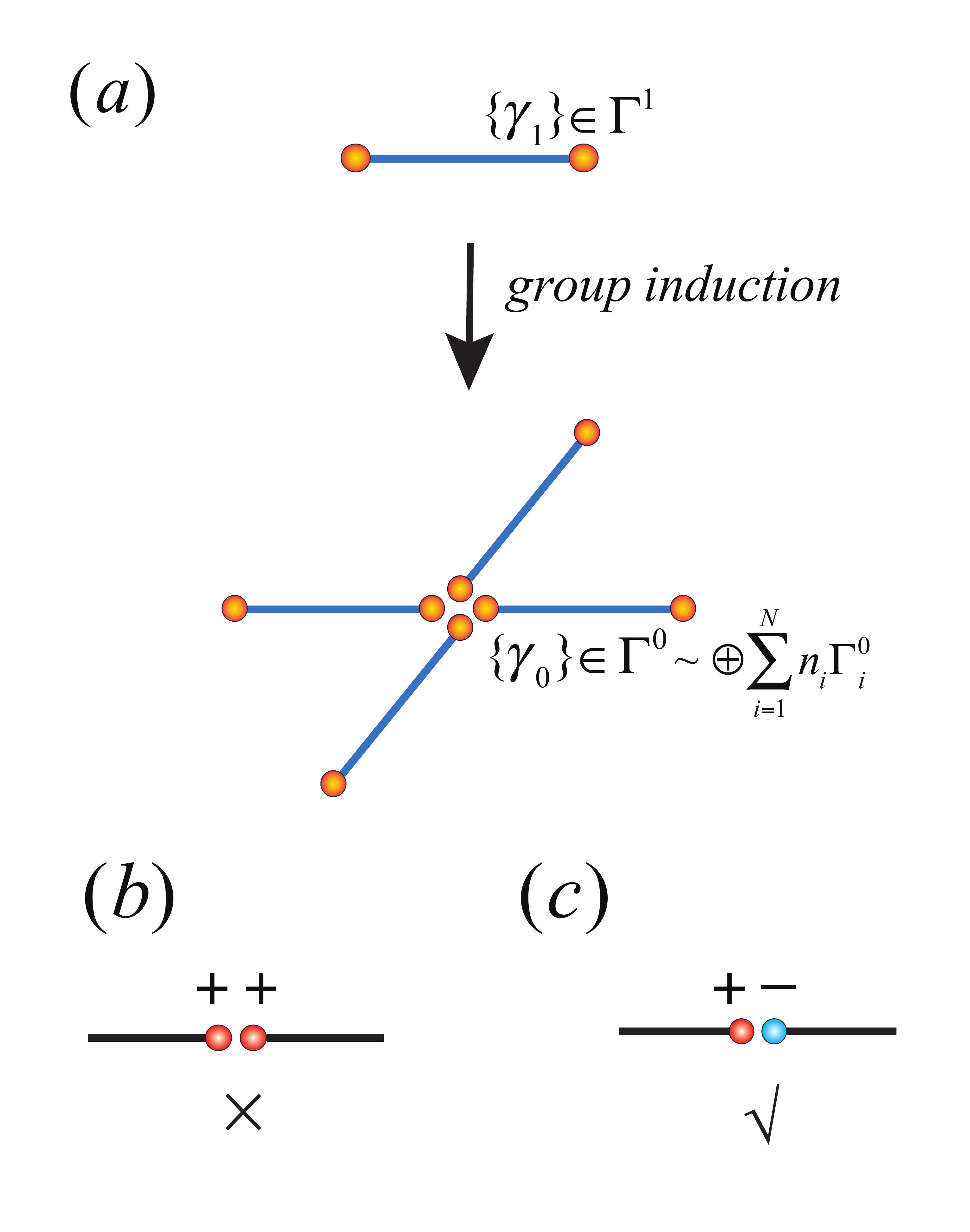}
	\caption{\label{group induction and gluing condition} 
	(a) Illustration of group induction.
	The 1-cells and 0-cell have onsite symmetry group $G^1$ and $G^0$, respectively. 
	The zero modes $\{ \gamma_1 \}$ belong to a projective representation $ \Gamma^1 $ of $G^1$, which locate at the 0D boundary of a single 1-cell. 
	There will be an induced projective representation $\Gamma^0$ of $G^0$, carried by the zero modes $\{ \gamma_0 \}$ from all the symmetry-related 1-cells which meet at the 0-cell.
	(c) and (d) Intuitive comprehension of the gluing condition.
	Two boundary modes of the same symmetry representation can not be gapped if they have the same topological number (c) and can be gapped if they have opposite topological numbers (d).}
\end{figure}

To determine whether 1-cells decorated with building blocks can be glued together, one has to carefully study the boundary theories of Majorana modes.
The crucial issue for gluing condition is to address the problem of group induction, which is illustrated in Fig.\ref{group induction and gluing condition}(a) and explained as follows.
Generally, there could be multiple 1-cells meeting at a 0-cell, and the onsite symmetry group of the 0-cell is larger than or equivalent to those of the surrounding 1-cells. 
Consider a set of symmetry-related 1-cells, each one having onsite symmetry group $G^1$, and the 0-cell that they share as a common end has onsite symmetry group $G^0$.
Then $G^1$ must be a subgroup of $G^0$, i.e., $G^1 \subseteq G^0$, and the symmetry-related 1-cells are in one-to-one correspondence to the representatives of the cosets $G^0/G^1$.
Therefore, once having a building block, i.e., choosing a representative among the symmetry-related 1-cells and decorating it by 1D TCS with 0D zero modes taking projective irreps of $G^1$, we can derive an induced projective representation of $G^0$ spanned by the entire set of zero modes from all the symmetry-related 1-cells.
In another word, the symmetry group that describes the zero modes at the 0-cell is enhanced from $G_0$ to $G_1$.
Mathematically, there is a standard formalism to derive induced representations, see Appendix.\ref{Appendix group induction}.
This group induction procedure is necessary since the boundary modes of 1-cells locate at 0-cells and therefore their corresponding onsite symmetry group is enhanced from $G^1$ to $G^0$. 
One can further imagine the 0D boundary modes are contributed by 1D TSCs with onsite symmetry group lifted from $G^1$ to $G^0$.

\begin{figure}[H]
	\centering
	\includegraphics[width=0.45\textwidth]{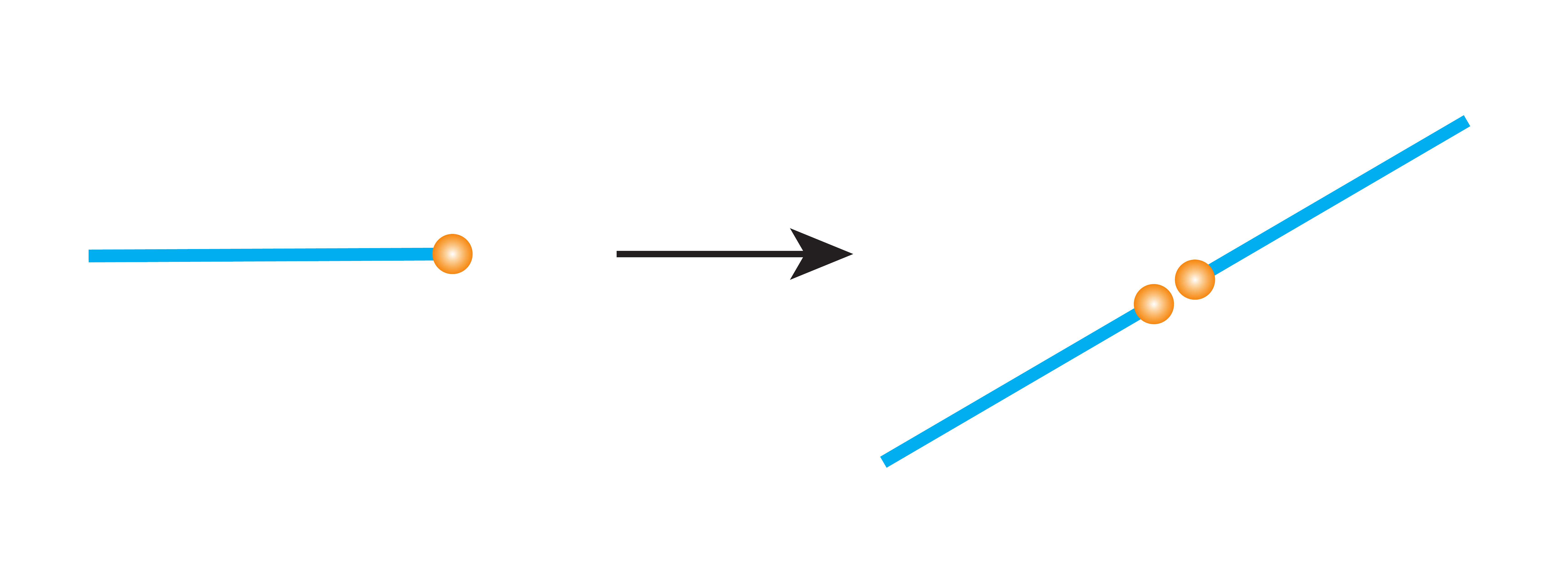}
	\caption{\label{2 1-cells related by inversion} 
	Two 1-cells related by inversion $\mathcal{I}$. Each 1-cell is decorated with 1D DIII TSC.}
\end{figure}
Here we give an intuitive example regarding inversion symmetry, as shown in Fig.\ref{2 1-cells related by inversion}.
For brevity, we ignore translation symmetries and consider only two 1-cells related by inversion symmetry $\mathcal{I}$ which intersect at the inversion center.
Each of them has trivial $C_c$, i.e., $G^1_c = \bm{1}$.
Therefore the 1-cell has a $\mathbb{Z}_2$ topological index, corresponding to the classification of DIII TCS in 1D.
However, due to the presence of $\mathcal{I}$, even putting two 1D DIII TSCs of $\mathbb{Z}_2 = 1$ together may not leave their boundary modes gapped out, although it is tempting to expect two $\mathbb{Z}_2=1$ states to result in a $\mathbb{Z}_2 = 0$ state.  
Consider $\chi_{\mathcal{I}} = 1$, which describes even parity superconductors.
We can write out the symmetry representation carried by the Majorana zero modes at the 0-cell as
\begin{align}
    & \mathcal{T} = \mu_0 i \sigma_2 \mathcal{K}
    \\
    & \mathcal{S} = \mu_0 \sigma_3
    \\
    & \mathcal{I} = \mu_1 \sigma_0
\end{align}
which satisfies $\mathcal{T}^2 = -1$, $\mathcal{S}^2 = 1$, $\mathcal{P}^2 = ( \mathcal{T} \mathcal{S})^2 = 1 $, $[ \mathcal{T}, \mathcal{I} ] = 0$, and $ [ \mathcal{S}, \mathcal{I} ] = 0 $.
To gap out these zero modes, the mass term $\mathcal{M}$ should satisfy 
\begin{align}
[ \mathcal{M}, \mathcal{T} ] = 0,  \;\;
[ \mathcal{M}, \mathcal{I} ] = 0,  \;\;
\{ \mathcal{M}, \mathcal{S} \} = 0 \;\
\end{align}
One can check there is no such mass term, therefore the zero modes fail to be gapped out.
In fact, since inversion is an onsite symmetry at the 0-cell,
the zero modes become inversion eigenstates labeled by parities $\pm 1$.
Here each parity sector is comprised of a Kramer's pair with topological number $\mathbb{Z}_2 = 1$, protecting the zero modes from being gapped.

Formally, the gluing condition can be defined as follows.
At each 0-cell, all the boundary modes contributed by 1-cells form an induced projective representation of $G^0$, which is usually reducible and can be decomposed into sectors labeled by projective irreps of $G^0_c$.
To gap out all the 0D boundary modes, within each sector, the net 1D topological number indicating the number of 0D boundary modes must vanish according to bulk-boundary correspondence, as illustrated in Fig.\ref{group induction and gluing condition}(b)-(c).
This means we need to determine the effective AZ classes as well as the corresponding topological classifications for the sectors belonging to irreps of $G^0_c$, as what we did for $G^1_c$, which can be done with the help of the Wigner test.

Mathematically, the space of decorations satisfying the gluing condition can be obtained by calculating the kernel of a linear map from all 1-cells to all 0-cells.
Suppose we have all the projective irreps of $G^1$ for all 1-cells, and we denote the $m$-th projective irrep for the $i$-th 1-cell as $\Gamma^{1}_{i,m}$.
Correspondingly, the generator of the states spanning $\Gamma^{1}_{i,m}$ is denoted as $\phi^{1}_{i,m}$, i.e., if the sector of $\Gamma^{1}_{i,m}$ has $\mathbb{Z}$ topological classification, then $\phi^{1}_{i,m}$ has $\mathbb{Z} = 1$.
Likewise, we denote the $n$-th projective irrep for the $j$-th 0-cell as $\Gamma^{0}_{j,n}$, and the generator as $\phi^{0}_{j,n}$.
By group induction, we get to know the mappings from all $ \Gamma^1 $'s to $ \Gamma^0 $'s, and the matrix elements from the linear space spanned by $\{ \phi^{1} \}$ to the space spanned by $\{ \phi^{0} \}$ are given by 
\begin{align}
\left(
        \begin{array}{cccc}
           & \vdots &  &   \\
          \cdots & c^{i,m}_{j,n} & \cdots & \cdots \\
           & \vdots &  &   \\
           & \vdots &  & 
         \end{array}
\right)
\end{align}
where $c^{i,m}_{j,n}$ is determined by the weight of $\Gamma^{0}_{j,n}$ in the decomposition of the induced projective representation from $\Gamma^{1}_{i,m}$, i.e., 
\begin{align}
    \Gamma^{1}_{i,m} \longrightarrow \Gamma^0 =  \bigoplus_{j,n} \; c^{i,m}_{j,n} \; \Gamma^{0}_{j,n}
\end{align}

\subsection{Bubble equivalence\label{subsection bubble equivalence}}

However, we have not arrived at our destination even we get all decorations allowed by gluing condition, because some of them could be unstable against a kind of real-space trivialization process called ``bubble equivalence''\cite{song2019topological,song2020real}.
In our case, ``bubbles'' are 2D objects inside 2-cells, created from a generic point and growing larger and larger, until their edges coincide with 1-cells.
Despite that its edges can be decorated with 1D TSCs, a bubble itself is intrinsically trivial as its interior is empty and it can shrink back into a point then vanish.
Because bubbles can create 1D TCSs on 1-cells by their edges, which could be equivalent to some decorations, one has to retract those unstable decorations.
As a result, the final topological classification is obtained by computing a quotient space $\mathcal{E}_1/\mathcal{E}_2$, where $\mathcal{E}_1$ stands for the space spanned by decorations satisfying gluing condition, and $\mathcal{E}_2$ for the space spanned by decorations trivialized by bubbles equivalence.

Like gluing condition, the bubble equivalence process also involves group induction. 
Consider a set of symmetry-related 2-cells (in our 2D case, at most two such bubbles, related by mirror or $C_2$) with onsite symmetry group $G^2$ intersecting at a 1-cell with onsite symmetry group $G^1$, which they share as a common edge.
Then we must have $G^2 \subseteq G^1$, as well as $G^2_c \subseteq G^1_c$.
Since a single bubble locates inside a 2-cell, it has the same onsite symmetry group as the 2-cell does, i.e., $G^2$, and the symmetry group of the entire set of symmetry-related bubbles should be enhanced to $G^1$. 
That is to say, the resulting states from the edges of surrounding bubbles on a 1-cell span an induced projective representation of $G^1$. 

\begin{figure}
	\centering
	\includegraphics[width=0.45\textwidth]{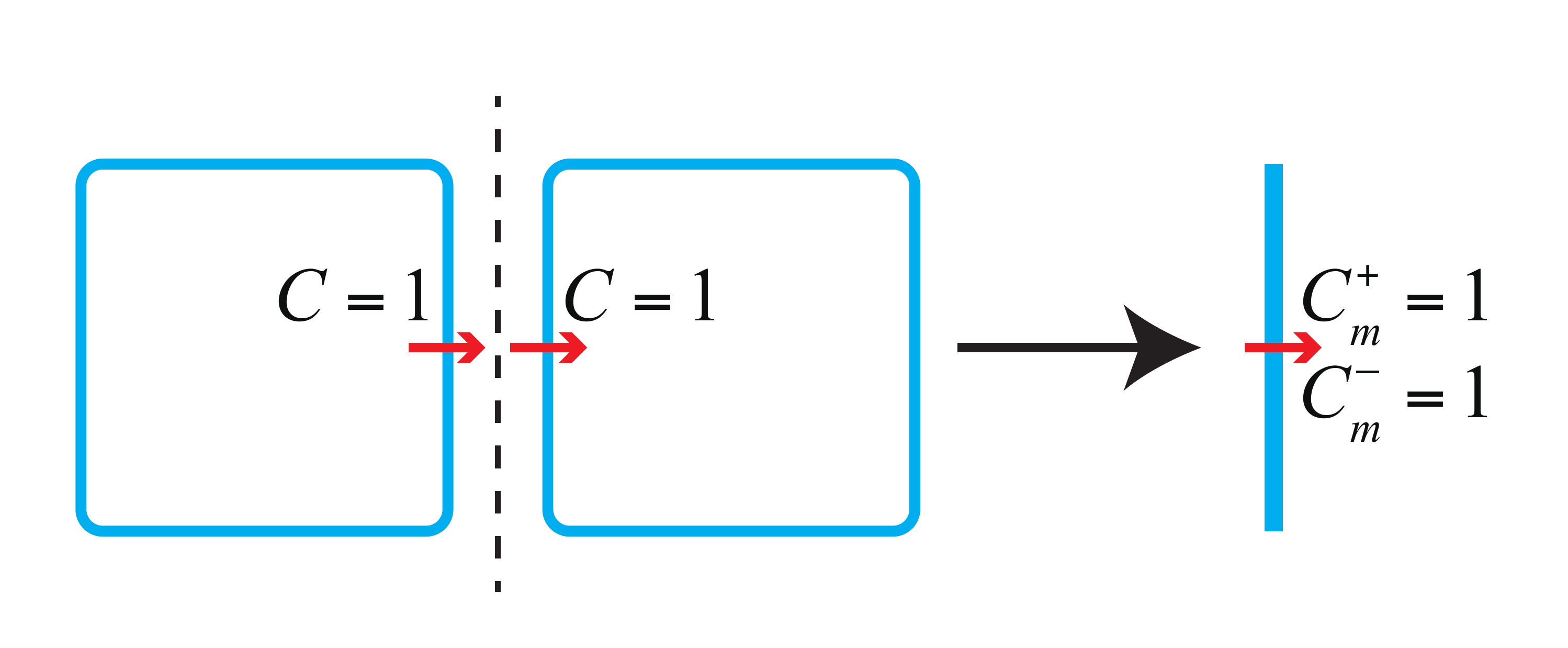}
	\caption{\label{TCI bubble equivalence} 
	Bubble equivalence process of magnetic TCI protected by mirror symmetry.
	On the 2-cell coinciding with the mirror plane, there are two bubbles related by mirror. 
	The arrow denotes the direction of Berry's curvature, thus the sign of the Chern number on the surface of bubbles, and both of the two bubbles have their surface carrying Chern number $C = 1$.
	When the two bubbles have their surface coincide with the 2-cell, the resulting states are $M$ eigenstates with $\pm i$, therefore bubble equivalence change the mirror Chern numbers on the 2-cell by $( C^{+}_m, C^{-}_m ) = ( 1, 1)$. 
    }
\end{figure}

However, to give a simple example of bubble equivalence to the readers, we turn to TCI classification in 3D, in the presence of mirror symmetry and in absence of time-reversal\cite{peng2021topological}, see Fig.\ref{TCI bubble equivalence}. 
The real-space construction can be realized by decorating a 2D mirror Chern insulator with mirror Chern numbers $( C^{+}_m, C^{-}_m )$ at the mirror plane.
Due to the absence of time-reversal, $C^{+}_m$ and $ -C^{-}_m$ are not necessarily equal, i.e., there could be nonvanishing net Chern number.
Then we show the decoration with $( C^{+}_m, C^{-}_m ) = \pm (1,1)$ can be trivialized by the generator of bubble equivalence.
Here the bubbles are 3D objects with their 2D surfaces carrying Chern numbers.
On the two sides of the mirror plane, there are two bubbles related by mirror reflection represented by an off-diagonal matrix, $ M = 
        \left(
        \begin{array}{cc}
          0 & -1  \\
          1 & 0  
         \end{array}
         \right) $, by noting the
two bubbles (as well as their surfaces) are interchanged under mirror reflection.
When the bubble surfaces coincide with the mirror plane, the resulting states become two mirror eigenstates with mirror eigenvalues $+i$ and $-i$, as can be seen by the diagonalization of $M$.
Meanwhile, since the sign of Chern number is invariant under mirror reflection by noting Berry curvature is a pseudovector, the two bubble surfaces carry the same Chern number $\pm 1$.
As a result, the two bubbles create a state of $(C^{+}_m, C^{-}_m) = \pm (1, 1)$ on the mirror 2-cell, thus trivialize all the decorations with $(C^{+}_m, C^{-}_m) = \pm (n, n)$ where $n \in \mathbb{Z}$.

Nevertheless, we find for 2D TCSCs in class DIII protected by wallpaper groups, any bubble equivalence has trivial effects, i.e., does not change the decorations on 1-cells, thus we need not consider such a process.
For a detailed analysis see Appendix\ref{Appendix trivial bubble equivalence in wallpaper groups}.

On the contrary, for layer groups, there can be nontrivial bubble equivalence due to additional symmetries.
In fact, nontrivial bubble equivalence happens if there is a vertical mirror reflection $M_z$, which is absent in wallpaper groups.

\begin{figure}
	\centering
	\includegraphics[width=0.5\textwidth]{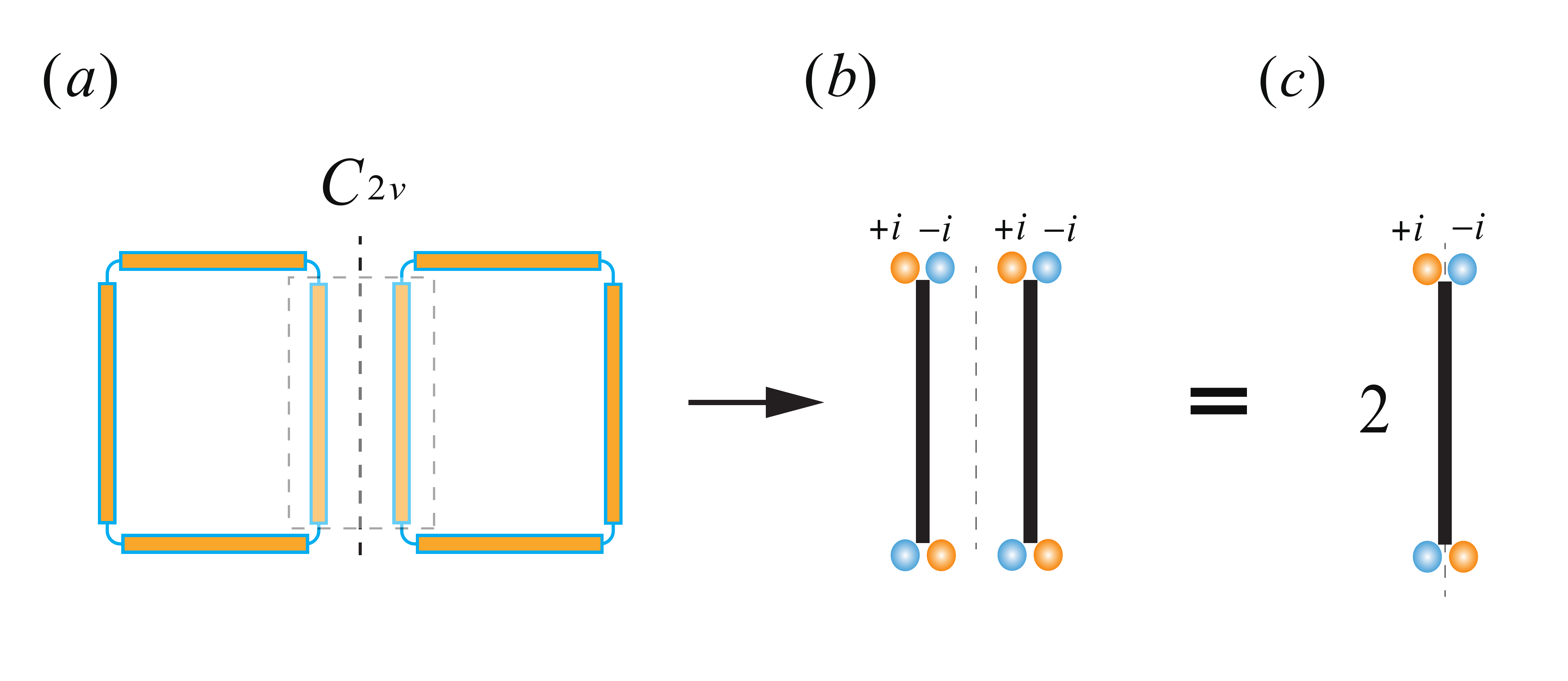}
	\caption{\label{nontrivial bubble equivalence}
	Nontrivial bubble equivalence in layer group $pm2m$.
	In (a) there are two bubbles related by $M_x$ ($C_{2y}$), each one has onsite symmetry $M_z$. 
	The resulting state by bubble equivalence is shown in (b).
	Each 1D state is a mirror-symmetric 1D TCS. 
	The blue and orange dots at the end denote the zero modes with $\mathcal{S} = \pm 1$, and their $M_z$ eigenvalues are also labeled to be $\pm i$. 
	Since the left and right states are related by $M_x$ which anti-commutes with both $M_z$ and $\mathcal{S}$, both the $M_z$ eigenvalue and $\mathcal{S}$ eigenvalue is reversed under $M_z$.
	Consequently, there left and right have the same mirror winding number $1$, and the total topological number changed by bubble equivalence is $2$.}
\end{figure}

Consider layer group $pm2m$ as an example. 
This group is generated by mirror reflection in $x$ direction, $M_x$, and vertical mirror reflection $M_z$.
The combination of them is a $C_{2y}$ rotation.
As a result, there are 1-cells that coincide with $C_{2v}$ axes at $x=0$ and $x=\frac{1}{2}$.
When the representation of gap function is chosen such that $\chi_{M_x} = -1$ and $\chi_{M_z} = 1$,
these two 1-cells can be decorated with 1D $C_{2v}$-symmetric TSC with $\mathbb{Z}$ classification.
The topological index is the mirror winding number regarding $M_z$, i.e., $w_{M_z} = \frac{1}{2} (w_{M_z}^{+} - w_{M_z}^{-}) $,
and the generator of decoration has $w_{M_z} = 1$.
Next we analysis the effect of bubbles on the $C_{2v}$ axis, as shown in Fig.\ref{nontrivial bubble equivalence}.
For brievity we only consider the $C_{2v}$ axis at $x=0$, and the effect of bubble equivalence on the $C_{2v}$ axis at $x=\frac{1}{2}$ is the same.
The bubbles have $C_c = \bm{M_z}$, and their edges are 1D mirror-symmetric TSCs with boundary zero modes described by symmetries
\begin{align}
    & \mathcal{T} = i \sigma_2 \mathcal{K}
    \\
    & \mathcal{S} = \sigma_3
    \\
    & M_z = i \sigma_3
\end{align}
When the two bubbles related by $M_x$ have their edges coinciding with the $C_{2v}$ axis, the $G_c$ of the resulting state contributed by the two bubble edges is enhanced from $\bm{M_z}$ to $\bm{C_{2v}}$, and by group induction the symmetries are represented as  
\begin{align}
    & \mathcal{T} = \mu_0 i \sigma_2 \mathcal{K}
    \\
    & \mathcal{S} = \mu_3 \sigma_3
    \\
    & M_z = \mu_3 i \sigma_3
    \\
    & M_x = i \mu_2 \sigma_0
\end{align}
which satisfy $[ M_z, \mathcal{S} ] = 0$ and $\{ M_x, \mathcal{S} \} =0$.
We can diagonalize $M_z$ and $\mathcal{S}$ simultaneously, leading to
\begin{align}
M_z = \left(
        \begin{array}{cccc}
          i & 0 & 0 & 0 \\
          0 & -i & 0 & 0 \\
          0 & 0 & i & 0 \\
          0 & 0 & 0 & -i           
         \end{array}
         \right),
         \; \;
\mathcal{S} = \left(
        \begin{array}{cccc}
          1 & 0 & 0 & 0 \\
          0 & -1 & 0 & 0 \\
          0 & 0 & 1 & 0 \\
          0 & 0 & 0 & -1           
         \end{array}
         \right)         
\end{align}
Obviously, the resulting state are described by two copies of $\bar{\Gamma}_5$, i.e., the only double-valued irrep of $\bm{C_{2v}}$, with $w_{M_z} = 1$, and thus $w_{M_z} = 2$ in total, which is exactly the double of the generator of $C_{2v}$-axis decoration.
Therefore we can see that the bubbles leave a state of $w_{M_z} = 2$ on the 1-cell, trivializing all the decorations with $w_{M_z} = 2 \mathbb{Z}$ and reducing the classification on the 1-cell from $\mathbb{Z}$ to $\mathbb{Z}_2$.

\section{Example: $p4$, $A_g$ representation\label{Section example P4}}

Here we use wallpaper group $p4$ with the representation of gap function $A_g$ ($\chi_{C_4} = 1$) to show the complete procedure of wire construction.
There is another example for wallpaper group in Appendix.\ref{Appendix example of wallpaper group}, as well as an example for layer group in Appendix.\ref{Appendix example of layer goup}.

\begin{figure}
	\centering
	\includegraphics[width=0.5\textwidth]{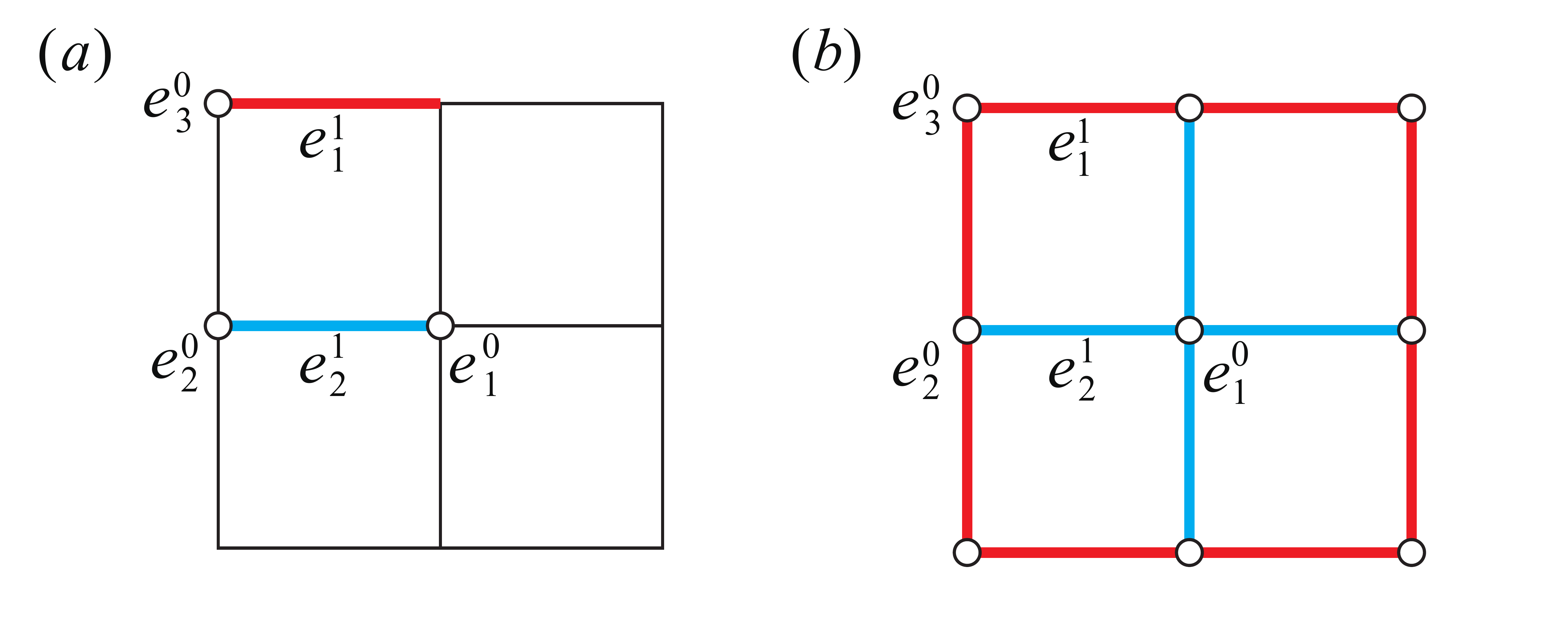}
	\caption{\label{cell complex of P4} The cell complex of wallpaper group $p4$.
	(a) There are two independent 1-cells, colored by blue and red, respectively, and denoted as $e^1_1$ and $e^1_2$. 
	There are three independent 0-cells, denoted as $e^0_1$, $e^0_2$ and $e^0_3$, among them $e^0_1$ and $e^0_3$ have $G^0_c = \bm{C_4}$, while $e^0_2$ has $G^0_c = \bm{C_2}$.
	We select the original point, i.e., the $C_4$ center, at $e^0_1$. 
	(b) We have all the 1-cells colored. }
\end{figure}

The cell complex of $p4$ is shown in Fig.\ref{cell complex of P4}.
All the three 1-cells have $G^1_c = \bm{1}$ and can be decorated with 1D DIII TSC with $\mathbb{Z}_2$ classification,
thus there are three independent decorations before considering gluing condition.

For convenience of analyzing gluing condition, we first look into the projective irreps of $C^0$, which in this case are actually equivalent to linear irreps since $\chi_{C_4} = 1$.
The irreps of $G_c^0$ are given in the following table.
Among the 0-cells, $e^0_1$ and $e^0_3$ have $G_c^0 = \bm{C_4}$, and $e^0_2$ has $G_c^0 = \bm{C_2}$.
The relevant double-valued irreps are presented in the following table.

\begin{table}[H]
\centering
    \begin{tabular}{cccccc|ccc}
        \hline \hline 
        & $\bar{\Gamma}_5$  & $\bar{\Gamma}_6$ & $\bar{\Gamma}_7$ & $\bar{\Gamma}_8$ & &  & $\bar{\Gamma}_3$ & $\bar{\Gamma}_4$
        \\
        \hline 
        $\bm{C_4}$
        & $e^{i\frac{3}{4}\pi}$ & $e^{-i\frac{\pi}{4}}$ & $e^{-i\frac{3}{4}\pi}$ & $e^{i\frac{\pi}{4}}$ & & $\bm{C_2}$ &
        $i$ & $-i$
        \\
        \hline \hline
    \end{tabular}
    \caption{\label{irreps of $C_4$ and $C_2$} 
     Double-valued irreducible representations of $\bm{C_4}$ and $\bm{C_2}$}
\end{table}          
        
Applying the Wigner test to the irreps of $\bm{C_4}$, we get
\begin{align}
     [ W^{\alpha}(\mathcal{T}),  W^{\alpha}(\mathcal{P}), W^{\alpha}(\mathcal{S}) ]  = [0,0,1]
\end{align}
for $\alpha = \bar{\Gamma}_{i,i=1,2,3,4}$, which means each irrep of $\bm{C_4}$ belongs to class AIII characterised by the 1D winding number $w \in \mathbb{Z}$, and each of them is closed under $\mathcal{S}$ but paired with its conjugate irrep, $\bar{\Gamma}_i^*$, under $\mathcal{T}$ and $\mathcal{P}$.
It is easy to see $\bar{\Gamma}_5$, $\bar{\Gamma}_7$ are paired, and $\bar{\Gamma}_6$, $\bar{\Gamma}_8$ are paired.
We denote the resulting projective irreps as $\bar{\Gamma}_{5 \oplus 7}$ and $\bar{\Gamma}_{6 \oplus 8}$, and
write out them explicitly as 
\begin{align}
       &  \mathcal{T} = i \sigma_2 \mathcal{K}
         \\
       &  \mathcal{S} =  \sigma_3
         \\
       &  C_4 = \left(
        \begin{array}{cc}
          e^{i\frac{3}{4}\pi} & 0   \\
          0 & e^{-i\frac{3}{4}\pi} 
         \end{array}
         \right)  = e^{i\frac{3}{4}\pi\sigma_3}
     \end{align}
for $\bar{\Gamma}_{5\oplus 7}$, and 
     \begin{align}
     &    \mathcal{T} = i \sigma_2 \mathcal{K}
         \\
      &   \mathcal{S} =  \sigma_3
         \\
     &    C_4 = \left(
        \begin{array}{cc}
          e^{i\frac{\pi}{4}} & 0   \\
          0 & e^{-i\frac{\pi}{4}} 
         \end{array}
         \right) = e^{i\frac{\pi}{4}\sigma_3}
\end{align}
for $\bar{\Gamma}_{6\oplus8}$.
It is obvious that the sectors belong to $\bar{\Gamma}_5$ and $\bar{\Gamma}_7$ have opposite winding numbers, as well as for the sectors belong to $\bar{\Gamma}_6$ and  $\bar{\Gamma}_8$.
We can define the independent topological index from the winding numbers regarding $C_4$ sectors, for $\bar{\Gamma}_{5\oplus7}$ 
\begin{align}
w_{C_4} = \frac{1}{2} ( w_{ e^{i\frac{3}{4}\pi}} - w_{ e^{-i\frac{3}{4}\pi}} )
\end{align}
and for $\bar{\Gamma}_{6\oplus8}$
\begin{align}
w_{C_4} = \frac{1}{2} ( w_{ e^{i\frac{\pi}{4}}} - w_{ e^{-i\frac{\pi}{4}}} ) 
\end{align}
Likewise, we apply the Wigner test to the irreps of $\bm{C_2}$ and get
\begin{align}
     [ W^{\alpha}(\mathcal{T}),  W^{\alpha}(\mathcal{P}), W^{\alpha}(\mathcal{S}) ]  = [0,0,1]
\end{align}
for $\alpha = \bar{\Gamma}_{i,i=3,4}$, which means each irrep of $\bm{C_2}$ also belongs to AZ class AIII characterised by the 1D winding number $w \in \mathbb{Z}$.
We denote the resulting projective irrep as $\bar{\Gamma}_{3 \oplus 4}$, and write it out explicitly as  
\begin{align}
     &  \mathcal{T} = i \sigma_2 \mathcal{K}
         \\
    &\mathcal{S} =  \sigma_3
         \\
    & C_2 = i \sigma_3
\end{align}
It is easy to see the two sectors of $C_2 = \pm i$ have opposite winding numbers.
We can define the independent topological index as 
\begin{align}
    w_{C_2} = \frac{1}{2} ( w_{+i} - w_{-i} )
\end{align} 
    
Now we are prepared to consider gluing condition for each 1-cell decoration.    
    
\begin{enumerate}
    
    \item $e^1_1$ decoration
    
    From the cell complex, it is easy to see that $e^1_1$ and its symmetry-related partners only intersect at the 0-cells $e^0_3$ and $e^0_2$, but not at $e^0_1$. Thus we only have to consider gluing condition at $e^0_3$ and $e^0_2$.
    
    \begin{enumerate}
        
    \item gluing condition at $e^0_3$
    
    We have symmetry representation for the zero modes at the boundary of $e^1_1$ as
    \begin{align}
        & \mathcal{T} = i\sigma_2 \mathcal{K}
        \\
        & \mathcal{S} = \sigma_3 \mathcal{K}        
    \end{align}
    By group induction, we have the induced representation at $e^0_1$:
    \begin{align}
       & \mathcal{T} = s_0 \mu_0 i \sigma_2 \mathcal{K}
        \\
      &  \mathcal{S} = s_0 \mu_0 \sigma_3
        \\        
      &  C_4 = \left(
        \begin{array}{cccc}
          0 & 1 & 0 & 0  \\
          0 & 0 & 1 & 0 \\
          0 & 0 & 0 & 1 \\
          -1 & 0 & 0 & 0
         \end{array}
         \right) 
         \otimes
         \sigma_0
    \end{align}    
    As $[ C_4, \mathcal{S} ] = 0$, we can simultaneously diagonalize them and get
    \begin{align}
       & C_4 = \sigma_0 \otimes
        \left(\begin{matrix}
        e^{i\frac{3}{4}\pi} & 0 & 0 & 0 
        \\
        0 & e^{-i\frac{3}{4}\pi} & 0 & 0 
        \\
        0 & 0 & e^{i\frac{\pi}{4}} & 0 
        \\
        0 & 0 & 0 & e^{-i\frac{\pi}{4}} 
        \end{matrix}\right)
        \\
      &  \mathcal{S} = \sigma_3 \otimes
        \left(\begin{matrix}
        1 & 0 & 0 & 0 
        \\
        0 & 1 & 0 & 0 
        \\
        0 & 0 & 1 & 0 
        \\
        0 & 0 & 0 & 1 
        \end{matrix}\right)
    \end{align}
    By decomposition, we can see there are two $\bar{\Gamma}_{5\oplus7}$'s and two $\bar{\Gamma}_{6 \oplus 8}$'s to describe the zero modes.
    Explicitly, the two $\bar{\Gamma}_{5\oplus7}$'s are given by
    \begin{align}
        C_4 =  \left(\begin{matrix}
        e^{i\frac{3}{4}\pi} & 0
        \\
        0 & e^{-i\frac{3}{4}\pi}
        \end{matrix}\right), 
        \;\;
        \mathcal{S} = 
        \left(\begin{matrix}
        1 & 0  
        \\
        0 & -1 
        \end{matrix}\right)     
    \end{align}
    and 
        \begin{align}
        C_4 =  \left(\begin{matrix}
        e^{i\frac{3}{4}\pi} & 0
        \\
        0 & e^{-i\frac{3}{4}\pi}
        \end{matrix}\right), 
        \;\;
        \mathcal{S} = 
        \left(\begin{matrix}
        -1 & 0  
        \\
        0 & 1 
        \end{matrix}\right)     
    \end{align}
    The first one has $w_{C_4}  = 1$ while the second one has $w_{C_4} = -1$, thus the net topological number by them is zero.
    Likewise, the two $\bar{\Gamma}_{6\oplus8}$'s are given by
    \begin{align}
        C_4 =  \left(\begin{matrix}
        e^{i\frac{\pi}{4}} & 0
        \\
        0 & e^{-i\frac{\pi}{4}}
        \end{matrix}\right), 
        \;\;
        \mathcal{S} = 
        \left(\begin{matrix}
        1 & 0  
        \\
        0 & -1 
        \end{matrix}\right)     
    \end{align}
    and 
        \begin{align}
        C_4 =  \left(\begin{matrix}
        e^{i\frac{\pi}{4}} & 0
        \\
        0 & e^{-i\frac{\pi}{4}}
        \end{matrix}\right), 
        \;\;
        \mathcal{S} = 
        \left(\begin{matrix}
        -1 & 0  
        \\
        0 & 1 
        \end{matrix}\right)     
    \end{align}
    with the first one having $w_{C_4}  = 1$ and the second one having $w_{C_4} = -1$, thus the net topological number is also zero.
    
    We see that both the sector of $\bar{\Gamma}_{5\oplus7}$ and the sector of $\bar{\Gamma}_{6\oplus8}$ have vanishing net topological numbers, thus the gluing condition at $e^0_3$ is met.

    \item gluing condition at $e^0_2$
    
    As what we did above, we have the induced representation at $e^0_2$ by group induction, i.e.,
    \begin{align}
      &  \mathcal{T} = \mu_0 i \sigma_2 \mathcal{K}
        \\
    &    \mathcal{S} = \mu_0 \sigma_3
        \\        
     &   C_2 = i \mu_2
         \sigma_0
    \end{align}    
    After simultaneous diagonalization of $C_2$ and $\mathcal{S}$, we have
    \begin{align}
        & C_2 = 
        \left(\begin{matrix}
        i & 0 & 0 & 0 
        \\
        0 & -i & 0 & 0 
        \\
        0 & 0 & i & 0 
        \\
        0 & 0 & 0 & -i 
        \end{matrix}\right)
        \\
        & \mathcal{S} =
        \left(\begin{matrix}
        1 & 0 & 0 & 0 
        \\
        0 & 1 & 0 & 0 
        \\
        0 & 0 & \text{-}1 & 0 
        \\
        0 & 0 & 0 & \text{-}1 
        \end{matrix}\right)
    \end{align}   
    This representation is reducible and can be decomposed in two $\bar{\Gamma}_{3\oplus4}$'s, with the first one given by 
    \begin{align}
        C_2 = 
        \left(\begin{matrix}
        i & 0  
        \\
        0 & -i 
        \end{matrix}\right), \;\;
        \mathcal{S} =
        \left(\begin{matrix}
        1 & 0 
        \\
        0 & -1 
        \end{matrix}\right)    
    \end{align}
    and the second one given by
    \begin{align}
        C_2 = 
        \left(\begin{matrix}
        i & 0  
        \\
        0 & -i 
        \end{matrix}\right), \;\;
        \mathcal{S} =
        \left(\begin{matrix}
        -1 & 0 
        \\
        0 & 1 
        \end{matrix}\right)    
    \end{align}   
    We can see the first one has $w_{C_2} = 1$ while the second one has $w_{C_2} = -1$, thus the net topological number is zero. 
    The vanishing topological number indicates the zero modes at $e^0_2$ can be gapped, therefore the gluing condition is satisfied. 
    
\end{enumerate}
    
    \item $e^1_2$ decoration
    
    This decoration is essentially the same with $e^1_1$, and the gluing condition is also satisfied at $e^0_2$ and $e^0_1$.
    
    \end{enumerate}

Through above analysis, we see that the decorations of $e^1_1$ and $e^1_2$ are allowed, both being $\mathbb{Z}_2$-type and serving as the two generators for all TCSCs by wire construction.
As bubble equivalence is trivial for wallpaper groups,
the final classification is $\mathbb{Z}_2^2$.

\section{wallpaper groups vs layer groups\label{section wallpaper groups vs layer groups}}

In the introduction, we mentioned that wallpaper groups (WGs) sometimes can not give a full description of symmetries for spin-$\frac{1}{2}$ particles in 2D, and for a complete description one should refer to layer groups (LGs).
We gave the example of $p2$ and $p\bar{1}$, which share the same lattice though, have different topological classifications of DIII TCSCs because $C_2$ and $\bar{I}$ are implemented differently on spinful electrons. 
In Sec.\ref{subsection  gluing condition}, we showed that $\bar{I}$ with $\chi_{I} = 1$ prohibits nontrivial decorations as gluing condition is not satisfied.
Below we show it is different for $C_2$.
\begin{figure}[H]
	\centering
	\includegraphics[width=0.45\textwidth]{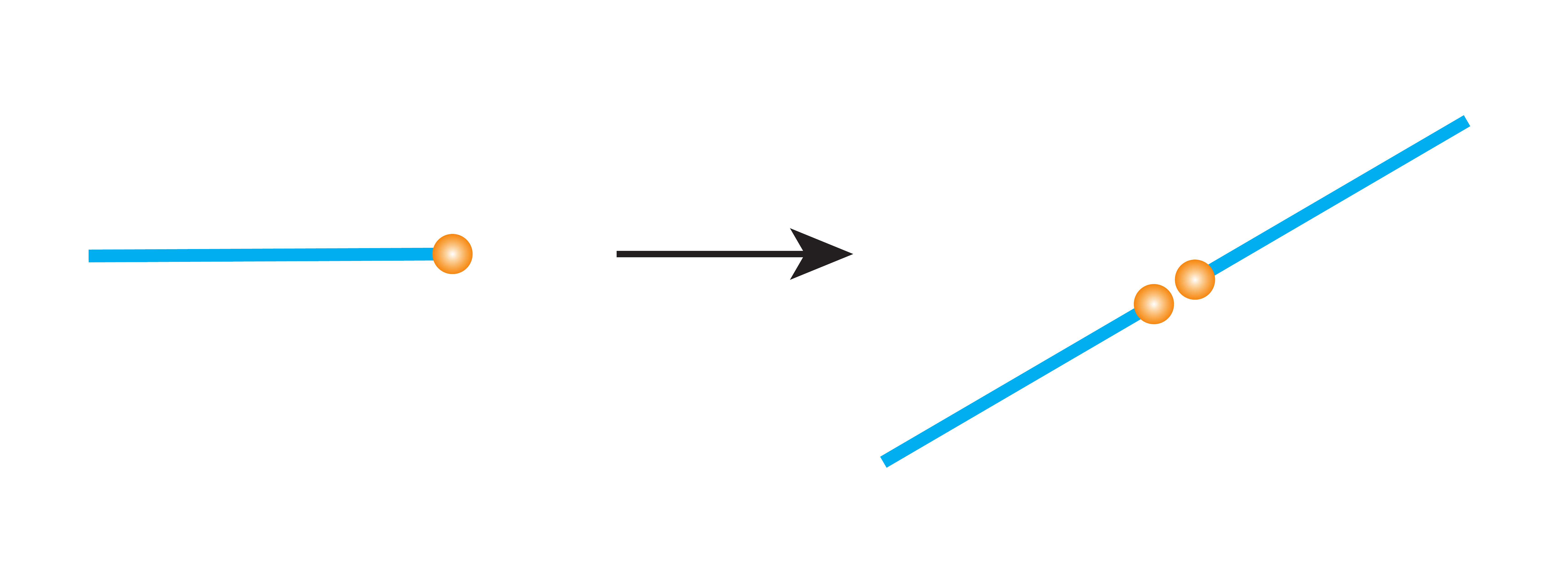}
	\caption{\label{WG vs LG} 
	Two 1-cells related by $\bar{1}$ or $C_2$. Each 1-cell is decorated with 1D DIII TSC.}
\end{figure}
For comparison, we consider $\chi_{C_2} = 1$.
The symmetries for the 0D zero modes contributed by the two $C_2$-related 1-cells are represented as
\begin{align}
    & \mathcal{T} = \mu_0 i \sigma_2 \mathcal{K}
    \\
    & \mathcal{S} = \mu_0 \sigma_3
    \\
    & C_2 = \mu_1 i \sigma_3
\end{align}
The mass term proportional to $\mu_2 \sigma_1$ commutes with $C_2$ and $\mathcal{T}$ while anti-commutes with $\mathcal{S}$, and therefore can be added to gap out these zero modes. 
As bubble equivalence is not needed for wallpaper groups, the success of gluing condition for $C_2$ case indicates nontrivial TCSC classification. 
By contrast, the failure of gluing condition for $\mathcal{I}$ case indicates trivial topological classification for even parity superconductors, and only odd-parity superconductors can be topologically nontrivial\cite{nelson2004odd,fu2010odd,sato2010topological}. 

It could be helpful for one to be clear with the relationships between WGs and LGs.
In fact, each LG has one or more corresponding WGs which share the same 2D lattice as this WG, and the number of LGs (80) is much larger than that of WGs (17).
Furthermore, among those LGs corresponding to the same WG, one of them shares completely the same group elements with the WG, and the remaining ones either have the same number of group elements although some elements differ, or can be generated from this WG by adding inversion or reflection (glide) in the vertical direction to the 2D plane.
For instance, for the WG $p4mm$ with all symmetry elements defined in $x\text{-}y$ plane, LG $p4mm$ is completely the same with it, LG $p422$ has the same elements except that mirrors are replaced by $C_2$ rotations, and LG $p4mmm$ is generated from $p4mm$ by adding mirror reflection in $z$ direction. 

It is worth noting that, $p4mm$ and $p422$ share the same results of TCSC classification by wire constrution, while $p4mmm$ is more complicated due to the inclusion of $M_z$ and its TCSC classification can not be inferred from that of $p4mm$.

\section{Classification results\label{Section classification results}}

Using wire construction, we obtain the classification of TCSCs protected by spacial symmetries in 2D, and we present the results as tables in Appendix.\ref{Appendix classification tables}.
Specifically, in TABLE.\ref{table classification WGs}, we show TCSC classification by wire construction for all WGs with all representations of gap function considered.
In TABLE.\ref{table WGs LGs}, we list the LGs which share the same wire constructions thus TCSC classifications as those of LGs.
In TABLE.\ref{table classification LGs}, we show the TCSC classification for a portion of LGs that can not be inferred directly from the results of WGs.

\section{Invariants and Boundary states\label{Section boundary states}}

When employing real-space method, one usually has the benefits that topological invariants and boundary states of the output states, i.e., those by real-space construction, are relatively easy to see.
One can usually define topological invariants in real-space, and for each decoration, identify these topological invariants.
In real-space construction of TCIs\cite{song2019topological}, for each crystalline symmetry operator $g$, its invariant is defined in the following way.
Choose an generic point $\mathcal{O}$ in a 3-cell, act it with $g$ and one gets the image point $g \mathcal{O}$. 
Then draw an arbitrary path connecting $\mathcal{O}$ and $g \mathcal{O}$, and the invariant is determined by the 2-cell decorations, which are actually 2D TIs or (mirror) Chern insulators, that the path crosses.  
Here in our wire construction for 2D TCSCs, one could define invariants in a similar way.

Apart from topological invariants, the TCSCs by wire construction can be characterized in terms of their anomalous boundary states, in one-to-one correspondence with their invariants.
We deal with decorations protected by single representative 2D spacial symmetries, including translation, $C_n$ rotation, mirror reflection, and glide.
For each independent generator of decorations, we figure out what its boundary states can be.
If there are more complicated symmetries, i.e., any kinds of combinations of them, the boundary states can be comprehended as a superimposition of those protected by single symmetry.

Through our careful analysis, we identify some boundary states which are not touched by previous works, as shown in the following two cases.
For boundary states protected by other symmetries, see Appendix.\ref{Appendix other boundary states}.

\subsubsection{wallpaper group $p1m1$, $\chi_{M} = -1$\label{ subsection boundary state Pm chi_M = -1 }}

\begin{figure}
	\centering
	\includegraphics[width=0.5\textwidth]{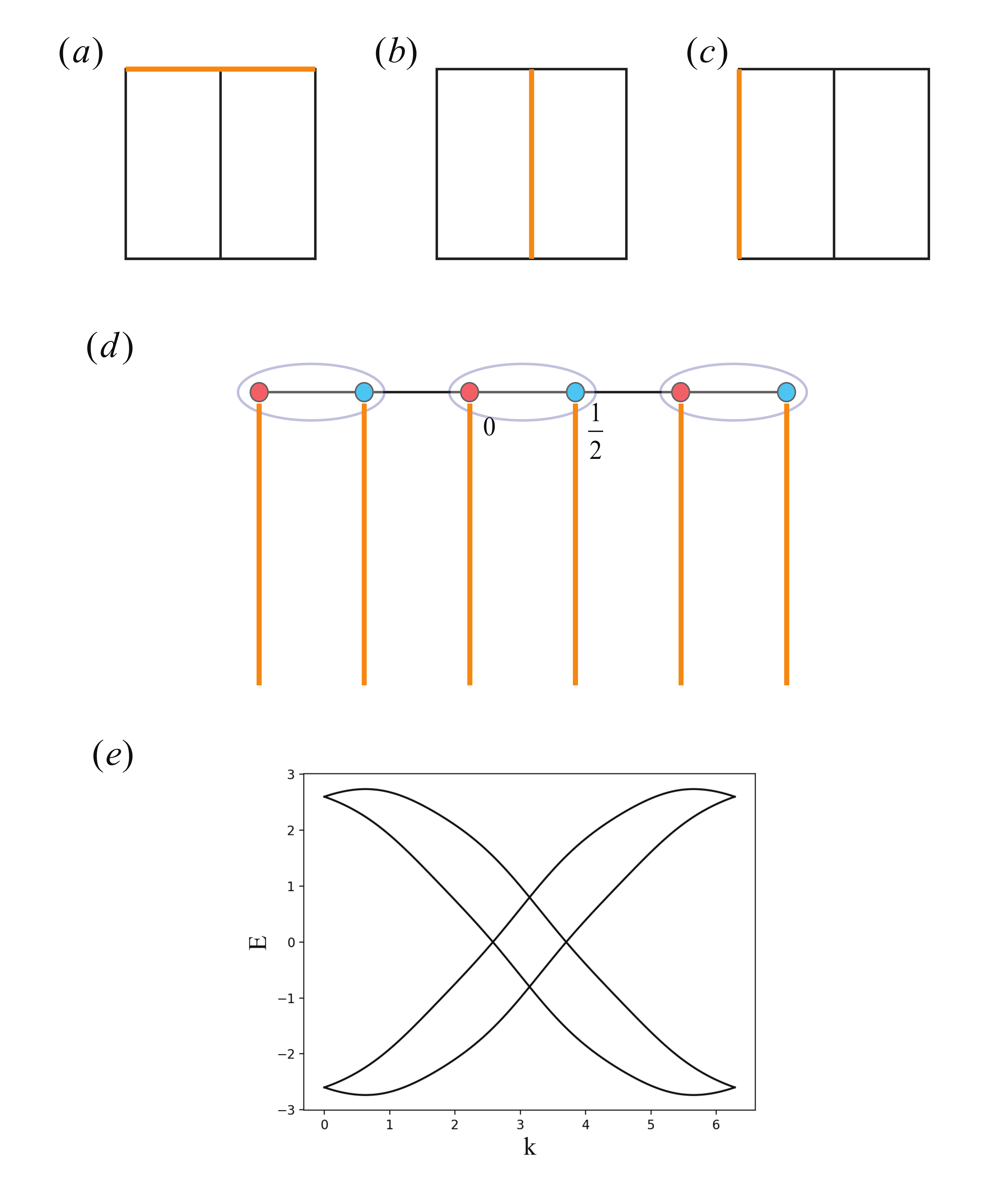}
	\caption{\label{boundary state Pm -1} 
	(a) The decoration protected by $y$-translation.
	(b)-(c) The decorations on mirror lines at $x=0$ and $x=-\frac{1}{2}$, respectively.
	(d) The boundary Majorana zero modes of the decoration by the superimposition of (b) and (c).
	The two zero modes in the same ellipse are in the same unit cell.  
	(e) Boundary spectrum along $k_x$ with the parameters set to be $d_1 = 0.3, \; e_1 = 0.3, \; f_1 = 1.3$ and $g_1 = 0.4$. 
	There are two generic gapless points, which are symmetric about $k_x = \pi$.}
\end{figure}

The generators of wire construction are shown in Fig.\ref{boundary state Pm -1}(a)-(c).
Here we focus on the boundary state of the decoration as a superposition of (b) and (c), which is kind of tricky.

Firstly one notes that since $\{ M, \mathcal{S} \} = 0 $, both the mirror-invariant lines in BZ, i.e., $k_x = 0$ and $k_y = \pi$, have effective AZ class D, with $\mathbb{Z}_2$ classification in 1D.
Both the two generators in Fig.\ref{boundary state Pm -1} (b) and (c), i.e., the one decorated with 1D mirror-symmetric TSC at $x=0$ and one decorated at $x=\frac{1}{2}$, have the same invariant $\mathbb{Z}_2 = 1$ at $k_x =0$ as well as at $k_x = \pi$.
Therefore, their superimposition, i.e., decorating at both $x = 0$ and $x =\frac{1}{2}$, has $\mathbb{Z}_2 = 0$ at $k_x = 0$ as well
as $\mathbb{Z} = 0$ at $k_x = \pi$.
Does the failure of description by the $\mathbb{Z}_2$ invariants in momentum space means this decoration is trivial? 
Otherwise, there should be another nonzero invariant accounting for this decoration.

To answer this question we carefully analyze the boundary state of this decoration.
Consider an edge respecting mirror symmetry as well as translation symmetry in $x$ direction, see Figure.\ref{boundary state Pm -1}(d).
We denote the Majorana zero modes as $\gamma_{i s} (n)$, where $i = a, b$ for the subsite in a unit cell at $x = N$ or $x= N + \frac{1}{2} $ where $N \in \mathbb{Z}$, $s = \uparrow, \downarrow$ for the ``spin'', and $n$ for the index of unit cell.
A generic tight-binding Hamiltonian can be written as 
\begin{align}
    \mathcal{H} = \sum_{n,n'} \sum_{i,j} \sum_{s,t} i F_{isjt}(n-n') \gamma_{is}(n) \gamma_{jt}(n') 
\end{align}
with $F$ being the hopping amplitude.
Before proceeding, we specify the symmetries of the Majorana zero modes.
We choose 
\begin{align}
    & \mathcal{T} = i \sigma_2 \mathcal{K}
    \\
    & \mathcal{P} = \mathcal{K}
    \\
    & M = \sigma_3  
\end{align}
Note here we take another convention with $M^2 = 1$ and $\{ M, \mathcal{T} \} = 0$, which is equivalent to the convention with $M^2 = -1$ and $[ M, \mathcal{T} ] = 0$ that we used before. 
Time-reversal symmetry requires
\begin{align}
    & F_{i\uparrow j\uparrow} (n-n') = -F_{i\downarrow j\downarrow} (n-n')
    \\
    & F_{i\uparrow j\downarrow} (n-n') = F_{i\downarrow j\uparrow} (n-n')
\end{align}
Mirror symmetry requires
\begin{align}
    & F_{i s i t} (n-n') = st F_{i s i t} (n'-n)
    \\
    & F_{a s b t} (n-n') = st F_{a s b t} (n'-n+1)    
\end{align}
Fermion statistics requires
\begin{align}
    F_{isjt} (n-n') = -F_{jtis} (n'-n)
\end{align}
As a result, the hopping amplitudes are constrained as
\begin{align}
    F_{i s i s}(n) & = 0
    \\
     F_{a\uparrow a\downarrow}(n)
    & = -F_{a\uparrow a\downarrow}(-n)
    = F_{a\downarrow a \uparrow}(n)
    \\
    & = -F_{a\downarrow a \uparrow}(-n) = d_n
    \\
     F_{b\uparrow b\downarrow}(n)
    & = -F_{b\uparrow b\downarrow}(-n)
    = F_{b\downarrow b \uparrow}(n)
    \\
    & = -F_{b\downarrow b \uparrow}(-n) = e_n   
    \\
     F_{a\uparrow b\uparrow}(n)
    & = F_{a\uparrow b\uparrow}(-n+1)
    = -F_{a\downarrow b\downarrow}(n)
    \\
    & = -F_{a\downarrow b\downarrow}(-n+1) = f_n
    \\
     F_{a\uparrow b\downarrow}(n)
    & = -F_{a\uparrow b\downarrow}(-n+1)
    = F_{a\downarrow b \uparrow}(n)
    \\
    & = -F_{a\downarrow b \uparrow}(-n+1) = g_n   
\end{align}
Because we choose $\mathcal{P} = \mathcal{K}$, the Majorana operators are real and $F$'s must be real as well. 
Keeping only the nearest hoppings, i.e., $n=1$, the Hamiltonian in momentum space reads
\begin{align}
    \mathcal{H} = \sum_k  
    \left(
        \begin{array}{cccc}
          \gamma_{a\uparrow}  &
          \gamma_{a\downarrow} &
          \gamma_{b\uparrow} &
          \gamma_{b\downarrow}
         \end{array}
         \right)
        H(k)
        \left(
        \begin{array}{c}
          \gamma_{a\uparrow}  \\
          \gamma_{a\downarrow} \\
          \gamma_{b\uparrow} \\
          \gamma_{b\downarrow}
         \end{array}
         \right)
\end{align}
where $H(k) = $
\begin{align}
    \left(
    \begin{array}{cccc}
    0 & -d_k & i f_1 (1+e^{ik}) & i g_1 (1-e^{ik})  
    \\
    -d_k & 0 & i g_1 (1-e^{ik}) & -i f_1 (1+e^{ik}) 
    \\
    -if_1 (1+e^{-ik}) & -i g_1 (1-e^{-ik}) & 0 & -e_k 
    \\
    -i g_1 (1-e^{-ik}) & i f_1 (1+e^{-ik}) & -e_k & 0
    \end{array}
    \right)
\end{align}
and 
\begin{align}
    d_k = d_1 \sin k
    \\
    e_k = e_1 \sin k
\end{align}
Below we show the spectrum is always gapless at some generic point between $k=0$ and $k=\pi$ (as well as between $k=\pi$ and $k = 2 \pi$, by symmetry).
First we set $k=0$ and we have
\begin{align}
    \text{det}(H(k=0)) = -16 f_1^4 \leq 0 
\end{align}
Then we set $k=\pi$ and we have 
\begin{align}
    \text{det}(H(k=\pi)) = 16 g_1^4 \geq 0
\end{align}
Since the determinant of $H(k)$ is a continuous function of $k$, according to the mean value theorem, there must exist at least one point $k_0$ between $0$ and $\pi$ satisfying $\text{det}( H(k=k_0) ) = 0$, which indicates the spectrum of $H(k_0)$ is gapless.

For illustration, we plot the boundary spectrum in Fig.\ref{boundary state Pm -1}(e) where $ 0 < k < 2 \pi$, with the parameters chosen as $d_1 = 0.3, \; e_1 = 0.3, \; f_1 = 1.3$ and $g_1 = 0.4$.

Through the above discussion, we showed the existence of gapless boundary modes, which confirms that the decoration by wires at both $x =0$ and $x= \frac{1}{2}$ is topologically nontrivial.
To further understand the nontrivial topology, we note a generic momentum at $k_x$ line has effective AZ class BDI, because one has $\mathcal{T}' = \mathcal{T} \cdot M_x$ and $\mathcal{P}' =\mathcal{P} \cdot M_x $, both of which leaving any $k_x$ unchanged and satisfying $\mathcal{T}'^2 = 1$, $\mathcal{P}'^2 = 1$. 
As class BDI has $\mathbb{Z}_2$ topological invariant in $0$ D, the existence of gapless mode at $k_x = k_0$ indicates different such $\mathbb{Z}_2$ values on the two sides of $k_0$.

\subsubsection{wallpaper group $p1g1$, $\chi_{g} = \pm 1$}

\begin{figure}
	\centering
	\includegraphics[width=0.6\textwidth]{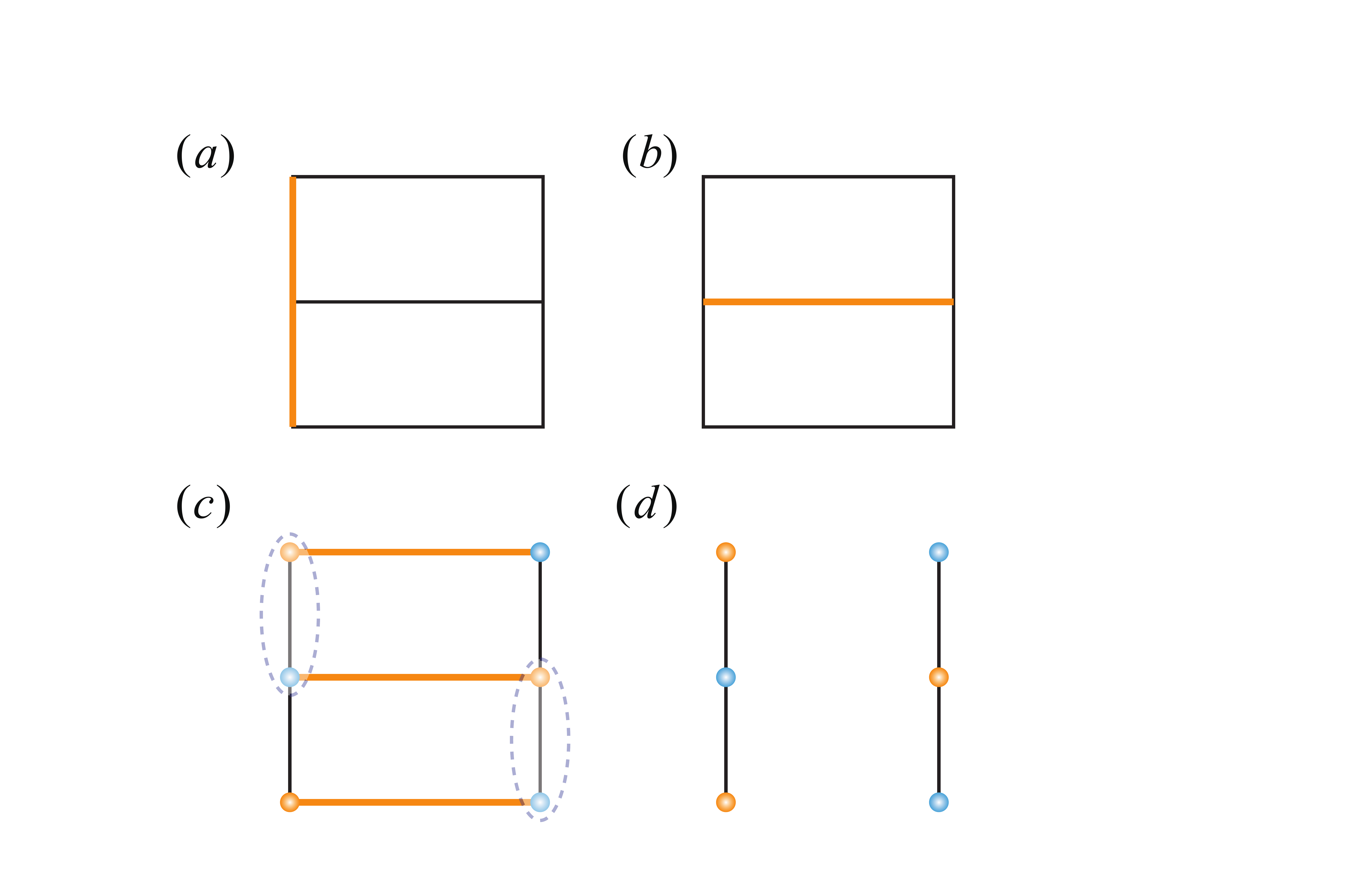}
	\caption{\label{boundary state P1g1} Decorations and boundary states of $p1g1$ with $\chi_{M} =\pm 1$. (a) The decoration protected by $x$-translation. 	(b) The decoration protected by glide symmetry. (c)-(d) When two boundaries related by glide are made, on each boundary the Majorana zero modes can be gapped. As a result, each boundary can be viewed as a 1D DIII TSC.}
\end{figure}

There are two decorations as the generators of wire construction for WG $p1g1$ with both $\chi_{g} = +1$ and $\chi_{g} = -1$, as shown in Fig.\ref{boundary state P1g1} (a)-(b).
Here we choose the glide to be $g = \{ M_x | 0\frac{1}{2} \}$, i.e., a combination of mirror reflection in $x$ direction and a half-integer translation in $y$ direction.
The first decoration has 1D DIII TCSs aligned along $x$ direction and is the one protected by $x$-translation.
Its boundary state is characterized by Majorana band along $k_x$ direction, as that for WG $p1$, see Appendix.\ref{Appendix other boundary states}.
Here we focus on the the second decoration, which is the one protected by glide, with 1D DIII TSCs aligned along $y$, decorated at both $y=n$ and $y=n+\frac{1}{2}$ where $n \in \mathbb{Z}$.

To maintain the glide symmetry, the boundary geometry can only be made into a strip finite in $x$ direction while infinite in $y$ direction, see Fig.\ref{boundary state P1g1}(c).
Note the left side and the right side are not independent but related by glide symmetry.
On each side, there are Kramer's pairs of Majorana zero modes from the terminated ends of 1D DIII TCS at both $y=n$ and $y=n+\frac{1}{2}$.
One can group two nearest Majorana Kramer's pairs together, e.g., the pair at $y=0$ and the pair at $y=\frac{1}{2}$, and gap them, as they are not related by any symmetries (two pairs of Majorana modes related by glide symmetry locate at different sides).
As a result, there are not any gapless states at the boundary.
To see why the anomaly of this decoration does not manifest at the boundary, one notes that
there is not any single edge which respects glide symmetry in 2D.
One can further imagine each side equivalently as a 1D DIII TCS, i.e., time-reversal symmetric (TRS) Kitaev chain, as suggested in Fig.\ref{boundary state P1g1}(d).
In each unit cell, there are Majorana Kramer's pair at both integer subsite and half-integer subsite.
Since an integer subsite on one side has its glide-related partner a half-integer subsite on the other side, the inter-cell pairing of Majorana modes on one side becomes intra-cell pairing on the other side.
Therefore, the two TRS Kitaev chains on left and right sides have topological invariant $\mathbb{Z}_2 = 1$ and $\mathbb{Z}_2 = 0$, respectively, or vice versa.
Suppose we could make the two TRS Kitaev chains contact with each other, then at their intersection, a domain wall is formed where the mass changes sign and gapless modes are supported.
However, they are related by glide symmetry and are always separated as long as the sample has finite width in $x$ direction.

\section{Discussion\label{Section discussion}}

In this work, we present a unified real-space framework to construct and classify TCSCs using 1D building blocks which we call wire construction.
We deal with 2D TCSCs of AZ class DIII, and our products are second-order TSCs as well as weak TSCs, i.e., those protected by translation symmetries. 
We show that our framework can be formulated mathematically with central issues as projective representations and group induction, and we use several examples to illustrate the detailed procedure of wire construction.
For an inclusive description of the symmetries for spinful electrons in 2D, we deal with both pure 2D systems which are described by wallpaper groups, and 2D systems embedded into 3D, which are described by layer groups. 
Furthermore, we discuss what kinds of boundary states these TCSCs by wire construction are characterized by. 

We comment again that our results are not complete classifications for TSCs in 2D, as the 2D strong TSC can not be obtained by wire construction.
To include the 2D strong phase for complete TSC classification, one should first figure out whether it is compatible with the crystalline symmetries, i.e., whether the $\mathbb{Z}_2$ invariant for 2D strong TSC is restricted to be zero by crystalline symmetries. 
For an incompatible example see the discussion of layer group $p11m$ in \cite{zhang2013topological}.
If compatible, one can further consider the group extension problem, as what was previously done in the full classification of TIs where both 3D strong TI and TCIs are included\cite{khalaf2018symmetry,song2019topological}. 
As the 2D strong TSC has $\mathbb{Z}_2$ classification itself, one should consider whether its double is still a trivial state in the presence of crystalline symmetries, or it becomes a second-order TCS as an ingredient in the complete TCS classification.
Such procedure is presented in \cite{chen2022topological} for wallpaper group $p2$, where the authors formulate the problem of complete classification as solving a short exact sequence.

For 3D TCSCs, wire construction can still be applied, with the products being third-order TCSs in 3D.
However, there is another problem to handle that is not encountered in 2D.
According to the discussions in \cite{song2020real}, there could be ``$(d+2)$-bubbles'' in real-space construction of TCSs, which suggests that for 1-cell decorations in 3D, not only 2-cell bubbles need to be considered, but also 3-cell bubbles should be involved.
That is to say, some 1-cell decorations, although stable under 2-cell bubble equivalence, may be trivialized by 3-cell bubbles.
Since this problem is beyond the scope of the present work, we leave it, together with the full TSC classification in 2D, to future works.

\section{Acknowledgements}

Bingrui Peng thanks Zhongyi Zhang and Zhida Song for helpful discussions, and thanks Yi Jiang for advices on the writing of this manuscript.
The work is supported by the Ministry of Science and Technology of China (Grant No. 2016YFA0302400) and Chinese Academy of Sciences (Grant No. XDB33000000).

\bibliography{ms}

\appendix

\section{Building blocks for layer groups\label{Appendix building blocks for layer groups}}

\begin{enumerate}
    
    \item $A_1$ representation
        
        For this representation, we have $\chi_{M_x} = 1$, $\chi_{M_y} = 1$, and $\chi_{C_2} = 1$.
        Applying Wigner test, we have 
        \begin{align}
            [ W^{\bar{\Gamma}_5} (\mathcal{T}), W^{\bar{\Gamma}_5} (\mathcal{P}), W^{\bar{\Gamma}_5} (\mathcal{S}) ] = [1, -1, 1]
        \end{align}
        which dictates $\Gamma_5$ belongs to class CI with trivial classification in 1D, and two $\bar{\Gamma}_5$'s are paired since $W^{\bar{\Gamma}_5}(\mathcal{P}) = -1$.
        We can write the projective irrep as 
        \begin{align}
            & \mathcal{T} = \mu_0 i \sigma_2 \mathcal{K}
            \\
            & \mathcal{S} = \mu_3 \sigma_3
            \\  
            & M_x = \mu_1 i \sigma_1
            \\
            & M_y = \mu_1 i \sigma_2
        \end{align}
        which satisfies $\mathcal{T}^2 = -1$, $\mathcal{S}^2 = 1$, $\mathcal{P}^2 = ( \mathcal{T} \mathcal{S} )^2 = 1$, $M_x^2 = -1$, $M_y^2 = -1$, $[ \mathcal{T}, M_x ] = 0$, $[ \mathcal{T}, M_y ] = 0$, $[ \mathcal{S}, M_x ] = 0$, and $[ \mathcal{S}, M_y ] = 0$.
        One can find mass term proportional to $\mu_0 \sigma_3$ to gap out the 0D zero modes described by these symmetries, verifying this building block is topologically trivial.
        
        \item $A_2$ representation
        
        For this representation, we have $\chi_{M_x} = -1$, $\chi_{M_y} = -1$, and $\chi_{C_2} = 1$.
        Applying Wigner test, we have 
        \begin{align}
            [ W^{\bar{\Gamma}_5} (\mathcal{T}), W^{\bar{\Gamma}_5} (\mathcal{P}), W^{\bar{\Gamma}_5} (\mathcal{S}) ] = [1, 1, 1]
        \end{align}
        which dictates $\bar{\Gamma}_5$ belongs to class BDI with $\mathbb{Z}$ classification in 1D, and it is not paired with itself.
        We can write the projective irrep as
        \begin{align}
            & \mathcal{T} = i \sigma_2 \mathcal{K}
            \\
            & \mathcal{S} = \sigma_3
            \\  
            & M_x = i \sigma_1
            \\
            & M_y = i \sigma_2
        \end{align}
        which satisfies $\mathcal{T}^2 = -1$, $\mathcal{S}^2 = 1$, $\mathcal{P}^2 = ( \mathcal{T} \mathcal{S} )^2 = 1$, $M_x^2 = -1$, $M_y^2 = -1$, $[ \mathcal{T}, M_x ] = 0$, $[ \mathcal{T}, M_y ] = 0$, $\{ \mathcal{S}, M_x \} = 0$, and $\{ \mathcal{S}, M_y \} = 0$.
        One can find no mass term to gap out the 0D zero modes described by these symmetries, as well as for the doubled system,
        verifying the $\mathbb{Z}$ classification for this building block. 
        Since $[ C_2, \mathcal{S} ] = 0$, one can define 1D winding number for $C_2 = \pm i$ sectors, i.e., $w_{C_2}^{\pm}$, and they satisfy $w_{C_2}^{+}= - w_{C_2}^{-}$. 
        We can define the independent topological index as $w_{C_2} = \frac{1}{2}(w_{C_2}^{+} - w_{C_2}^{-})$.
        
        \item $B_1$ representation
        
        For this representation, we have $\chi_{M_x} = 1$, $\chi_{M_y} = -1$, and $\chi_{C_2} = -1$.
        Applying Wigner test, we have 
        \begin{align}
            [ W^{\bar{\Gamma}_5} (\mathcal{T}), W^{\bar{\Gamma}_5} (\mathcal{P}), W^{\bar{\Gamma}_5} (\mathcal{S}) ] = [1, 1, 1]
        \end{align}
        which dictates $\bar{\Gamma}_5$ still belongs to class BDI with $\mathbb{Z}$ classification in 1D.
        We can write the projective irrep as
        \begin{align}
            & \mathcal{T} = i \sigma_2 \mathcal{K}
            \\
            & \mathcal{S} = \sigma_1
            \\  
            & M_x = i \sigma_1
            \\
            & M_y = i \sigma_2
        \end{align}
        which satisfies $\mathcal{T}^2 = -1$, $\mathcal{S}^2 = 1$, $\mathcal{P}^2 = ( \mathcal{T} \mathcal{S} )^2 = 1$, $M_x^2 = -1$, $M_y^2 = -1$, $[ \mathcal{T}, M_x ] = 0$, $[ \mathcal{T}, M_y ] = 0$, 
        $[ \mathcal{S}, M_x ] = 0$ and $\{ \mathcal{S}, M_y \} = 0$.
        Since $[ M_x, \mathcal{S} ] = 0$, the $M_x = \pm i$ sectors have 1D winding number $w_{M_x}^{\pm}$, and they satisfy $w_{M_x}^{+}= - w_{M_x}^{-}$. 
        we can define the independent topological index as $w_{M_x}= \frac{1}{2}(w_{M_x}^{+} - w_{M_x}^{-})$. 
        
        \item $B_2$
        
        For this representation, we have $\chi_{M_x} = -1$, $\chi_{M_y} = 1$ and $\chi_{C_2} = -1$.
        Wigner test shows that 
        \begin{align}
            [ W^{\bar{\Gamma}_5} (\mathcal{T}), W^{\bar{\Gamma}_5} (\mathcal{P}), W^{\bar{\Gamma}_5} (\mathcal{S}) ] = [1, 1, 1]
        \end{align}
        which dictates $\bar{\Gamma}_5$ belongs to class BDI again.
        We can write the projective irrep as
        \begin{align}
            & \mathcal{T} = i \sigma_2 \mathcal{K}
            \\
            & \mathcal{S} = \sigma_2
            \\  
            & M_x = i \sigma_1
            \\
            & M_y = i \sigma_2
        \end{align}
        which satisfies $\mathcal{T}^2 = -1$, $\mathcal{S}^2 = 1$, $\mathcal{P}^2 = ( \mathcal{T} \mathcal{S} )^2 = 1$, $M_x^2 = -1$, $M_y^2 = -1$, $[ \mathcal{T}, M_x ] = 0$, $[ \mathcal{T}, M_y ] = 0$, $\{ \mathcal{S}, M_x \} = 0$ and $[ \mathcal{S}, M_y ] = 0$.
        Since $[ M_y, \mathcal{S} ] = 0$, we can define the independent topological index as $w_{M_y} = \frac{1}{2}(w_{M_y}^{+} - w_{M_y}^{-})$.         
\end{enumerate}

\section{Group induction\label{Appendix group induction}}

According to group theory, the representations of a group can be derived based on the representations of its subgroups, which are called induced representations.  
One application of induced representations in solid-state physics is the theory of elementary band representations\cite{evarestov2012site,bradlyn2017topological,cano2018building,cano2021band}, where one can derive representations of space groups 
from representations of site symmetry groups of the orbitals at Wyckoff positions. 
In real-space construction, all cells are equivalent to Wyckoff positions in terms of symmetry. 
For example, the 1-cells coinciding with $C_n$ axis are equivalent to Wyckoff positions along $C_n$ axis. 
In our wire construction, for a cell-complex respecting a given crystalline symmetry group $G$, the whole 1-cells consist of different sets of symmetry-related ones, and each set forms induced representations of $G$ from the representations of the onsite-symmetry group $ G^1$.

\begin{figure}[H]
	\centering
	\includegraphics[width=0.5\textwidth]{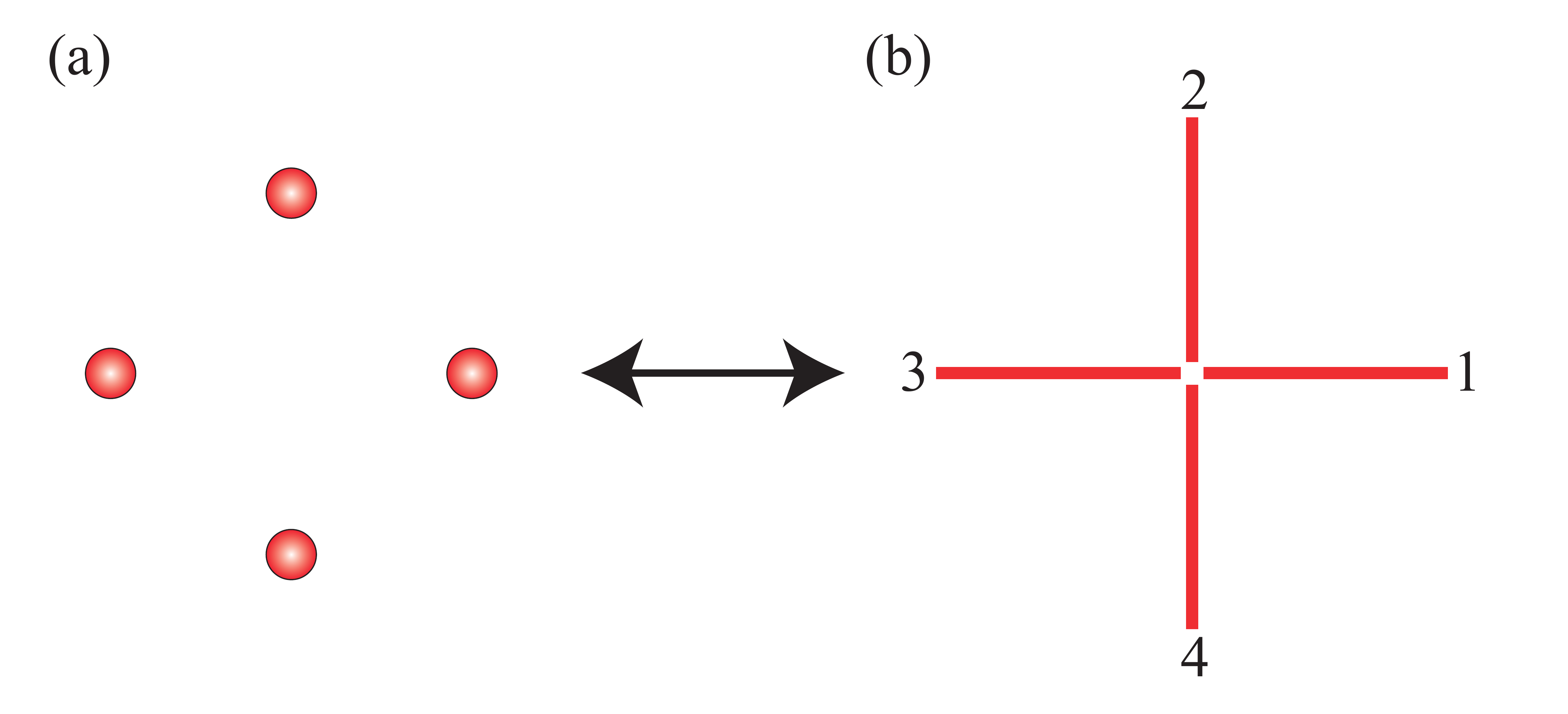}
	\caption{\label{group inducation appendix} 
    A set of $C_4$-related Wyckoff positions, as shown in (a), is equivalent to a set of $C_4$-related 1-cells, as shown in (b), in terms of symmetry. }
\end{figure}

Here we briefly review the procedure for group induction, and more elaborate discussions can be found in Ref.\cite{cano2018building}. 
Suppose there is a set of Wyckoff positions with multiplicity $N$. 
We denote the representative position as $q$, then other positions locate at the orbit of $q$, i.e., $\{ g q | g \in G \}$, where $G$ is the whole crystalline symmetry group. 
We label all the N Wyckoff positions as $q_{\alpha} = g_{\alpha} q, \alpha = 1, 2, ..., N $, and $q_1 = q$ in particular.
The collection of symmetry operations that keep the Wyckoff position $q_{\alpha}$ invariant form the site symmetry group of $q_{\alpha}$, denoted as $G_{q_{\alpha}}$.
If we assign an irrep $\rho$ of $G_q$, i.e., place a set of local orbitals $\{ \alpha_i \}$ which span $\rho$ at $q$, then we have an induced representation $\bar{\rho}$ of $G$, spanned by the union of orbitals at all the symmetry-related Wyckoff positions, given by
\begin{align}
    \bar{\rho} (h) a_{i \alpha} = \sum_{i'= 1}^{n_q}
    \rho_{i' i} ( g_{\beta}^{-1} \{ E|- t_{\beta \alpha} \} h g_{\alpha} ) a_{i' \beta}
\end{align}
where $h \in G$, $n_q$ denotes the number of $\{ \alpha_i \}$, i.e., the dimension of $\rho$, and $a_{i \alpha}$ denotes the $i$-th orbital at Wyckoff position $q_{\alpha}$.
For each $\alpha$, the determination of $\beta$ is given by the following way.
Firstly one makes the coset decomposition of $G$, i.e.
\begin{align}
    G=\bigcup_{k=1}^{N} g_{k} ( G_q
    \ltimes \mathbb{Z}^{d} )
\end{align}
where $\mathbb{Z}^{d}$ indicates translation group in $d$-dimension.
Then $g_{\beta}$ is determined by finding the coset which includes $h g_{\alpha}$, i.e., there exists some $g \in G_q$ such that 
\begin{align}
    h g_{\alpha} = \{ E | t_{\beta \alpha} \} g_{\beta} g 
\end{align}
with $t_{\alpha \beta}$ given by
\begin{align}
    t_{\alpha \beta} = h q_{\alpha} - q_{\beta}
\end{align}

We can apply the above formulas to our wire construction, where the representations become projective and group elements include $ \mathcal{T}$, $\mathcal{P}$ and $\mathcal{S}$ apart from crystalline operations.
We give the example of group $G$ generated by $C_4$, $\mathcal{T}$, and $\mathcal{S}$.

First we note for point group, (B3) reduces to
\begin{align}
    h g_{\alpha} = g_{\beta} g
\end{align}
and (B1) reduces to
\begin{align}
    \bar{\rho} (h) a_{i \alpha} = \sum_{i'= 1}^{n_q}
   \rho_{i' i} ( g_{\beta}^{-1}  h g_{\alpha} ) a_{i' \beta}
   = \sum_{i'= 1}^{n_q} \rho_{i' i} ( g ) a_{i' \beta}
\end{align}
as translations are now absent.
As shown in Figure.\ref{group inducation appendix}, we label the four 1-cells as $1,2,3,4$ in the counterclockwise direction, and the $C_4$ center, which is also the original point, is denoted as $O$.
The site symmetry group of each 1-cell is trivial, and the onsite symmetry group is generated by $\mathcal{T}$ and $\mathcal{S}$.
We can write the representation as
\begin{align}
    & \mathcal{T} = i \sigma_2 \mathcal{K}
    \\
    &\mathcal{S} = \sigma_3 \mathcal{K}
\end{align}
Then we use the formula (B6) to derive the induced representation at the 0-cell $O$.
We first deal with $C_4$, under which the four 1-cells transform as 
\begin{align}
1 \rightarrow 2, \; 2 \rightarrow 3, \; 3 \rightarrow 4, \; 4 \rightarrow 1
\end{align}
 i.e., in a permutation manner.
We can assign the coset representatives as 
$g_i = C_4^{i-1}$, i.e., 
\begin{align}
g_1 = \bm{1}, \; g_2 = C_4, \; g_3 = C_4^2 = C_2, \; g_4 = C_4^3
\end{align}
where $\bm{1}$ stands for the identity operation.
It is easy to find for $g_{\alpha} = g_i$, we have $g_{\beta} = g_{i+1}$, where $i \in \mathbb{Z}_4$.
Therefore, for $h = C_4$, (120) reads 
\begin{align}
g = g^{-1}_{\beta} C_4 g_{\alpha} = g^{-1}_{i+1} C_4 g_i
\end{align}
which leads to $g = \bm{1}$ for $i = 1,2,3$ while $g = -\bm{1}$ for $i = 4$, by noting $C_4^4 = -\bm{1}$.
We can see that the induced representation of $C_4$ at the 0-cell can be derived based on the representation of the identity operation, $\bm{1}$, at the 1-cell, perhaps up to a minus sign. 
As a result, we have 
\begin{align}
C_4 = \left(
        \begin{array}{cccc}
          0 & 1 & 0 & 0  \\
          0 & 0 & 1 & 0 \\
          0 & 0 & 0 & 1 \\
          -1 & 0 & 0 & 0
         \end{array}
         \right) 
         \otimes \sigma_0    
\end{align}
Then we deal with $\mathcal{T}$.
Since $\mathcal{T}$ acts locally and always commutes with $C_4$, one can directly write the induced representation of $\mathcal{T}$ with a thought
\begin{align}
\mathcal{T} = \left(
        \begin{array}{cccc}
          1 & 0 & 0 & 0  \\
          0 & 1 & 0 & 0 \\
          0 & 0 & 1 & 0 \\
          0 & 0 & 0 & 1
         \end{array}
         \right) 
         \otimes i \sigma_2 \mathcal{K}    
\end{align}
Lastly for $\mathcal{S}$, one has to consider $\chi_{C_4} = \pm 1$, i.e., both $[ C_4, \mathcal{S} ] = 0 $ and $\{ C_4, \mathcal{S} \} = 0$.

\begin{enumerate}
    
    \item $[ C_4, \mathcal{S} ] = 0 $
    
    In this case, it is similar with $\mathcal{T}$, and we have
    \begin{align}
        \mathcal{S} = \left(
        \begin{array}{cccc}
          1 & 0 & 0 & 0  \\
          0 & 1 & 0 & 0 \\
          0 & 0 & 1 & 0 \\
          0 & 0 & 0 & 1
         \end{array}
         \right) 
         \otimes \sigma_3       
    \end{align}
    
    \item $\{ C_4, \mathcal{S} \} = 0 $
    
    In this case, we should be a little careful, since $C_4$ reverses the eigenvalue of $\mathcal{S}$.
    Note $\mathcal{S}$ does not change the position of 1-cells, so we always have $g_{\beta} = g_{\alpha}$.
    Specifically, for $g_{\alpha} = g_{i, i =1, 2, 3, 4}$, we have
    \begin{align}
    & g = g^{-1}_i \mathcal{S} g_i 
    = ( C_4^{i-1} )^{-1} \mathcal{S} C_4^{i-1} 
    \\
    & = ( C_4^{i-1} )^{-1} C_4^{i-1} \mathcal{S} (-1)^{i-1} 
    = \mathcal{S} (-1)^{i-1} 
    \end{align}
    where the extra factor $(-1)^{i-1}$ comes from $\{ C_4, \mathcal{S}\} = 0 $. 
    Thus we have the induced representation.
    \begin{align}
        \mathcal{S} = \left(
        \begin{array}{cccc}
          1 & 0 & 0 & 0  \\
          0 & -1 & 0 & 0 \\
          0 & 0 & 1 & 0 \\
          0 & 0 & 0 & -1
         \end{array}
         \right) 
         \otimes \sigma_0       
    \end{align}    
    
\end{enumerate}

\section{Trivial bubble equivalence in wallpaper groups\label{Appendix trivial bubble equivalence in wallpaper groups}}

Below we show that for wallpaper groups, bubble equivalence always has trivial effects.

Firstly we note in 2D, any 1-cell is always shared by two 2-cells, which means when considering bubble equivalence, a 1-cell has always two bubbles on its two sides.
Recall that there are two types of 1-cells for wallpaper groups, one with $G^1_c = \bm{1}$ and the other with $G^1_c = \bm{M}$.
A 1-cell of the first type has $\mathbb{Z}_2$  
classification, and the resulting states by two bubbles can only have $\mathbb{Z}_2 = 0$.
For the second type, the trivial effects of bubble equivalence are not obvious, and we make careful analysis as follows.

\begin{enumerate}

    \item $\chi_M = 1 $
    
    According to Sec.\ref{subsubsection building blocks for WGs}, the classification on the 1-cell is $\mathbb{Z}$, and the corresponding topological index can be defined as mirror winding number $w_M = \frac{1}{2} ( w^{+} - w^{-} )$, which indicates the number of zero modes at the 0D boundary.  
    On the two sides of the 1-cell, two bubbles are related by mirror $M$.
    Each bubble edge is a 1D DIII TCS with boundary zero modes described by symmetries
    \begin{align}
        & \mathcal{T} = i \sigma_2 \mathcal{K}
        \\
        & \mathcal{S} = \sigma_3 
    \end{align}   
    When coinciding with the 1-cell, by group induction, the resulting state from the two mirror-related bubble edges have zero modes described by 
    \begin{align}
        & \mathcal{T} = \mu_0 i \sigma_2 \mathcal{K}
        \\
        & \mathcal{S} = \mu_0 \sigma_3  \\
        & M = i \mu_2 \sigma_0
    \end{align}
    Since $[ M, \mathcal{S} ] = 0$, we can diagonalize them simultaneously, leading to 
    \begin{align}
        \mathcal{S} = 
        \left(
        \begin{array}{cccc}
          1 & 0 & 0 & 0  \\
          0 & 1 & 0 & 0 \\
          0 & 0 & -1 & 0 \\
          0 & 0 & 0 & -1
         \end{array}
         \right),
        M = 
        \left(
        \begin{array}{cccc}
          i & 0 & 0 & 0  \\
          0 & -i & 0 & 0 \\
          0 & 0 & i & 0 \\
          0 & 0 & 0 & -i
         \end{array}
         \right)        
    \end{align}    
    which can be decomposed into $M = +i$ sector with $w = 0$
    \begin{align}
        \mathcal{S} 
        = \left(
        \begin{array}{cc}
          1 & 0   \\
          0 & -1  
         \end{array}
         \right), 
        M = \left(
        \begin{array}{cc}
          i & 0   \\
          0 & i  
         \end{array}
         \right)   
    \end{align}
    and  
    $M = -i$ sector with $w = 0$
    \begin{align}
        \mathcal{S} 
        = \left(
        \begin{array}{cc}
          1 & 0   \\
          0 & -1  
         \end{array}
         \right), 
        M = \left(
        \begin{array}{cc}
          -i & 0   \\
          0 & -i  
         \end{array}
         \right)   
    \end{align}
    Since both $w^{+}$ and $w^{-}$ are zero, $w_M = \frac{1}{2} (w^{+} - w^{-})$ is also zero.
    The vanishing topological number means the zero modes can be gapped out and the resulting 1D state from the two bubble edges is trivial.
    Consequently, the classification on the 1-cell is unchanged under bubble equivalence.
    
    \begin{figure}
	\centering
	\includegraphics[width=0.5\textwidth]{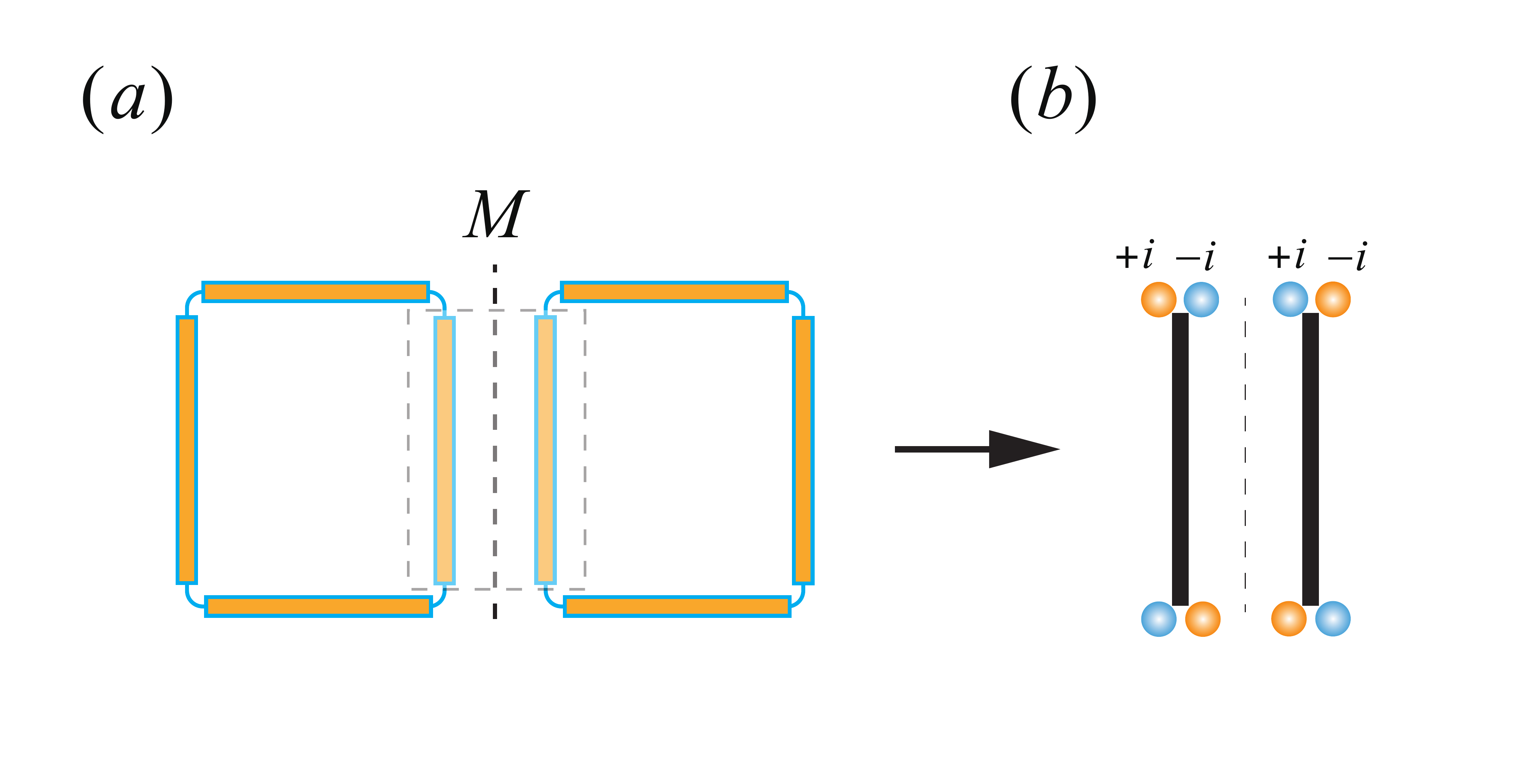}
	\caption{\label{trivial bubble equivalence}
	The trivial effect of bubble equivalence on the classification of mirror 1-cell with $\chi_{M} = 1$.
	(a) There are two bubbles related by $M$, with each bubble edge being 1D DIII TCS. 
	(b) When the two bubble edges coincide with the mirror 1-cell, the resulting states become mirror-symmetric 1D TCSs, with boundary zero modes characterized by $M$ eigenvalues $\pm i$ as well as $\mathcal{S}$ eigenvalues $\pm 1$, and the orange and blue dots in the figure denote zero modes different $\mathcal{S}$ eigenvalues. 
	One of them has topological number $w_M = \frac{1}{2} ( w_{M= +i} - w_{M = -i} ) = 1$, while the other one has $w_M = -1$. 
	As a result, the total topological number of them is zero.}
    \end{figure}
    
    \item $\chi_M = -1$
    
    According to Sec.\ref{subsubsection building blocks for WGs}, the topological classification on the 1-cell is $\mathbb{Z}_2$.
    Similar to the 1-cells with $C^1_c = \bm{1}$, bubble equivalence can not change the topological index by $1$ thus has trivial effects on the 1-cell classification.
    
\end{enumerate}

\section{Example: wallpaper group $p2mm$ with $B_1$ representation\label{Appendix example of wallpaper group}}

Here we show the wire construction of TCSCs protected by wallpaper group $p2mm$ with complete procedure.
We choose the representation of gap function $B_1$, with $\chi_{M_x} = 1$ and $\chi_{M_y} = -1$.

\begin{figure}
	\centering
	\includegraphics[width=0.4\textwidth]{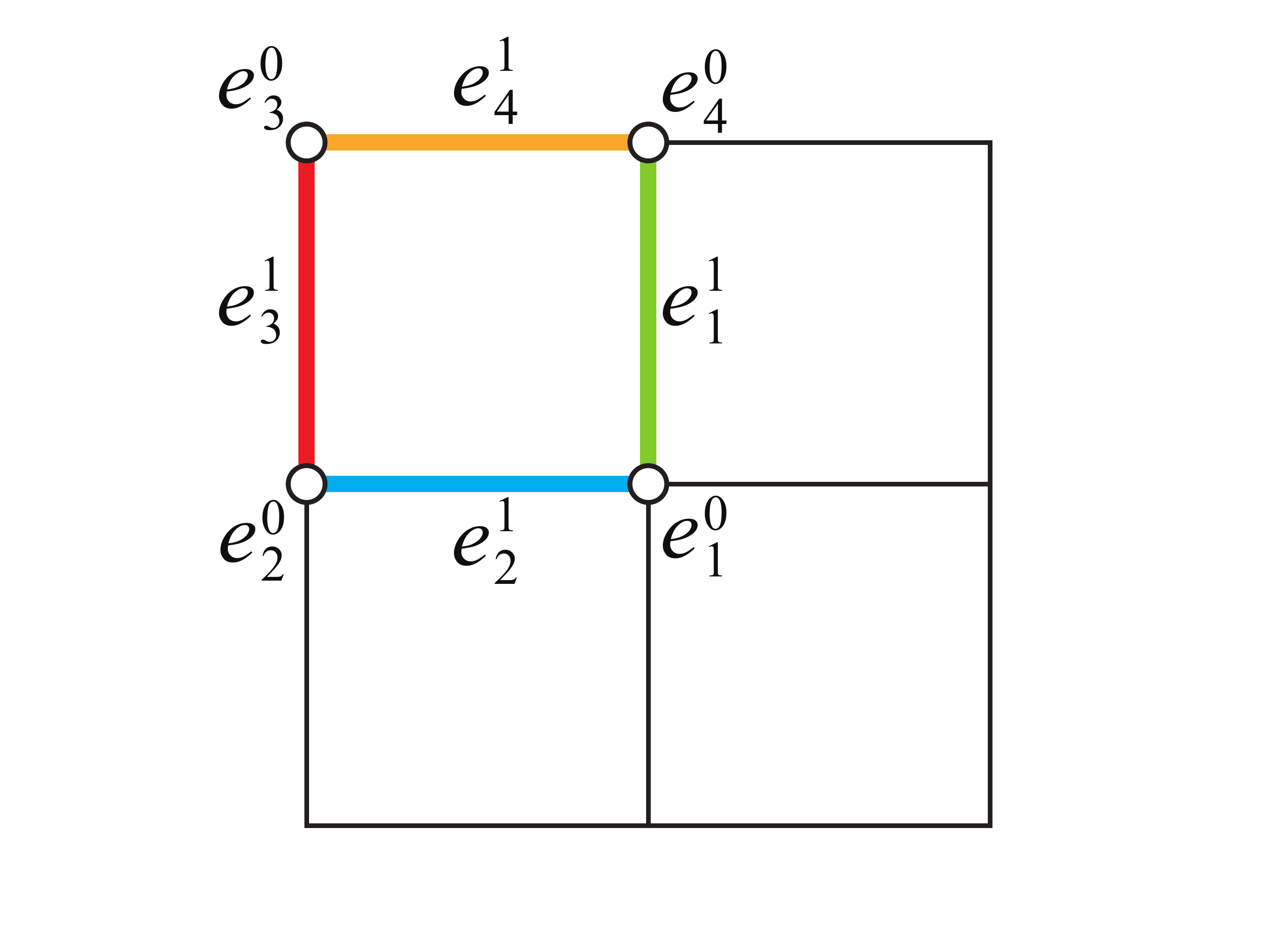}
	\caption{\label{cell complex of p2mm} The cell complex of wallpaper group $p2mm$.
	The only independent 2-cell is $e^2_1$.
	There are three independent 1-cells, denoted as $e^1_{i,i=1,2,3}$, among which $e^1_{i,i=1,3}$ have $G^1_c = \bm{M_x}$ and $e^1_{i,i=2,4}$ have $G^1_c = \bm{M_y}$.
	There are four independent 0-cells, denoted as $e^0_{i,i=1,2,3,4}$, all of which have $G^0_c = \bm{C_{2v}}$. }
\end{figure}

This wallpaper group is generated by mirror reflections in $x$ and $y$ directions, i.e., $M_x$ and $M_y$, and their combination is $C_{2z}$. 
The cell complex of $p2mm$ is shown in Figure.\ref{cell complex of p2mm}.
Since all 1-cells coincide with mirror lines, they have either $G^1_c = \bm{M_x}$ or $G^1_c = \bm{M_y}$.
According to Sec.\ref{subsubsection building blocks for WGs}, the building blocks are 1D mirror TSCs, with $\mathbb{Z}$ invariant if $\chi_{M} = 1$ or $\mathbb{Z}_2$ invariant if $\chi_{M} = -1$. 
Meanwhile, all 0-cells have $G^c_0 = \bm{C_{2v}}$, with the only double-valued irrep being $\bar{\Gamma}_5$.

We first analysis the projective irreps of $G^0$, which describe the symmetries of zero modes at $e^0$'s.
Apply Wigner test to $\bar{\Gamma}_5$, we have
\begin{align}
     [ W^{\bar{\Gamma}_5}(\mathcal{T}),  W^{\bar{\Gamma}_5}(\mathcal{P}), W^{\bar{\Gamma}_5}(\mathcal{S}) ]  = [1,1,1]
\end{align}
which means $\bar{\Gamma}_5$ belongs to class BDI with $\mathbb{Z}$ classification in 1D.
The projective irrep can be written as
\begin{align}
    & \mathcal{T} = i \sigma_2 \mathcal{K}
    \\
    & \mathcal{S} =  \sigma_1
    \\
    & M_x = i \sigma_1
    \\
    & M_y = i \sigma_2
\end{align}
which satisfies $[ M_x, \mathcal{S} ] = 0 $ and  $\{ M_y, \mathcal{S} \} = 0 $.
Since $[ M_x, \mathcal{S} ] = 1$, we can define the topological index as $w_{M_x} = \frac{1}{2} ( w_{M_x}^{+} - w_{M_x}^{-} ) $.

Then we can analyze the gluing condition.
     
\begin{enumerate}
     
     \item $ e^1_1 $ ($e^1_3$) decoration
     
     The decorations on $e^1$ and $e^3$ are essentially the same, and we take the one on $e^1$.
     The 1D TSC with $\mathbb{Z}$ classification decorated on $e^1_1$ ($e^1_3$) has 0D zero modes described by 
     \begin{align}
         & \mathcal{T} = i \sigma_2 \mathcal{K}
         \\
         & \mathcal{S} =  \sigma_1
         \\
         & M_x = i \sigma_1
     \end{align}    
     By group induction, the zero modes at $e^0_1$ ($e^0_4$) are described by
      \begin{align}
         & \mathcal{T} = \mu_0 i \sigma_2 \mathcal{K}
         \\
         & \mathcal{S} =  \mu_3 \sigma_1
         \\
         & M_x = \mu_3 i \sigma_1
         \\
         & M_y = i \mu_0 \sigma_0          
      \end{align}
      Diagonalizing $M_x$ and $\mathcal{S}$ simultaneously, we
      have 
      \begin{align}
        M_x = \left(
        \begin{array}{cccc}
          i & 0 & 0 & 0  \\
          0 & -i & 0 & 0 \\
          0 & 0 & i & 0 \\
          0 & 0 & 0 & -i
         \end{array}
         \right), 
         \; \; 
        \mathcal{S} = \left(
        \begin{array}{cccc}
          1 & 0 & 0 & 0  \\
          0 & -1 & 0 & 0 \\
          0 & 0 & 1 & 0 \\
          0 & 0 & 0 & -1
         \end{array}
         \right) 
      \end{align}
      One can see the topological index is
      \begin{align}
            w_{M_x} = \frac{1}{2} (2 - (-2))= 2
      \end{align}
      which is nonzero and suggests the gluing condition is failed at $e^0_1$ and $e^0_4$.
      Therefore the decorations on $e^1_1$ and $e^1_3$ are not allowed.
      
     \item $e^1_2$ ($e^1_4$) decoration
     
     Then we analysis the decoration on $e^1_2$ ($e^1_4$).
     The gluing condition for these two decorations are essentially the same,
     and we take $e^1_2$.
      
     Since $e^1_2$ has $G^1_c = \bm{M_y}$ with $\chi_{M_y} = -1$, the decoration on has $\mathbb{Z}_2$ classification, with 0D zero modes described by
     \begin{align}
         & \mathcal{T} = i \sigma_2 \mathcal{K}
         \\
         & \mathcal{S} =  \sigma_3
         \\
         & M_y = i \sigma_2
     \end{align}    
     By group induction, the zero modes at $e^0_1$ ($e^0_2$) are described by
      \begin{align}
         & \mathcal{T} = \mu_0 i \sigma_2 \mathcal{K}
         \\
         & \mathcal{S} =  \mu_0 \sigma_3
         \\
         & M_x = i \mu_2 \sigma_0
         \\
         & M_y = \mu_3 i \sigma_2          
      \end{align}
      Diagonalizing $M_x$ and $\mathcal{S}$ simultaneously, we
      have 
      \begin{align}
        M_x = \left(
        \begin{array}{cccc}
          i & 0 & 0 & 0  \\
          0 & -i & 0 & 0 \\
          0 & 0 & i & 0 \\
          0 & 0 & 0 & -i
         \end{array}
         \right), 
         \;\;
        \mathcal{S} = \left(
        \begin{array}{cccc}
          1 & 0 & 0 & 0  \\
          0 & 1 & 0 & 0 \\
          0 & 0 & -1 & 0 \\
          0 & 0 & 0 & -1
         \end{array}
         \right) 
      \end{align}
    One can see the topological number is
    \begin{align}
        w_{M_x} = \frac{1}{2}(1-1+1-1) = 0
    \end{align}
    which means the gluing condition is met and the decoration of $e^1_2$ ($e^1_4$) is allowed.
    
\end{enumerate}
      
As a result, decorations on $e^1_2$ and $e^1_4$ are the generators for all decorations.
As they are both $\mathbb{Z}_2$-type, the classification is $\mathbb{Z}_2^2$.

\section{Example: layer group $pm2m$ with $A_1$ and $B_1$ representations\label{Appendix example of layer goup}}

Here we present an example of layer group, $pm2m$.
There are four representations of gap function, and here we deal with two of them, $A_1$ and $B_1$.

This LG is generated by adding mirror reflection in $z$ direction, $M_z$, to the WG $p1m1$.
Since $p1m1$ has $M_x$, the product $M_z \cdot M_x$ generates a $C_2$ rotation in $y$ direction, $C_{2y}$.
The cell complex of $pm2m$ is the same as that of WG $p1m1$.

\begin{figure}[H]
	\centering
	\includegraphics[width=0.45\textwidth]{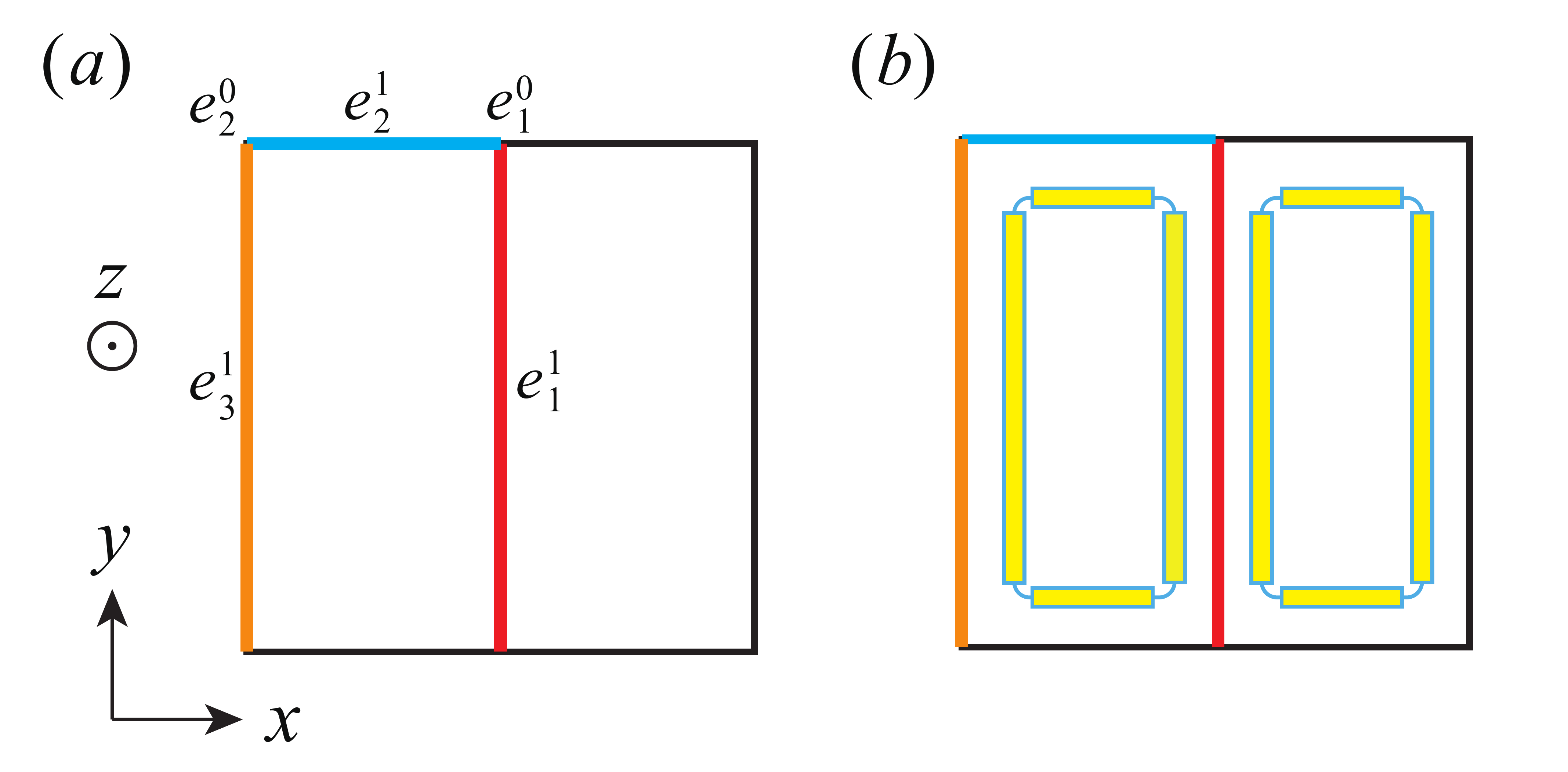}
	\caption{\label{appendix cell complex Pm2m} 
    (a) Cell complex of $pm2m$. (b) Bubbles of $pm2m$.}
\end{figure}

Note all the $d$-cells ($d = 0,1,2$) have $M_z$ as onsite symmetry.
Besides, $e^0_{i,i=1,2}$ and $e^1_{i,i=1,3}$ have $M_x$ as onsite symmetry thus they have $G^c = \bm{C_{2v}}$.
Therefore we have the particular type of building blocks for layer groups to be decorated on $e^1_1$ and $e^1_3$, which have $G^1_c = \bm{C_{2v}}$.
Nevertheless, we need not explicitly analyze gluing condition involving them.
Actually, at $e^0_1$ ($e^0_2$), the two relevant
$e^1_1$'s ($e^1_2$'s) are related by $y$ translation and one has its tail connected by the other's head, which makes the gluing condition trivially satisfied, and the decorations of $e^1_1$ and $e^1_3$ are always allowed. 
As a result, for gluing condition we only need to explicitly consider the two $e^1_2$'s related by $C_{2y}$ at $e^0_1$ ($e^0_2$).

\begin{enumerate}
    
    \item $A_1$ representation
    
    For this representation of gap function, one has $\chi_{M_x} = 1$, $\chi_{M_z} = 1$, and $\chi_{C_{2y}} = 1$.
    The projective irreps for the 1-cells and 0-cells are shown in Table \ref{table 1-cells for A_1 representation} and Table \ref{table 0-cells for A_1 representation}, respectively. 
    
    \begin{table}[H]
    \centering
    \begin{tabular}{c|cccccc}
    \hline\hline	
    \textbf{1cell} & $\bm{G^c_1}$ & \textbf{IR} & \textbf{EAZ} & \textbf{classification} 
    \\
    \hline
     $e^1_1$, $e^1_3 $ & $C_{2v}$ & $\bar{\Gamma}_5$ & CI  & $ \text{N/A} $ 
    \\
    \hline
     $e^1_2$ & $M_z$ & $\bar{\Gamma}_3$, $\bar{\Gamma}_4$ & AIII & $\mathbb{Z}$ 
    \\
    \hline\hline
    \\
    \textbf{1cell} & \textbf{PIR name}  & \textbf{PIR} & \textbf{TI}
    \\
    \hline
    \multirow{4}{*}{$e^1_1$, $e^1_3$}  &  
    \multirow{4}{*}{ $ \bar{\Gamma}_5 $ } &
    $ \mathcal{T} = \mu_0 i \sigma_2 \mathcal{K} $ & \multirow{4}{*}{$ \text{N/A} $}
    \\
     &  & $ \mathcal{S} = \mu_3 \sigma_3 $ & 
     \\
     &  & $ M_x = \mu_1 i \sigma_1 $ &
     \\
     &  & $ M_z =  \mu_0 i \sigma_3 $ &
    \\
    \hline
    \multirow{3}{*}{$e^1_2$}  &  
    \multirow{3}{*}{ $\bar{\Gamma}_{3\oplus4} $ } &
    $ \mathcal{T} = i \sigma_2 \mathcal{K} $ & \multirow{3}{*}{$ w_{M_z} $ }
    \\
     &  & $ \mathcal{S} = \sigma_3 $ & 
     \\
     &  & $ M_z =  i \sigma_3 $ &
    \\
    \hline\hline
    \end{tabular}
    \caption{\label{table 1-cells for A_1 representation} 
    1-cells for $A_1$ representation. 
    Here we use \textbf{IR} for the abbreviation of irreducible representation, \textbf{PIR} for projective irreducible representations, \textbf{TI} for topological index. 
    $\text{N}/\text{A}$ means trivial classification or topological index.
    Here the topological index $w_{M_z}$ is defined as $w_{M_z} = \frac{1}{2} ( w_{+i} - w_{-i}) $.}
    \end{table}

    \begin{table}[H]\label{0-cells for A1 representation of Pm2m}
    \centering
    \begin{tabular}{c|cccccc}
    \hline\hline	
    \textbf{0cell} & $\bm{G^c_0}$ & \textbf{IR} & \textbf{EAZ} & \textbf{classification} 
    \\
    \hline
     $e^0_1$, $e^0_2 $ & $C_{2v}$ & $\bar{\Gamma}_5$ & CI  & $\text{N/A}$ 
    \\
    \hline\hline
    \\
    \textbf{0cell} & \textbf{PIR name}  & \textbf{PIR} & \textbf{TI}
    \\
    \hline
    \multirow{4}{*}{$e^0_1$, $e^0_2$}  &  
    \multirow{4}{*}{ $ \bar{\Gamma}_5 $ } &
    $ \mathcal{T} = \mu_0 i \sigma_2 \mathcal{K} $ & \multirow{4}{*}{$ \text{N/A} $}
    \\
     &  & $ \mathcal{S} = \mu_3  \sigma_3 $ & 
     \\
     &  & $ M_x = \mu_1 i \sigma_1 $ &
     \\
     &  & $ M_z =  \mu_0 i \sigma_3 $ &
    \\
    \hline\hline
    \end{tabular}
    \caption{\label{table 0-cells for A_1 representation} 
    0-cells for $A_1$ representation.    
    Here we use \textbf{IR} for the abbreviation of irreducible representation, \textbf{PIR} for projective irreducible representations, \textbf{TI} for topological index. 
    $\text{N}/\text{A}$ means trivial classification or topological index.}
    \end{table}    
    
    According to the above tables, the only double-valued irrep of $C_{2v}$, i.e., $\bar{\Gamma}_5$, belongs to class CI with trivial classification in 1D.
    This means
    the building blocks on $e^1_1$ and $e^1_3$ with $G^1_c = \bm{C_{2v}}$ are trivial.
    What's more, the gluing condition is trivially satisfied at $e^0_1$ and $e^0_2$ with $G^0_c = \bm{C_{2v}}$. 
    Therefore one has 1-cell decoration of $e^1_2$ allowed by gluing condition.
    
    Then we consider bubble equivalence.
    However, the two bubbles on the two sides of $e^1_2$ are related by translation in $y$ direction and they have trivial effects on $e^1_2$.
    
    As a result, the final classification is $\mathbb{Z}$, contributed by $e^1_2$ decoration.

    \item $B_1$ representation
    
    For this representation of gap function, one has $\chi_{M_x} = -1$, $\chi_{M_z} = 1$ and $\chi_{C_{2y}} = -1$.
    
    The projectvie irreps for the 1-cells and 0-cells are shown in Table \ref{table 1-cells for B_1 representation} and Table \ref{table 0-cells for B_1 representation}, respectively.
    
    \begin{table}[H]
    \centering
    \begin{tabular}{c|cccccc}
    \hline\hline	
    \textbf{1cell} & $\bm{G^c_1}$ & \textbf{IR} & \textbf{EAZ} & \textbf{classification} 
    \\
    \hline
     $e^1_1$, $e^1_3 $ & $C_{2v}$ & $\bar{\Gamma}_5$ & BDI  & $\mathbb{Z}$ 
    \\
    \hline
     $e^1_2$ & $M$ & $\bar{\Gamma}_3$, $\bar{\Gamma}_4$ & AIII & $\mathbb{Z}$ 
    \\
    \hline\hline
    \\
    \textbf{1cell} & \textbf{PIR name}  & \textbf{PIR} & \textbf{TI}
    \\
    \hline
    \multirow{4}{*}{$e^1_1$, $e^1_3$}  &  
    \multirow{4}{*}{ $ \bar{\Gamma}_5 $ } &
    $ \mathcal{T} = i \sigma_2 \mathcal{K} $ & \multirow{4}{*}{$ w_{M_z } $}
    \\
     &  & $ \mathcal{S} = \sigma_3 $ & 
     \\
     &  & $ M_x = i \sigma_1 $ &
     \\
     &  & $ M_z =  i \sigma_3 $ &
    \\
    \hline
    \multirow{3}{*}{$e^1_2$}  &  
    \multirow{3}{*}{ $\bar{\Gamma}_{3\oplus4} $ } &
    $ \mathcal{T} = i \sigma_2 \mathcal{K} $ & \multirow{3}{*}{$ w_{M_z} $}
    \\
     &  & $ \mathcal{S} = \sigma_3 $ & 
     \\
     &  & $ M_z =  i \sigma_3 $ &
    \\
    \hline\hline
    \end{tabular}
    \caption{\label{table 1-cells for B_1 representation} 
    1-cells for $B_1$ representation. }
    \end{table}

    \begin{table}[H]
    \centering
    \begin{tabular}{c|cccccc}
    \hline\hline	
    \textbf{0cell} & $\bm{G^c_0}$ & \textbf{IR} & \textbf{EAZ} & \textbf{classification} 
    \\
    \hline
     $e^0_1$, $e^0_2 $ & $C_{2v}$ & $\bar{\Gamma}_5$ & BDI  & $\mathbb{Z}$ 
    \\
    \hline\hline
    \\
    \textbf{0cell} & \textbf{PIR name}  & \textbf{PIR} & \textbf{TI}
    \\
    \hline
    \multirow{4}{*}{$e^0_1$, $e^0_3$}  &  
    \multirow{4}{*}{ $ \bar{\Gamma}_5 $ } &
    $ \mathcal{T} = i \sigma_2 \mathcal{K} $ & \multirow{4}{*}{$ w_{M_z}$}
    \\
     &  & $ \mathcal{S} = \sigma_3 $ & 
     \\
     &  & $ M_x = i \sigma_1 $ &
     \\
     &  & $ M_z =  i \sigma_3 $ &
    \\
    \hline\hline
    \end{tabular}
    \caption{\label{table 0-cells for B_1 representation} 
    0-cells for $B_1$ representation. }
    \end{table}    
    
    From the above tables, we know the building blocks on $e^1_{i,i= 1,2,3}$ are all nontrivial.
    
    Then we consider the gluing condition at $e^0_1$ ($e^0_2$), which involves $e^1_2$.
    The induced representation from $\bar{\Gamma}_{3 \oplus 4}$ at $e^1_2$ is given by
    \begin{align}
        & \mathcal{T} = \mu_0 i \sigma_2 \mathcal{K}
        \\
        & \mathcal{S} = \mu_3 \sigma_3
        \\
        & M_x = i \mu_2 \sigma_0
        \\
        & M_z = \mu_3 i \sigma_3
    \end{align}
    By diagonalizing $\mathcal{S}$ and $M_z$ together, we obtain
    \begin{align}
        M_z = \left(
        \begin{array}{cccc}
          i & 0 & 0 & 0  \\
          0 & -i & 0 & 0 \\
          0 & 0 & i & 0 \\
          0 & 0 & 0 & -i
         \end{array}
         \right), \;
         \mathcal{S} = \left(
        \begin{array}{cccc}
          1 & 0 & 0 & 0  \\
          0 & -1 & 0 & 0 \\
          0 & 0 & 1 & 0 \\
          0 & 0 & 0 & -1
         \end{array}
         \right) 
    \end{align}
    which equals to two $\bar{\Gamma}_5$'s with the same $\mathcal{S}$, i.e., the same topological number $w_{M_z} = \frac{1}{2}( w_{+i} - w_{-i} ) =1$.
    This leads to the non-vanishing total topological number $w_{M_z} = 2$, which protects the 0D modes from being gapped out.
    Therefore the gluing condition is failed, which means decoration on $e^1_2$ is not allowed.
    
    Then one has to analyze whether the decorations on $e^1_1$ and $e^1_3$, which are trivially allowed the gluing condition, are stable against bubble equivalence. 
    For a single bubble, its edge can be 1D mirror-symmetric TSC with 0D zero modes described by 
    \begin{align}
        & \mathcal{T} = i \sigma_2 \mathcal{K}
        \\
        & \mathcal{S} = \sigma_3
        \\      
        & M_z = i \sigma_3 
    \end{align}
    By group induction, the two $M_x$-related bubble edges contribute 1D states on $e^1_1$ ($e^1_3$) with 0D zero modes described by 
    \begin{align}
        & \mathcal{T} = \mu_0 i \sigma_2 \mathcal{K}
        \\
        & \mathcal{S} = \mu_0 \sigma_3 
        \\     
        & M_x = i \mu_2 \sigma_0
        \\
        & M_z = \mu_3 i \sigma_3
    \end{align}    
    This induced projective representation can be decomposed into two $\bar{\Gamma}_5$'s with the same $\mathcal{S}$, which is exactly double of the irrep carried by $e^1_1$ ($e^1_3$) decoration.
    Thus we can see that the bubble equivalence generates double $e^1_1$ decoration and double $e^1_3$ decoration simultaneously, and the final classification is given by $\mathbb{Z}^2 / 2 \mathbb{Z} = \mathbb{Z} \oplus \mathbb{Z}_2$.

\end{enumerate}

\section{Boundary states protected protected by other symmetries\label{Appendix other boundary states}}

In Sec.\ref{Section boundary states} of the main text, we discussed some TSCS boundary states.
Here we give detailed analyses for other ones.

\subsection{wallpaper group $p1$\label{Appendix boundary state of P1}}

This wallpaper group has only translation symmetries.
The two decorations are simply 1D DIII TSCs stacking along $x/y$ directions.
Here we choose the decoration in $x$ direction for discussion, and the decoration in $y$ direction is essentially the same. 

\begin{figure}[h]
	 \centering
	    \includegraphics[width=0.5\textwidth]{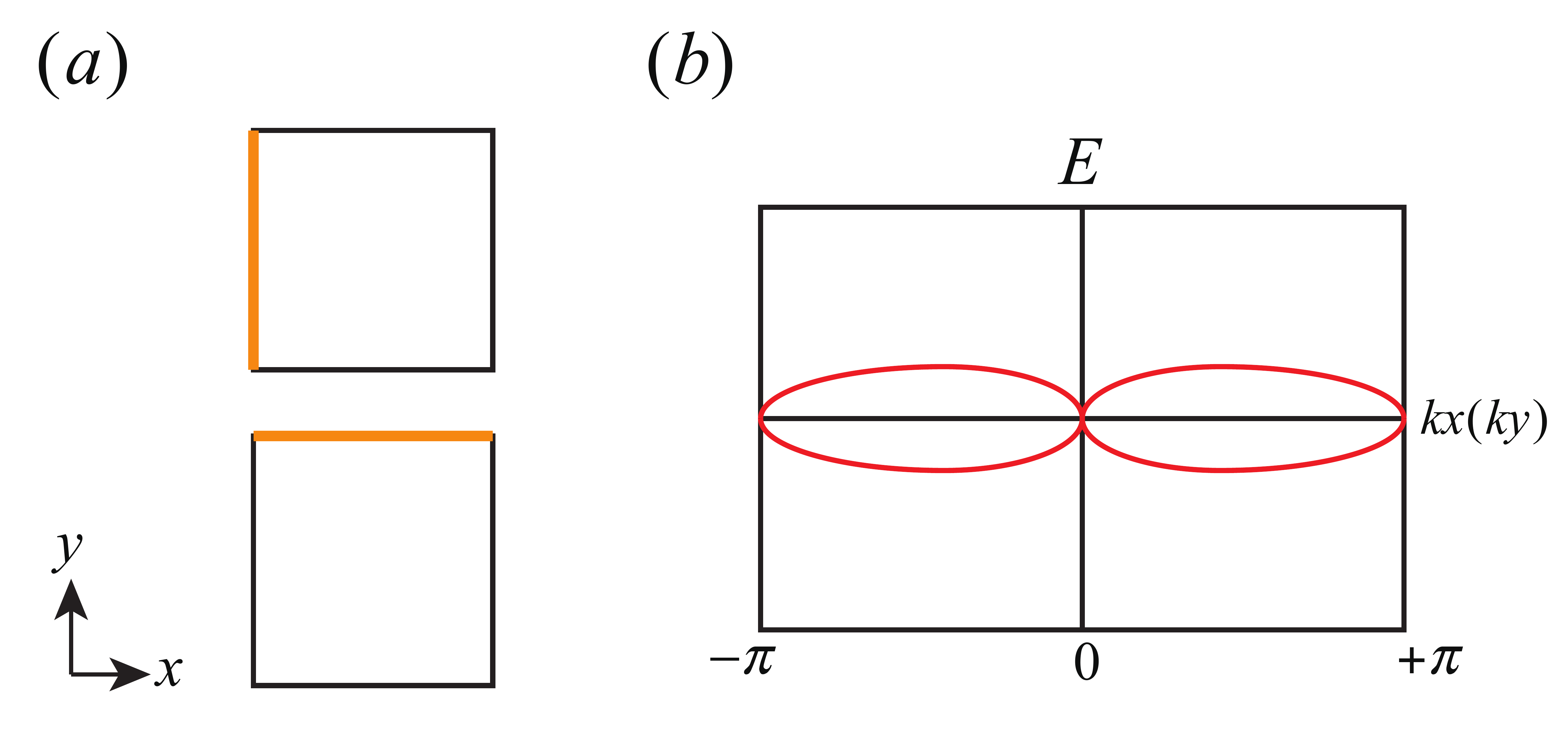}
	    \caption{\label{boundary state P1} (a)Decorations of $p1$, which are 1D DIII TSCs aligned along $x$ or $y$ direction. 
	    (b)Boundary spectrum along $x$ or $y$.}
\end{figure}

The real-space topological invariant, i.e., the $\mathbb{Z}_2$ topological number of the 1D DIII TSC, can be transformed into momentum space, denoted as $\nu$.
It can be evaluated at $k_x=0$ or $k_x = \pi$ line\cite{sato2017topological}
\begin{align}
    & (-1)^{\nu} = e^{i \gamma^{\text{I} (\text{II})}} 
    \\
    & \gamma^{
    \text{I}(\text{II})} = \int^{\pi}_{-\pi} d k_y \mathcal{A}^{\text{I}(\text{II})}_{(-)} (k_x = 0/ \pi, \; k_y)
\end{align}
where $\mathcal{A}^{\text{I}}$ and $\mathcal{A}^{\text{II}}$ are the Berry connection of the Kramer's pair formed by the occupied states.
When a boundary termination is made  preserving $x$ translation, the array of 0D Majorana modes form Majorana band in $x$ direction, as shown in Fig.\ref{boundary state P1}.
At $k_x = 0 $ and $k_x = \pi$, 
the effective symmetry class is DIII, since these two momenta are invariant under $\mathcal{T}$, $\mathcal{P}$ and $\mathcal{S}$.
The $\nu = 1$ invariant grantees a Kramer's pair of Majorana zero modes,
while at generic momentum, the zero modes are absent and a gap opens.

\subsection{wallpaper group $p2$, $\chi_{C_2} = \pm 1$}

\begin{figure}
	 \centering
	    \includegraphics[width=0.45\textwidth]{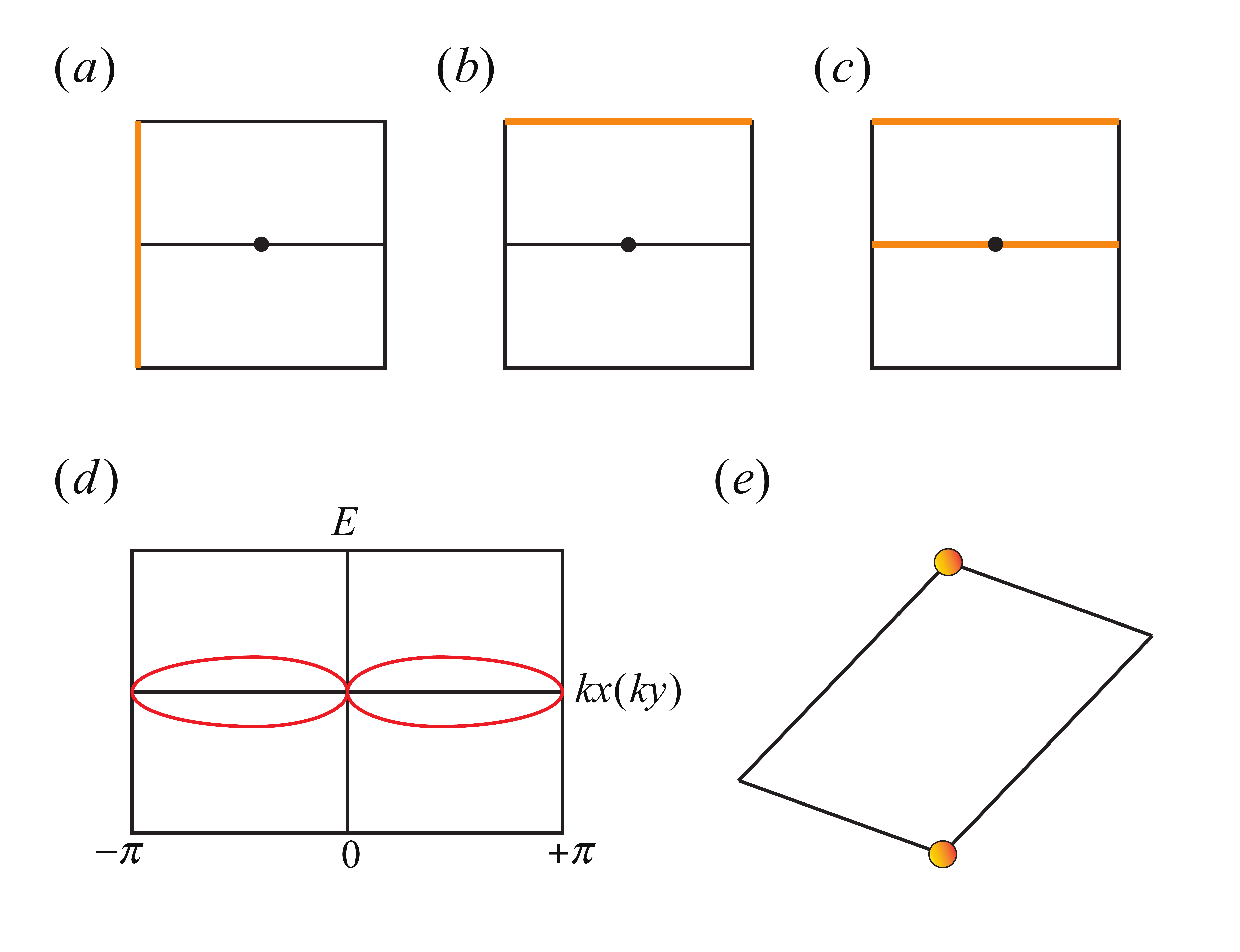}
	    \caption{\label{boundary state P2}Decorations and boundary states of $p2$. (a)The decoration with $x$-translation invariant. (b)The decoration with $y$-translation invariant. (c)The decoration with zero translation invariants but nonzero $C_2$ rotation invariant. (d) The boundary spectrum of either (a) or (b), which is the same as that of $p1$. (e) Majorana corner modes of decoration (c) with boundary geometry respecting $C_2$ symmetry, where we use a single circle to represent a Kramer's pair of Majorana modes.}
\end{figure}

We can see the decorations of $p1$ are still compatible with $p2$, as shown in Fig.\ref{boundary state P2}(a)-(b), and the corresponding boundary states are same as those of $p1$, as shown in Fig.\ref{boundary state P2}(d).
Besides, there is an additional decoration protected by $C_2$ rotation, see Fig.\ref{boundary state P2}(c).
Since this decoration has two 1D DIII TSCs in a unit cell, the translation invariant is zero due to its $\mathbb{Z}_2$ essence, and the corresponding boundary state, i.e., Majorana band formed in $x$/$y$ direction, is absent.
Instead, in a $C_2$-symmetric boundary termination, there are two $C_2$-related Majorana Kramer's pairs as corner modes, as shown in Fig.\ref{boundary state P2}(e).

We remark that the authors in Ref.\cite{chen2022topological} did more elaborate works for DIII TSCs protected by $p2$, with the overlapped part in agreement with our results.

\subsection{wallpaper group $p1m1$, $\chi_{M} = 1$}

In Sec.\ref{ subsection boundary state Pm chi_M = -1 } of the main text, we discussed the boundary state protected by $p1m1$ with $\chi_M = -1$. 
Here we focus on the $\chi_M = 1$ case.

\begin{figure}
	 \centering
	    \includegraphics[width=0.5\textwidth]{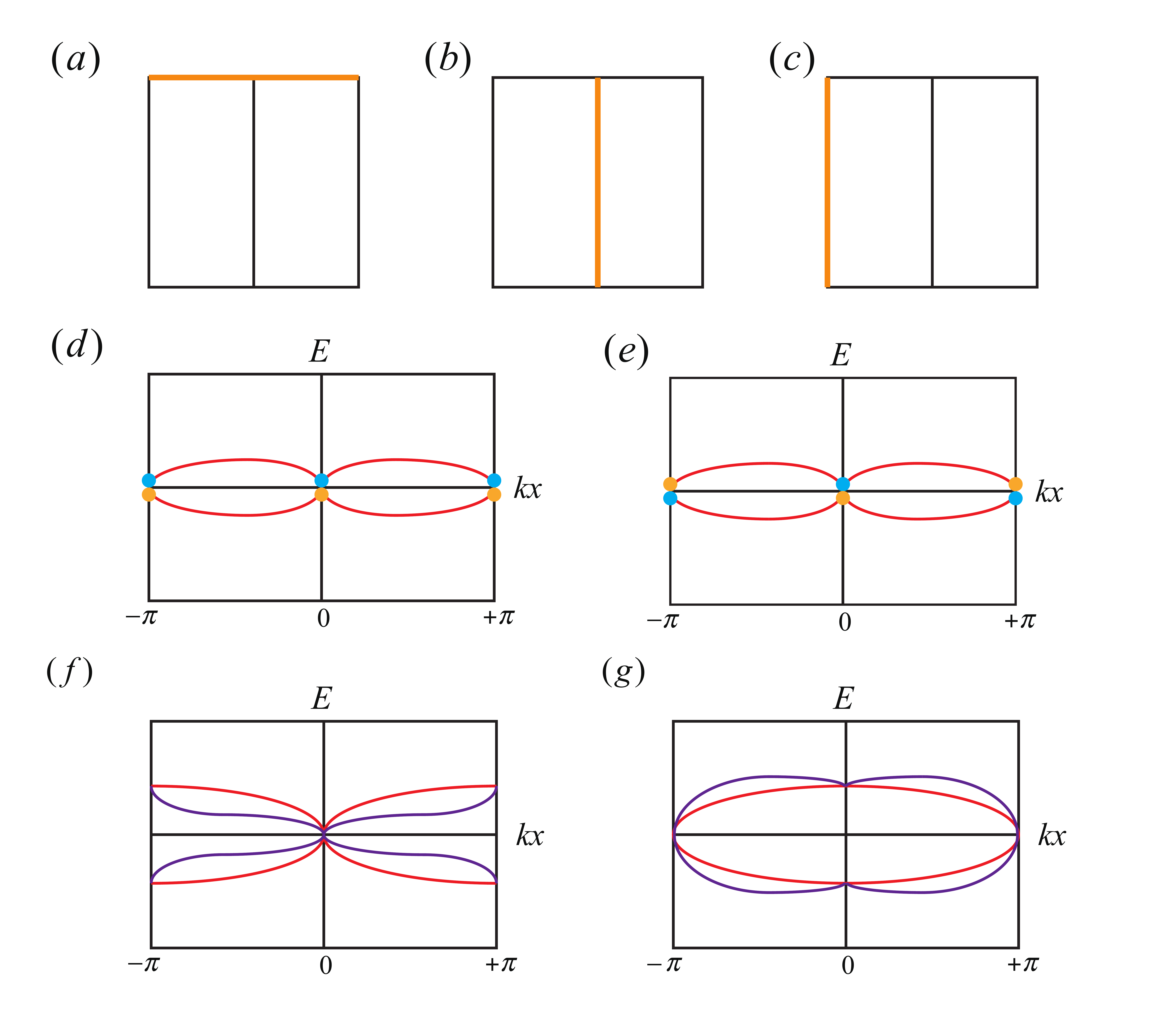}
	    \caption{\label{boundary state Pm chi_M = 1} Decorations and boundary states of $p1m1$ with $\chi_{M} = 1$. (a) Decoration protected by $y$-translation. (b)-(c) Decorations protected by mirror together with $x$-translation, with 1D building blocks placed at mirror line $x=0$ and $x=\frac{1}{2}$, respectively.
	    (d) Boundary spectrum along $k_x$ of the decoration in (b). (e) Boundary spectrum along $k_x$ of the decoration in (c). Note we use blue and orange dots to represent zero modes protected by different mirror winding numbers.
	    (f)-(g) Boundary spectrum along $k_x$ of the states from the combinations of (b) and (c). }
\end{figure}

The first decoration, as shown in Fig.\ref{boundary state Pm chi_M = 1} (a), is a stacking of 1D DIII TSCs along $y$ direction, which has the same boundary state as that of $p1$.
The other two decorations are protected by mirror $M_x$ and $x$-translation, and here we focus on them.

These two decorations are built by mirror-symmetric 1D TSCs, whose topological invariant is the mirror winding number $w_{M} = \frac{1}{2} ( w_{+} - w_{-} ) $. 
The momentum-space mirror winding numbers can be derived from the real-space ones.
Consider assigning mirror eigenstates with $M = \pm i$ at $x = n$ where $n \in \mathbb{Z}$, i.e.,
\begin{align}
    M \phi_x = \pm i \phi_{-x}
\end{align}
which is the case for the decoration in Fig.\ref{boundary state Pm chi_M = 1}(b).
After Fourier transformation, the momentum state satisfies
\begin{align}
    M \phi_k & = M ( \sum_{n} \phi_n e^{i k n} ) 
    = \sum_n e^{i k n} ( M \phi_n )
    \\
    & = \sum_n e^{i k n} (\pm i) \phi_{-n}
    = \pm i \sum_n e^{-i k n} \phi_n
    \\
    & = \pm i \sum_n e^{i k n} e^{-i 2 k n} \phi_n
\end{align}
At both $k=0$ and $k=\pi$, we have $ M \phi_k = \pm i \phi_k $, i.e., both $\phi_{k=0}$ and $\phi_{k=\pi}$ are mirror eigenstates with $M = \pm i$.

Similarly, if one puts mirror eigenstates at $x = n + \frac{1}{2}$ with $M = \pm i$, which is the case for the decoration in Fig.\ref{boundary state Pm chi_M = 1}, the momentum state $\phi_k$ satisfies
\begin{align}
    M \phi_k & = M ( \sum_{n} \phi_{n+\frac{1}{2}} e^{i k (n+\frac{1}{2})} ) 
    \\
    & = \sum_n e^{i k (n+\frac{1}{2})} ( M \phi_{n+\frac{1}{2}} )
    \\
    & = \sum_n e^{i k (n+\frac{1}{2})} (\pm i) \phi_{-(n+\frac{1}{2})}
    \\
    & = \pm i \sum_n e^{-i k (n+\frac{1}{2})} \phi_{n+\frac{1}{2}}
    \\
    & = \pm i \sum_n e^{i k (n+\frac{1}{2})} e^{-i 2 k (n+\frac{1}{2})} \phi_{n+\frac{1}{2}}
\end{align}
At $k=0$, we have $M \phi_k = \pm i \phi_k$, while at $k=\pi$, we have $ M \phi_k = \mp i \phi_k $, i.e., $\phi_{k=0}$ is a mirror eigenstate with $M = \pm i$ while $\phi_{k=\pi}$ is a mirror eigenstate with $M = \mp i$.

Then we apply the above results to analyze the boundary states. 
The decoration in Fig.\ref{boundary state Pm chi_M = 1}(b) has mirror line at $x=0$ decorated by 1D mirror-symmetric TSC with $w_{M} = 1$, i.e., $w_{+i} = +1$ and $w_{-i} = -1$.
It is straightforward to figure out its boundary state, as plotted in Fig.\ref{boundary state Pm chi_M = 1}(d).
At $k_x = 0$ and $k_x = \pi$, the mirror winding numbers $w_M$ are both $1$, protecting a pair of zero modes.
At generic momentum, there is a gap opened.
Similarly, the decoration in Fig.\ref{boundary state Pm chi_M = 1}(c), which has mirror line at $x=\frac{1}{2}$ decorated by 1D mirror-symmetric TSC with $w_{M} = +1$, has boundary state shown in Fig.\ref{boundary state Pm chi_M = 1}(e).
At $k_x = 0$, the mirror winding number $w_M$ is $+1$, while at $k_x = \pi$, $w_M$ becomes $-1$ since the two mirror sectors are exchanged.
At both the two momenta, there are a pair of zero modes.
Different with the boundary state in Fig.\ref{boundary state Pm chi_M = 1}, here the zero modes at $k_x=0$ and $k_x = \pi$ are protected by opposite $w_M$. 

We further consider combinations of the two decorations.
First we put 1D mirror-symmetric TSC of $w_M = +1$ at both $x= 0 $ and $x= \frac{1}{2} $.
It is easy to see the momentum mirror winding numbers are given as 
\begin{align}
    w_M(k_x = 0) = 1 + 1 = +2 
\end{align}
and 
\begin{align}
w_M(k_x = \pi) = 1-1 =0
\end{align}
Consequently, at $k_x = 0$ there are two pairs of zero modes, while at $k_x = \pi$, there is no zero mode but a gap opened, as shown in Fig.\ref{boundary state Pm chi_M = 1}(f).
Next we put 1D mirror-symmetric TSC of $w_M = +1$ at $x= 0 $, while 1D mirror-symmetric TSC of $w_M = -1$ at $x= \frac{1}{2} $.
The momentum mirror winding numbers are given as 
\begin{align}
w_M(k_x = 0) = 1 - 1 = 0
\end{align}
and 
\begin{align}
w_M(k_x = \pi) = 1 + 1 = +2
\end{align}
Contrary to the first combination, there are no zero modes at $k_x = 0$ but two pairs of zero modes at $k_x = \pi$, as shown in Fig.\ref{boundary state Pm chi_M = 1}(g).

\subsection{Layer group symmetries}

As we have mentioned previously, the generating symmetry operations of LGs are those of WGs plus mirror or glide in the vertical direction.
For brevity, we only consider the boundary states protected solely by vertical mirror and vertical glide. 

For the LG generated by vertical mirror reflection, i.e. $p11m$, its decorations are generated by stacking 1D mirror-symmetric TSCs with $\chi_{M_z} = \pm 1$ along $x$ or $y$ directions, which are similar as those of WG $p1$, with the difference of building blocks.
The only thing to note is that, for $\chi_{M_z} = 1$, the decorations are $\mathbb{Z}$-type, while for $\chi_{M_z}$, the decorations are $\mathbb{Z}_2$-type.
At $k_{x/y} = 0$ and $k_{x/y} = \pi$, there are the same number of Majorana zero modes, as also discussed in Ref.\cite{zhang2013topological}.

\begin{figure}
	 \centering
	    \includegraphics[width=0.5\textwidth]{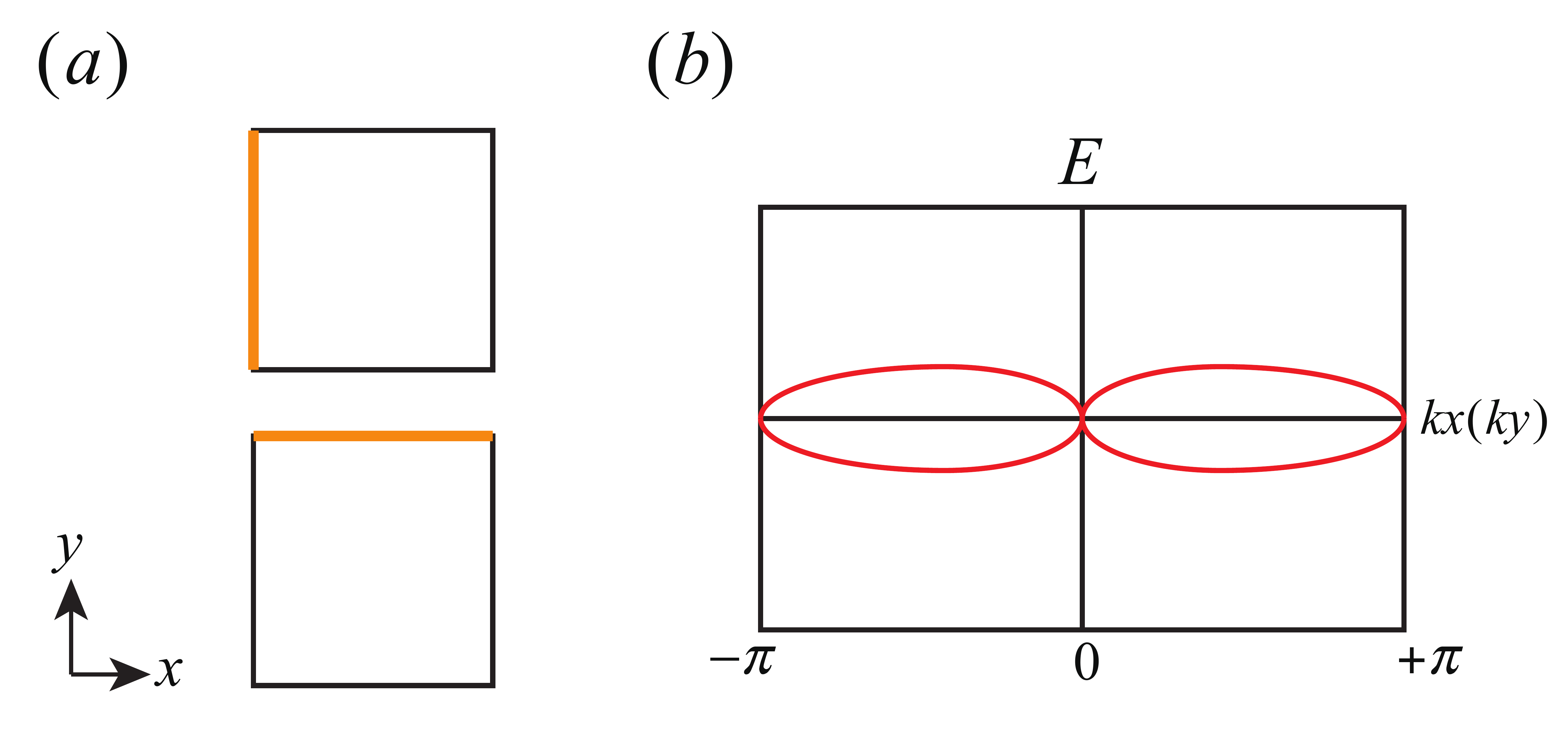}
	    \caption{\label{surface_state_P11m} The decorations and boundary states for $p11m$ are similar with those of $p1$. Only note that for $\chi_{M}$, the decorations are $\mathbb{Z}$-type instead if $\mathbb{Z}_2$-type, which means there can be $n \in \mathbb{Z}$ pairs of Majorana zeros.}
\end{figure}

The LG generated by vertical glide $\{ M_z | t_x t_y 0 \}$, where $t_x, t_y = 0, \frac{1}{2}$, is $p1g1$.
The TCSC boundary states are already studied in Ref.\cite{shiozaki2016topology}.

\section{Tables for classification results\label{Appendix classification tables}}

\begin{table}[H]
	\centering
	\begin{tabular}{cccccc}
\hline\hline	
\textbf{WG} &  \textbf{Rep} & \textbf{classification} & \textbf{decorations}
\\
\hline
 $p1$ & $A$ & $\mathbb{Z}_2^2$  
     &  $\{e^1_1\}$, $\{e^1_2\}$  &   &  
\\
\hline
 $p2$ & $A$ & $\mathbb{Z}_2^3$  
     &  $\{e^1_1\}$, $\{e^1_2\}$, $\{e^1_3\}$  &   &  
\\
 $p2$ & $B$ & $\mathbb{Z}_2^3$  
     &  $\{e^1_1\}$, $\{e^1_2\}$, $\{e^1_3\}$  &   &  
\\
\hline
 $p1m1$ & $A'$ & $\mathbb{Z}^2 \times \mathbb{Z}_2$  
     & $\{e^1_1\}$, $\{e^1_3\}$, $\{e^1_2\}$   &   &  
\\
 $p1m1$ & $A''$ & $\mathbb{Z}_2^3$
     &  $\{e^1_1\}$, $\{e^1_2\}$, $\{e^1_3\}$  &   & 
\\
\hline
 $p1g1$ & $A'$ &   $\mathbb{Z}_2^2$
     &  $\{e^1_1\}$, $\{e^1_2\}$  &   &  
\\
 $p1g1$ & $A''$ &  $\mathbb{Z}_2^2$
     &  $\{e^1_1\}$, $\{e^1_2\}$  &   & 
\\
\hline
 $c1m1$ & $A'$ &   $\mathbb{Z} \times \mathbb{Z}_2$
     & $\{e^1_1\}$, $\{e^1_2\}$   &   &  
\\
 $c1m1$ & $A''$ &  $\mathbb{Z}_2^2$
     & $\{e^1_1\}$, $\{e^1_2\}$   &   & 
\\
\hline
 $p2mm$ & $A_1$ & $\mathbb{Z}^4$
     &  $\{e^1_1\}$, $\{e^1_2\}$, $\{e^1_3\}$, $\{e^1_4\}$  &   &  
\\
 $p2mm$ & $A_2$ & $\mathbb{Z}_2^4$
     & $\{e^1_1\}$, $\{e^1_2\}$, $\{e^1_3\}$, $\{e^1_4\}$   &   & 
\\
 $p2mm$ & $B_1$ & $\mathbb{Z}_2^2$
     & $\{e^1_2\}$, $\{e^1_4\}$   &   &
\\
 $p2mm$ & $B_2$ & $\mathbb{Z}_2^2$
     & $\{e^1_1\}$, $\{e^1_3\}$   &   & 
\\
\hline
 $p2mg$ & $A_1$ & $\mathbb{Z}_2^2 \times \mathbb{Z} $
     & $\{e^1_1\}$, $\{e^1_2\}$, $\{e^1_3\}$  &   &  
\\
 $p2mg$ & $A_2$ & $\mathbb{Z}_2^3$
     &  $\{e^1_1\}$, $\{e^1_2\}$, $\{e^1_3\}$ & 
\\
 $p2mg$ & $B_1$ &  $\mathbb{Z}_2^3$
     & $\{e^1_1\}$, $\{e^1_2\}$, $\{e^1_3\}$   &   &
\\
 $p2mg$ & $B_2$ & $ \mathbb{Z}_2^2 \times \mathbb{Z}$
     &  $\{e^1_1\}$, $\{e^1_2\}$, $\{e^1_3\}$  &   & 
\\
\hline
 $p2g g$ & $A_1$ & $\mathbb{Z}_2^2$
     &  $\{e^1_1\}$, $\{e^1_2\}$  &   &  
\\
 $p2g g$ & $B_1$ & $\mathbb{Z}_2^2$
     &  $\{e^1_1\}$, $\{e^1_2\}$  &   & 
\\
 $p2g g$ & $A_2$ & $\mathbb{Z}_2^2$
     &  $\{e^1_1\}$, $\{e^1_2\}$  &   &
\\
 $p2g g$ & $B_2$ & $\mathbb{Z}_2^2$
     &  $\{e^1_1\}$, $\{e^1_2\}$  &   & 
\\
\hline
 $c2mm$ & $A_1$ & $\mathbb{Z}^2 \times \mathbb{Z}_2$
     &  $\{e^1_1\}$, $\{e^1_2\}$, $\{e^1_3\}$  &   &  
\\
 $c2mm$ & $B_1$ & $\mathbb{Z}_2^2$
     &  $\{e^1_1\}$, $\{e^1_3\}$  &   & 
\\
 $c2mm$ & $A_2$ & $\mathbb{Z}_2^3$
     &  $\{e^1_1\}$, $\{e^1_2\}$, $\{e^1_3 \}$  &   &
\\
 $c2mm$ & $B_2$ & $\mathbb{Z}_2^2$
     & $\{e^1_2\}$, $\{e^1_3\}$   &   & 
\\
\hline
 $p4$ & $A$ & $\mathbb{Z}_2^2$
     &  $\{e^1_1\}$, $\{e^1_2\}$  &   & 
\\
 $p4$ & $B$ & $\mathbb{Z}_2^2$
     &  $\{e^1_1\}$, $\{e^1_2\}$  &   & 
\\
\hline
 $p4mm$ & $A_1$ & $\mathbb{Z}^3$
     &  $\{e^1_1\}$, $\{e^1_2\}$, $\{e^1_3\}$  &   & 
\\
 $p4mm$ & $B_1$ & $\mathbb{Z}^2 \times \mathbb{Z}_2$
     &  $\{e^1_1\}$, $\{e^1_2\}$, $\{e^1_3\}$  &   & 
\\
 $p4mm$ & $A_2$ & $\mathbb{Z}_2^3$
     &  $\{e^1_1\}$, $\{e^1_2\}$, $\{e^1_3\}$  &   & 
\\
 $p4mm$ & $B_2$ & $\mathbb{Z}_2^2 \times \mathbb{Z}$
     & $\{e^1_1\}$, $\{e^1_2\}$, $\{e^1_3\}$   &   & 
\\
\hline
 $p4gm$ & $A_1$ & $\mathbb{Z}_2 \times \mathbb{Z}$
     &  $\{e^1_1\}$, $\{e^1_2\}$  &   & 
\\
 $p4gm$ & $B_1$ & $ \mathbb{Z}_2^2$
     &  $\{e^1_1\}$, $\{e^1_2\}$  &   & 
\\
 $p4gm$ & $A_2$ & $ \mathbb{Z}_2^2$
     &  $\{e^1_1\}$, $\{e^1_2\}$  &   & 
\\
 $p4gm$ & $B_2$ & $\mathbb{Z}_2 \times \mathbb{Z} $
     &  $\{e^1_1\}$, $\{e^1_2\}$  &   & 
\\
\hline
 $p3$ & $A_1$ & $\text{N/A}$
     &  $\text{N/A}$  &   & 
\\
\hline
 $p3m1$ & $A_1$ & $\mathbb{Z}$
     &  $\{e^1_1 + e^1_2 + e^1_3\}$  &   & 
\\
 $p3m1$ & $B_1$ & $\mathbb{Z}_2$
     &  $\{e^1_1 + e^1_2 + e^1_3\}$  &   & 
\\
\hline
 $p31m$ & $A_1$ & $\mathbb{Z}$
     &  $\{e^1_1 + e^1_2 + e^1_3\}$  &   & 
\\
 $p31m$ & $B_1$ & $\mathbb{Z}_2$
     &  $\{e^1_1 + e^1_2 + e^1_3\}$  &   & 
\\
\hline
 $p6$ & $A$ & $\mathbb{Z}_2$
     &  $\{e^1_1 + e^1_2\}$  &   & 
\\
 $p6$ & $B$ & $\mathbb{Z}_2$
     & $\{e^1_1 + e^1_2\}$   &   & 
\\
\hline
 $p6mm$ & $A_1$ & $\mathbb{Z}^2$
     &  $\{e^1_1 + e^1_2 \}$,  $\{e^1_3\}$  &   & 
\\
 $p6mm$ & $B_1$ & $\mathbb{Z}_2$
     &  $\{e^1_3\}$  &   & 
\\
 $p6mm$ & $A_2$ & $\mathbb{Z}_2^2$
     &  $\{e^1_1 + e^1_2 \}$, $\{ e^1_3 \}$  &   & 
\\
 $p6mm$ & $B_2$ & $\mathbb{Z}_2$
     &  $\{e^1_1 + e^1_2 \}$  &   & 
\\
\hline\hline						
\end{tabular}
\caption{\label{table classification WGs} 
TCSCs by wire construction for WGs. The First column are WG names. 
The second column are representations of the gap function. 
The third column are classifications.
The fourth column are 1-cell decorations in correspondence with classifications.
$\text{N/A}$ means trivial classification and decorations. 
For instance, for $p6mm$ with $A_1$ representation, there are two types of independent decorations, both with $\mathbb{Z}$ classification.
The generator for the first is constructed by decorating $e^1_1$ and $e^1_2$ simultaneously, while the generator for the latter is constructed by decorating $e^1_3$.
Readers can refer to the cell-complex that we give in Appendix.\ref{Appendix cell complex WGs} to see what these 1-cells are.}
\end{table}

\begin{table}[H]
	\centering
	\begin{tabular}{c|ccccc}
\hline\hline	
\textbf{WG} &  \textbf{LG} 
\\
\hline
 $p1$ & \color{blue}{$p1$} &   
     &    &   &  
\\
 $p2$ & \color{blue}{$p112$} &  
     &    &   &  
\\
 $p1m1$ & $p211$, \color{blue}{$pm11$}, &  
     &    &   &  
\\
 $p1g1$ & $p 2_1 11$, \color{blue}{$p b 11$} &   
     &    &   &  
\\
 $c1m1$ & $c211$, \color{blue}{$cm11$} &   
     &    &   &  
\\
 $p2mm$ & $p222$, \color{blue}{$pmm2$} & 
     &    &   &  
\\
 $p2mg$ & $p2_1 22$, \color{blue}{$pma2$} & 
     &    &   &  
\\
 $p2g g$ & $p2_1 2_1 2$, \color{blue}{$pba2$} & 
     &    &   &  
\\
 $c2mm$ & $c222$, \color{blue}{$cmm2$} & 
     &    &   &  
\\
 $p4$ & $p\text{-}4$, \color{blue}{$p4$}
     &    &   & 
\\
 $p4mm$ & $p422$, $p\text{-}42m$, $p\text{-}4m2$, \color{blue}{$p4mm$} & 
     &    &   & 
\\
 $p4gm$ & $p42_1 2$, $p\text{-}4 2_1 m$, $p\text{-}4b2$, \color{blue}{$p4bm$} & 
     &    &   & 
\\
 $p3$ & \color{blue}{$p3$} & 
     &    &   & 
\\
 $p3m1$ & $p312$, \color{blue}{$p3m1$} & 
     &    &   & 
\\
 $p31m$ & $p321$, \color{blue}{$p31m$} & 
     &    &   & 
\\
 $p6$ & \color{blue}{$p6$} & 
     &    &   & 
\\
 $p6mm$ & $p622$, \color{blue}{$p6mm$} & 
     &    &   & 
\\
\hline\hline						
\end{tabular}
\caption{\label{table WGs LGs} 
LGs whose wire construction results are the same as corresponding WGs.
The blue one is the LG that shares completely the same group elements with the corresponding WG.}
\end{table}

\begin{table}[H]
	\centering
	\begin{tabular}{cccccc}
\hline\hline	
\textbf{LG} &  \textbf{Rep} & \textbf{classification} & \textbf{decorations}
\\
\hline
 $p\text{-}1$ & $A_g$ & $\text{N/A}$  
     &  $\text{N/A}$  &   &  
\\
 $p\text{-}1$ & $A_u$ & $\mathbb{Z}_2^3$  
     &  $\{e^1_1\}$, $\{e^1_2\}$, $\{e^1_3\}$  &   &  
\\
\hline
 $p\text{-}3$ & $A_g$ & $\text{N/A}$
     &  $\text{N/A}$  &   & 
\\
 $p\text{-}3$ & $A_u$ & $\mathbb{Z}_2$
     &  $\{e^1_1 + e^1_2\}$  &   & 
\\
\hline
 $p11m$ & $A'$ & $\mathbb{Z}^2$
     &  $\{e^1_1\}$, $\{e^1_2\}$  &   & 
\\
 $p11m$ & $A''$ & $\mathbb{Z}^2_2$
     &  $\{e^1_1\}$, $\{e^1_2\}$  &   & 
\\
\hline
 $p11a$ & $A'$ & $\mathbb{Z}_2^2$
     &  $\{e^1_1\}$, $\{e^1_2\}$  &   & 
\\
 $p11a$ & $A''$ & $\mathbb{Z}_2^2$
     &  $\{e^1_1\}$, $\{e^1_2\}$  &   & 
\\
\hline
 $p112/m$ & $A_g$ & $\text{N/A}$
     &  $\text{N/A}$  &   & 
\\
 $p112/m$ & $A_u$ & $\mathbb{Z}^3_2$
     &  $\{e^1_1\}$, $\{e^1_2\}$, $\{e^1_3\}$  &   & 
\\
 $p112/m$ & $B_g$ & $\text{N/A}$
     &  $\text{N/A}$  &   & 
\\
 $p112/m$ & $B_u$ & $\mathbb{Z}^2 \times \mathbb{Z}_2$
     &  $\{e^1_1\}$, $\{e^1_2\}$, $\{ e^1_1 + e^1_3\}$  &   & 
     \\
\hline
 $pm2m$ & $A_1$ & $\mathbb{Z}$
     &  $\{e^1_2\}$  &   & 
\\
 $pm2m$ & $A_2$ & $\mathbb{Z}^2 \times \mathbb{Z}_2$
     &  $\{e^1_1\}$, $\{e^1_3\}$, $\{ e^1_2 \}$  &   & 
\\
 $pm2m$ & $B_1$ & $\mathbb{Z} \times \mathbb{Z}_2$
     & $\{e^1_1\}$, $\{ e^1_1 + e^1_3\}$   &   & 
\\
 $pm2m$ & $B_2$ & $\mathbb{Z}^2 \times \mathbb{Z}_2$
     &  $\{e^1_1\}$, $\{e^1_3\}$, $\{ e^1_2 \}$  &   & 
\\
\hline
 $p3/m$ & $A'$ & $\text{N/A}$
     &  $\text{N/A}$  &   & 
\\
 $p3/m$ & $A''$ & $\text{N/A}$
     &  $\text{N/A}$  &   & 
\\
\hline
 $p4/m$ & $A_g$ & $\text{N/A}$
     &  $\text{N/A}$  &   & 
\\
 $p4/m$ & $A_u$ & $\mathbb{Z}_2^2$ 
     &  $\{e^1_1\}$, $\{e^1_2\}$  &   & 
\\
 $p4/m$ & $B_g$ & $\mathbb{Z}_2$
     &  $\{e^1_1 + e^1_2\}$  &   & 
\\
 $P4/m$ & $B_u$ & $\mathbb{Z}_2^2$
     & $\{e^1_1\}$, $\{e^1_2\}$   &   & 
\\
\hline
 $p6/m$ & $A_g$ & $\text{N/A}$
     &  $\text{N/A}$  &   & 
\\
 $p6/m$ & $A_u$ & $\mathbb{Z}_2$
     &  $\{e^1_1 + e^1_2\}$  &   & 
\\
 $p6/m$ & $B_g$  & $\text{N/A}$
     &  $\text{N/A}$  &   & 
\\
 $p6/m$ & $B_u$ & $\mathbb{Z}_2$
     &  $\{e^1_1 + e^1_2\}$  &   & 
\\
\hline\hline						
\end{tabular}
\caption{\label{table classification LGs} Wire construction results for some LGs, which can not be directly inferred from those of corresponding WGs.
To avoid confusion, we remark that for LG $pm2m$, whose group elements are $\bm{1}$, $M_x$, $C_{2y}$ and $M_z$, the representations of gap function are given as follows.
$A_1$: $\chi_{M_x} = 1, \chi_{M_z} = 1$, $A_2$: $\chi_{M_x} = 1, \chi_{M_z} = -1$, $B_1$: $\chi_{M_x} = -1, \chi_{M_z} = 1$, $B_2$: $\chi_{M_x} = -1, \chi_{M_z} = -1$.
To see what these 1-cells are, readers can refer to Appendix.\ref{Appendix cell complex LGs}, where we plot the cell complex for each LG in this table.}
\end{table}

\newpage

\section{Cell complex for wallpaper groups \label{Appendix cell complex WGs}}

\begin{figure}[H]
	 \centering
	 \includegraphics[width=0.3\textwidth]{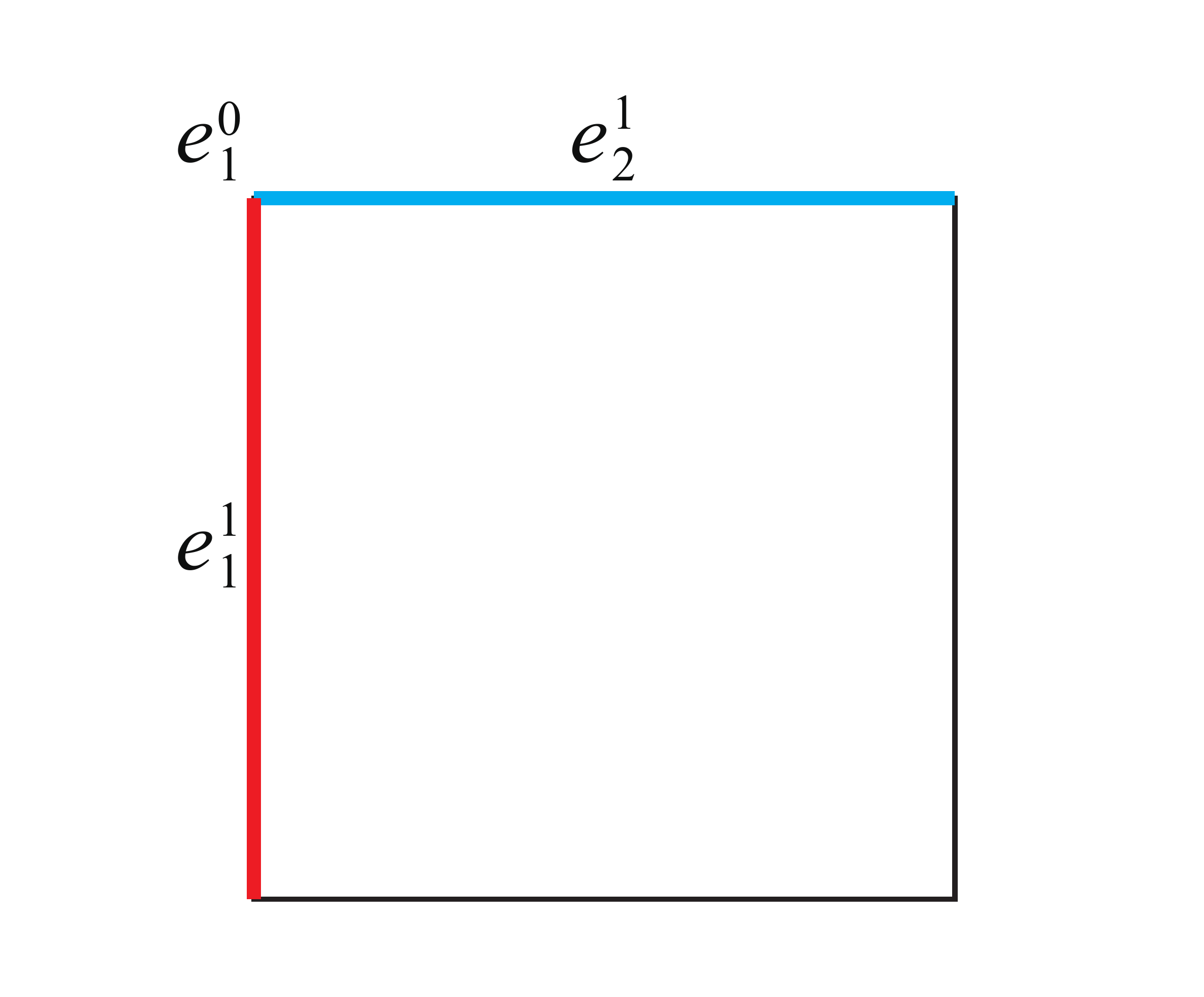}
	 \caption{\label{appendix_cell_complex_p1} WG $p1$  }
\end{figure}

\begin{figure}[H]
    \centering
	    \includegraphics[width=0.3\textwidth]{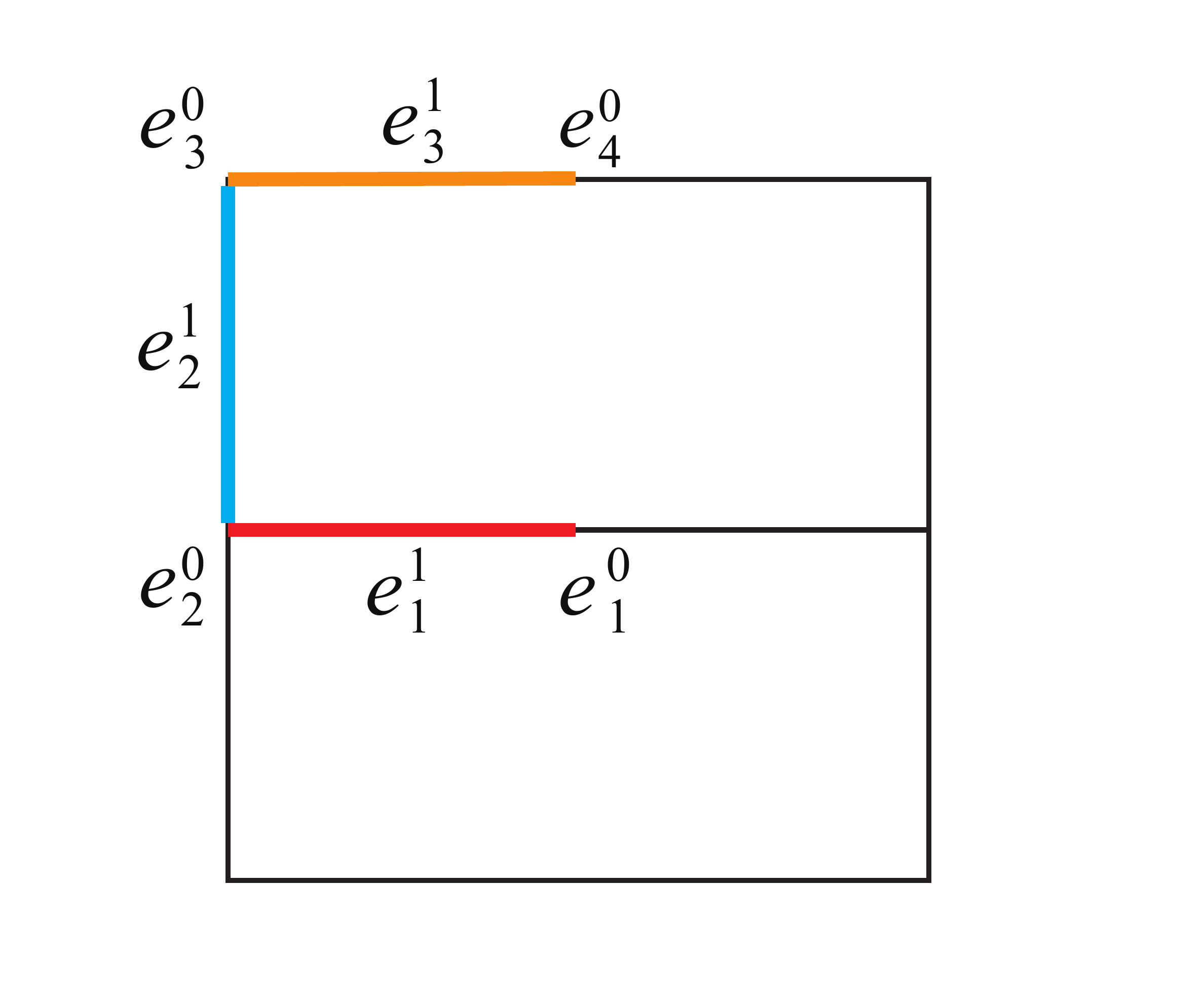}
	    \caption{\label{appendix_cell_complex_p2} 
	WG $p2$}
\end{figure}

\begin{figure}[H]
	\centering
	    \includegraphics[width=0.3\textwidth]{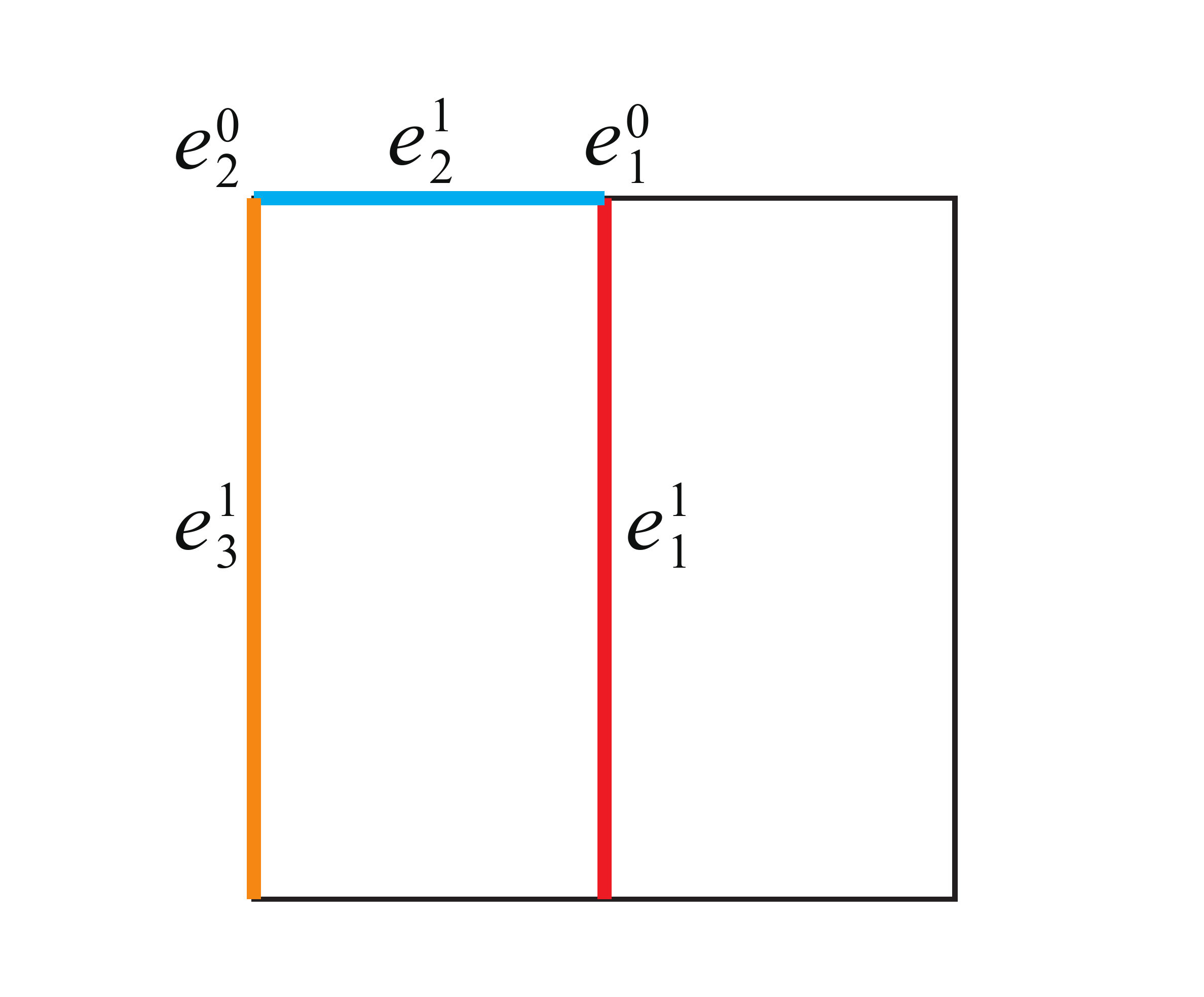}
	    \caption{\label{appendix_cell_complex_p1m1} 
	WG $p1m1$}
\end{figure}

\begin{figure}[H]
	\centering
	 \includegraphics[width=0.3\textwidth]{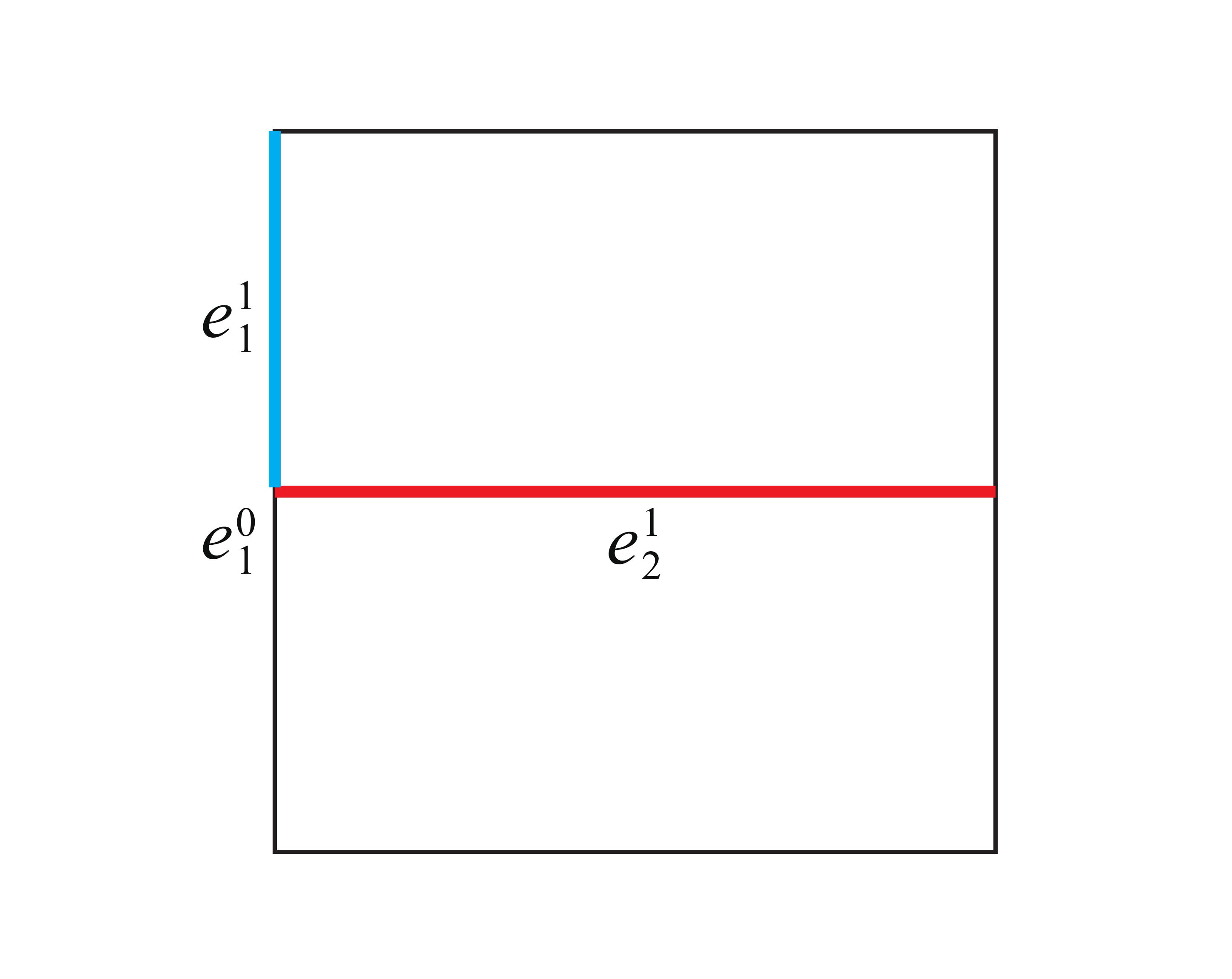}
	 \caption{\label{appendix_cell_complex_p1g1} 
	WG $p1g1$}
\end{figure}

\begin{figure}[H]
	\centering
	\includegraphics[width=0.33\textwidth]{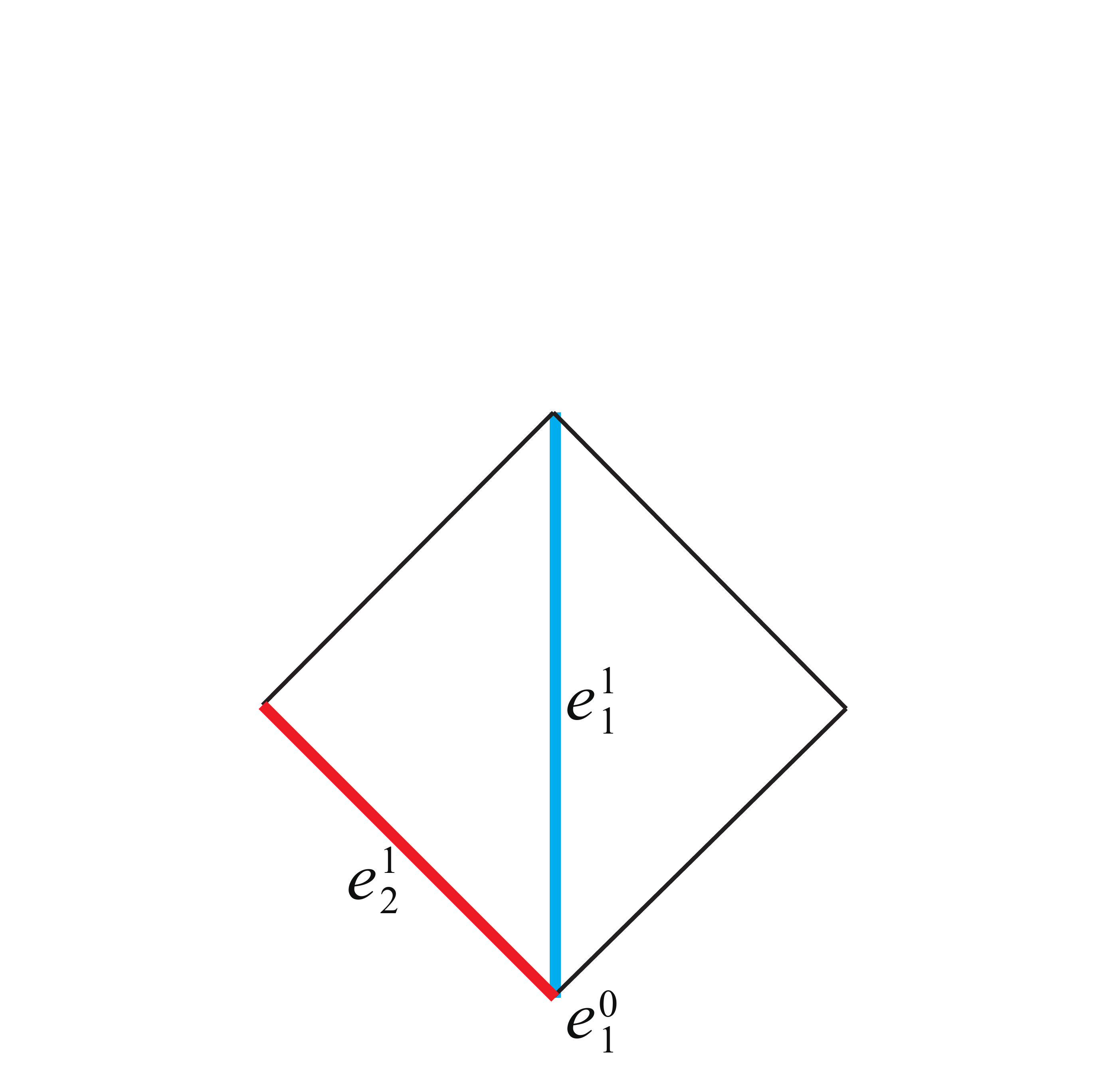}
	\caption{\label{appendix_cell_complex_c1m1} 
	cell complex of WG $c1m1$.}
\end{figure}

\begin{figure}[H]
    \centering
	\includegraphics[width=0.3\textwidth]{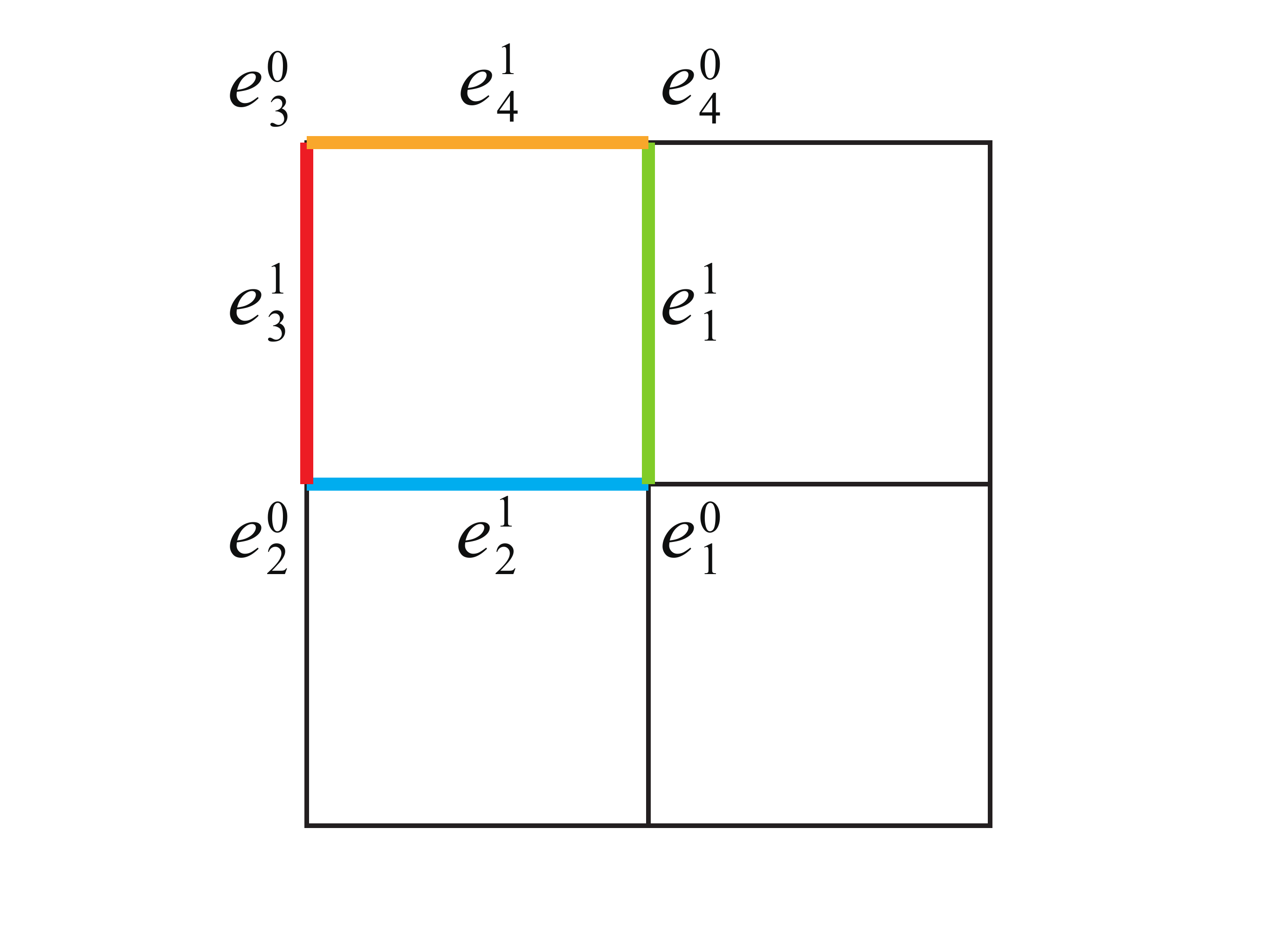}
	\caption{\label{appendix_cell_complex_p2mm} 
	WG $p2mm$}
\end{figure}

\begin{figure}[H]
	\centering
	    \includegraphics[width=0.3\textwidth]{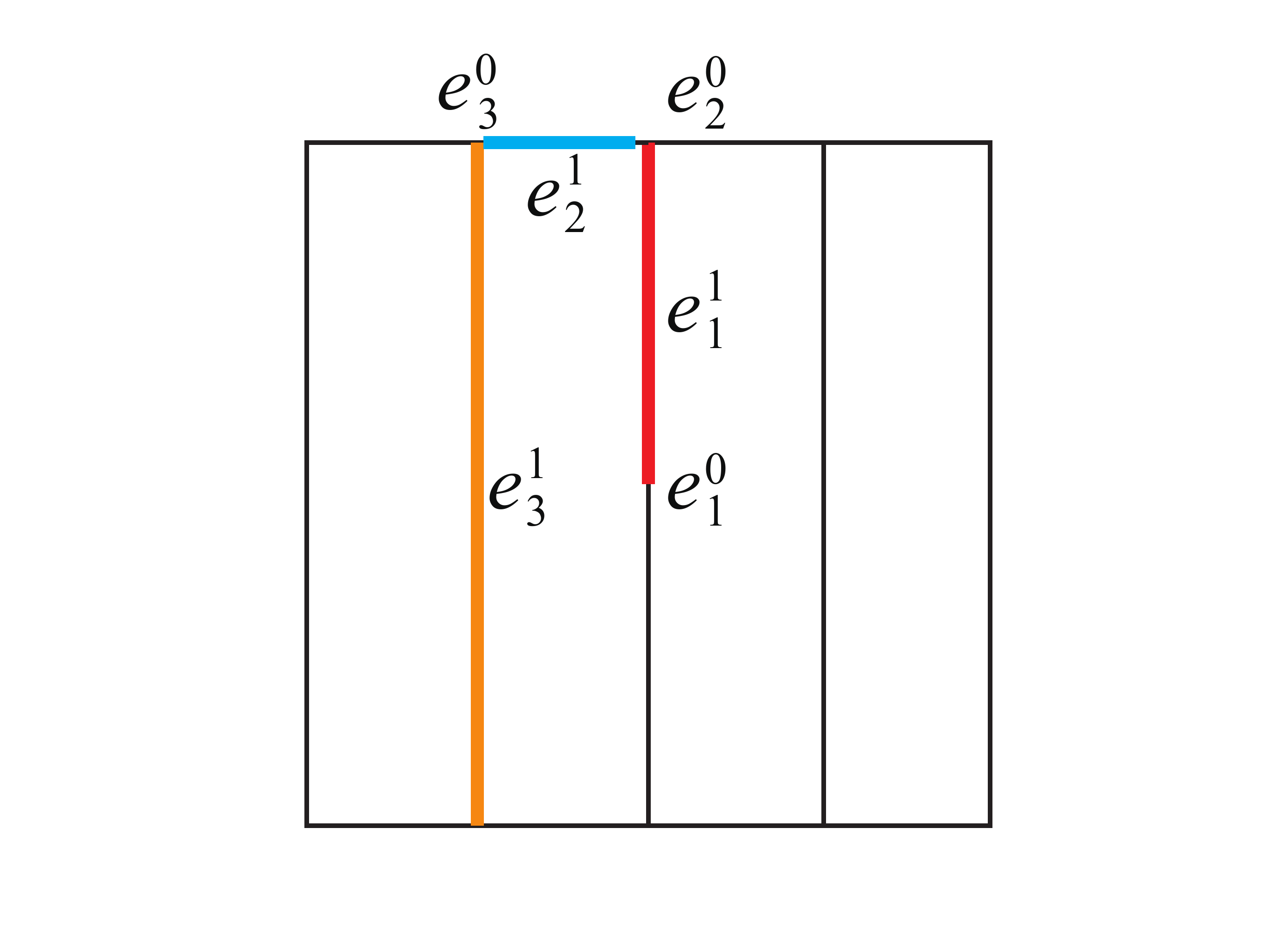}
	    \caption{\label{appendix_cell_complex_p2mg} 
	cell complex of WG $p2mg$.}
\end{figure}

\begin{figure}[H]
	\centering
	 \includegraphics[width=0.3\textwidth]{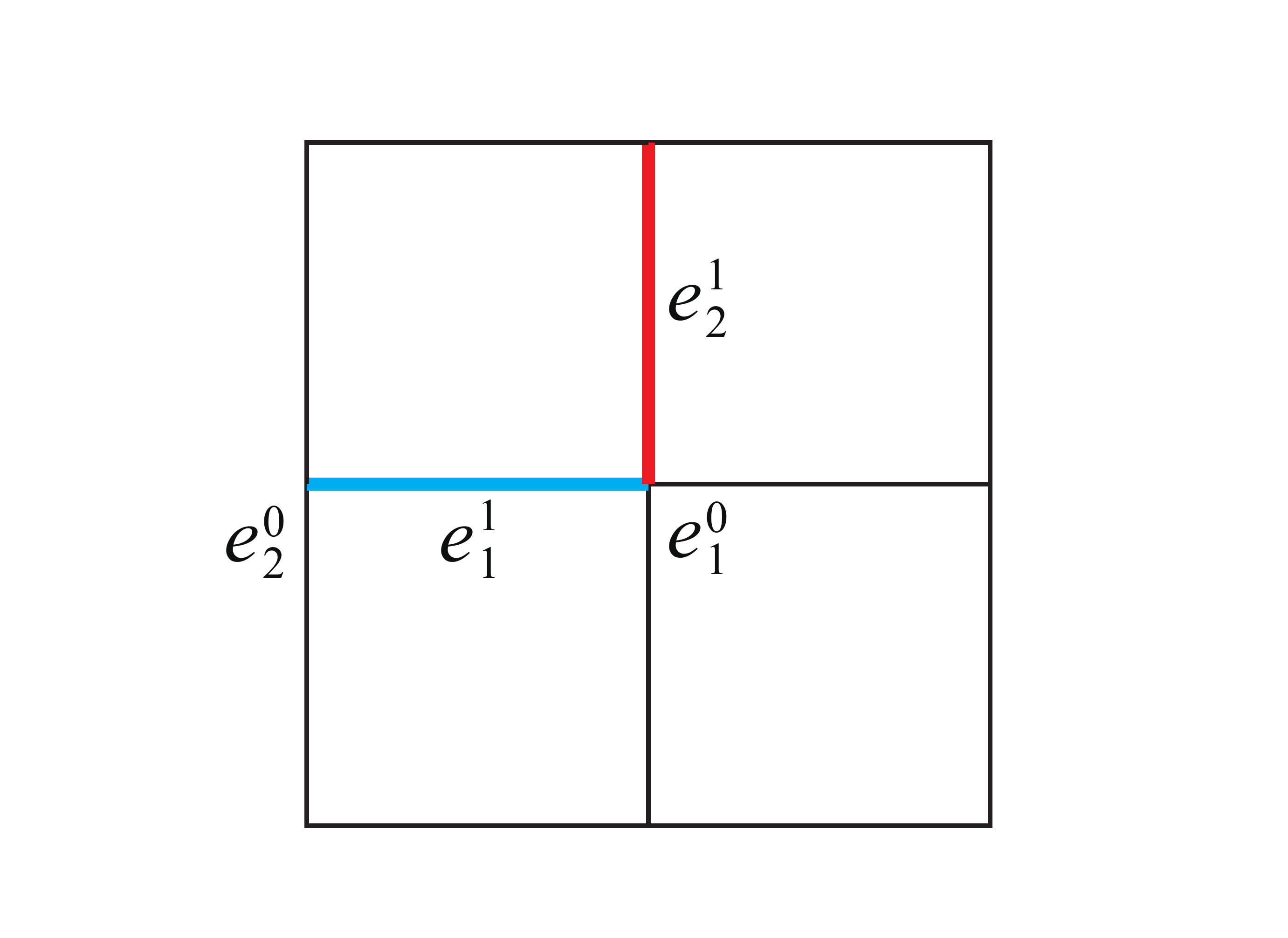}
	    \caption{\label{appendix_cell_complex_p2gg} 
	cell complex of WG $p2gg$.}
\end{figure}

\begin{figure}[H]
    \centering
	 \includegraphics[width=0.33\textwidth]{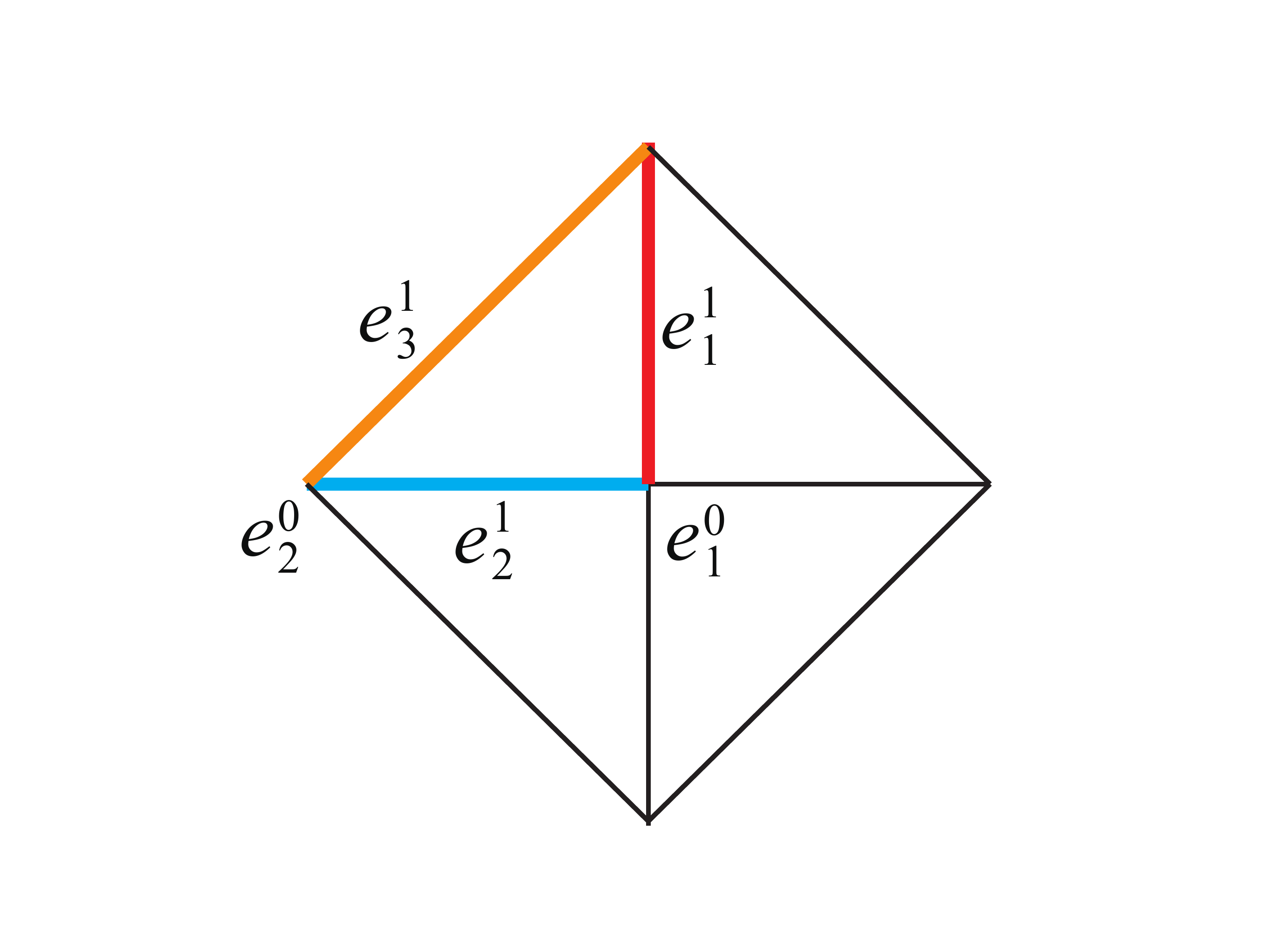}
	 \caption{\label{appendix_cell_complex_c2mm} 
	cell complex of WG $c2mm$.}
\end{figure}

\begin{figure}[H]
	\centering
	 \includegraphics[width=0.3\textwidth]{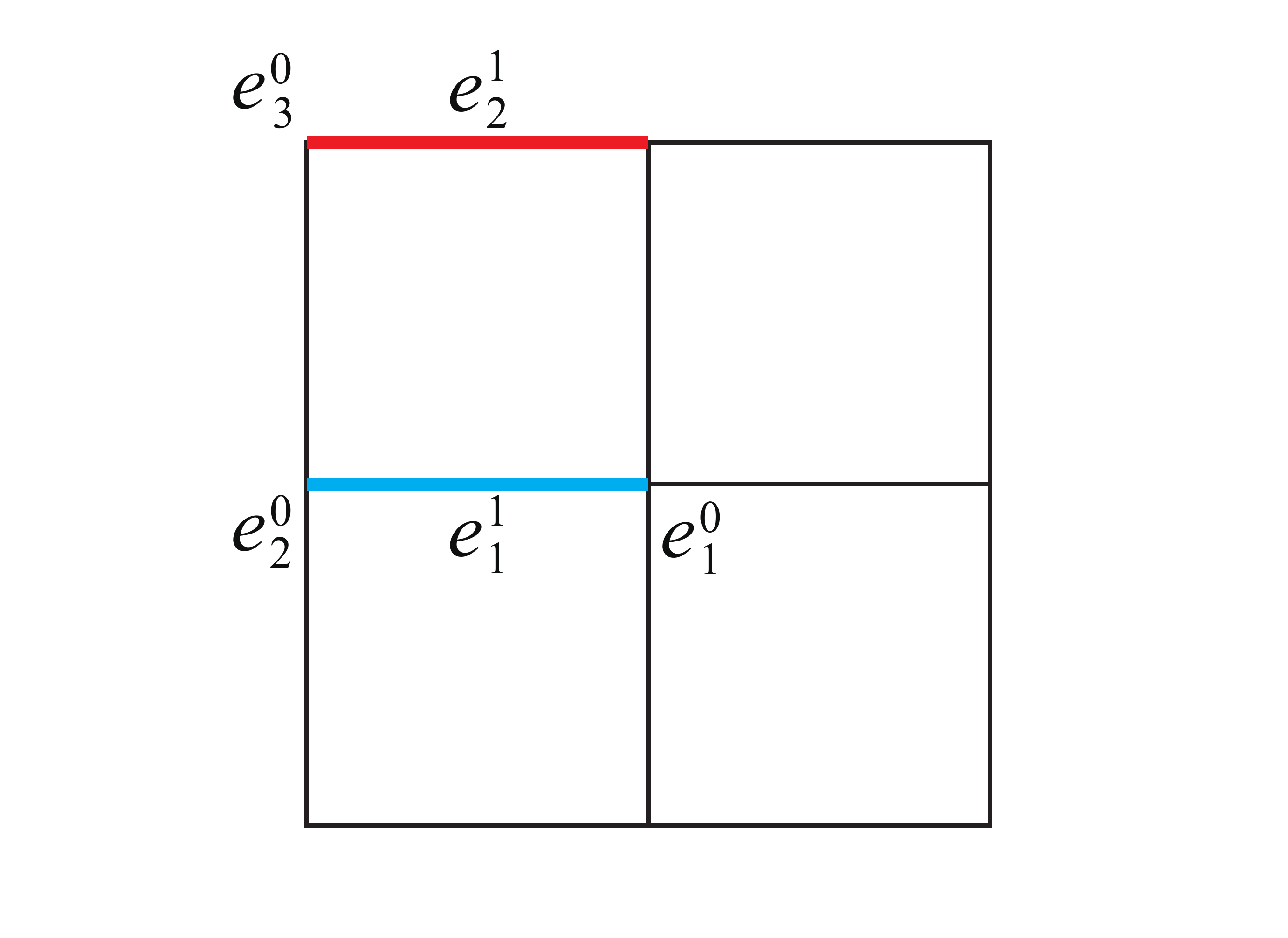}
	 \caption{\label{appendix_cell_complex_p4}WG $p4$}
\end{figure}

\begin{figure}[H]	    
	 \centering
	 \includegraphics[width=0.3\textwidth]{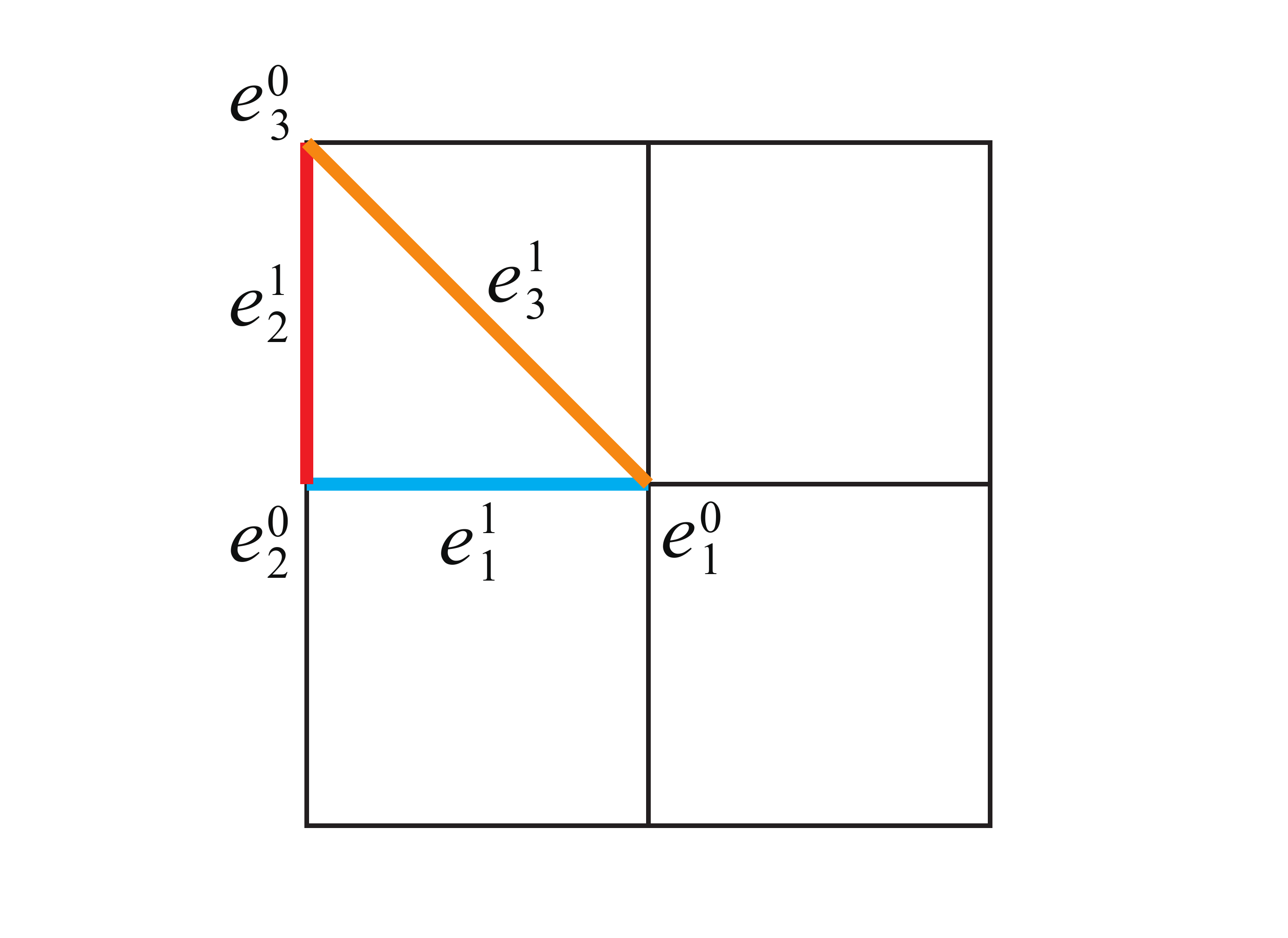}
	 \caption{\label{appendix_cell_complex_p4mm}WG $p4mm$.}
\end{figure}

\begin{figure}[H]	    
	\centering
	    \includegraphics[width=0.3\textwidth]{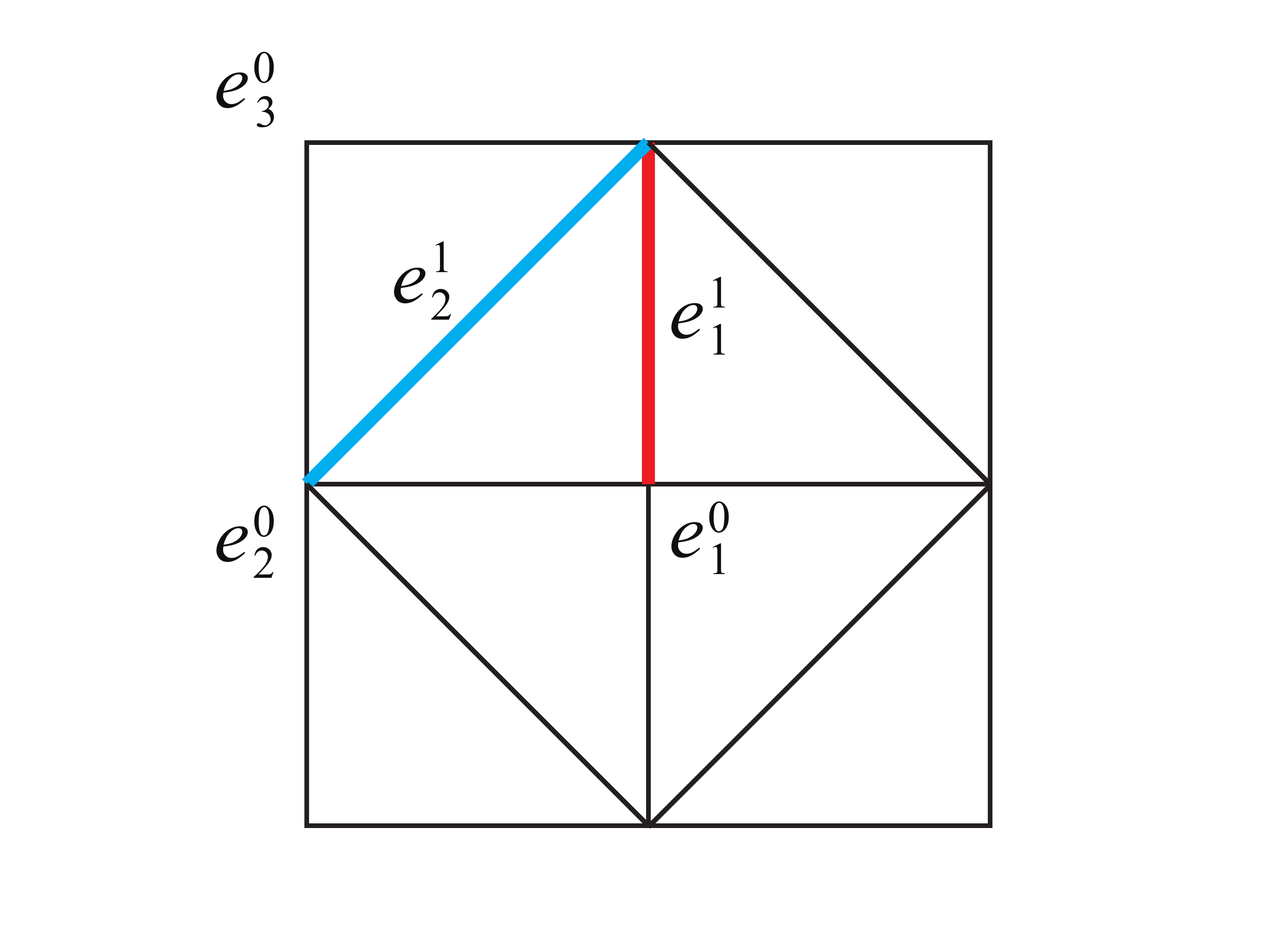}
	    \caption{\label{appendix_cell_complex_p4gm}WG $p4gm$.}
\end{figure}

\begin{figure}[H] 
	 \centering
	    \includegraphics[width=0.3\textwidth]{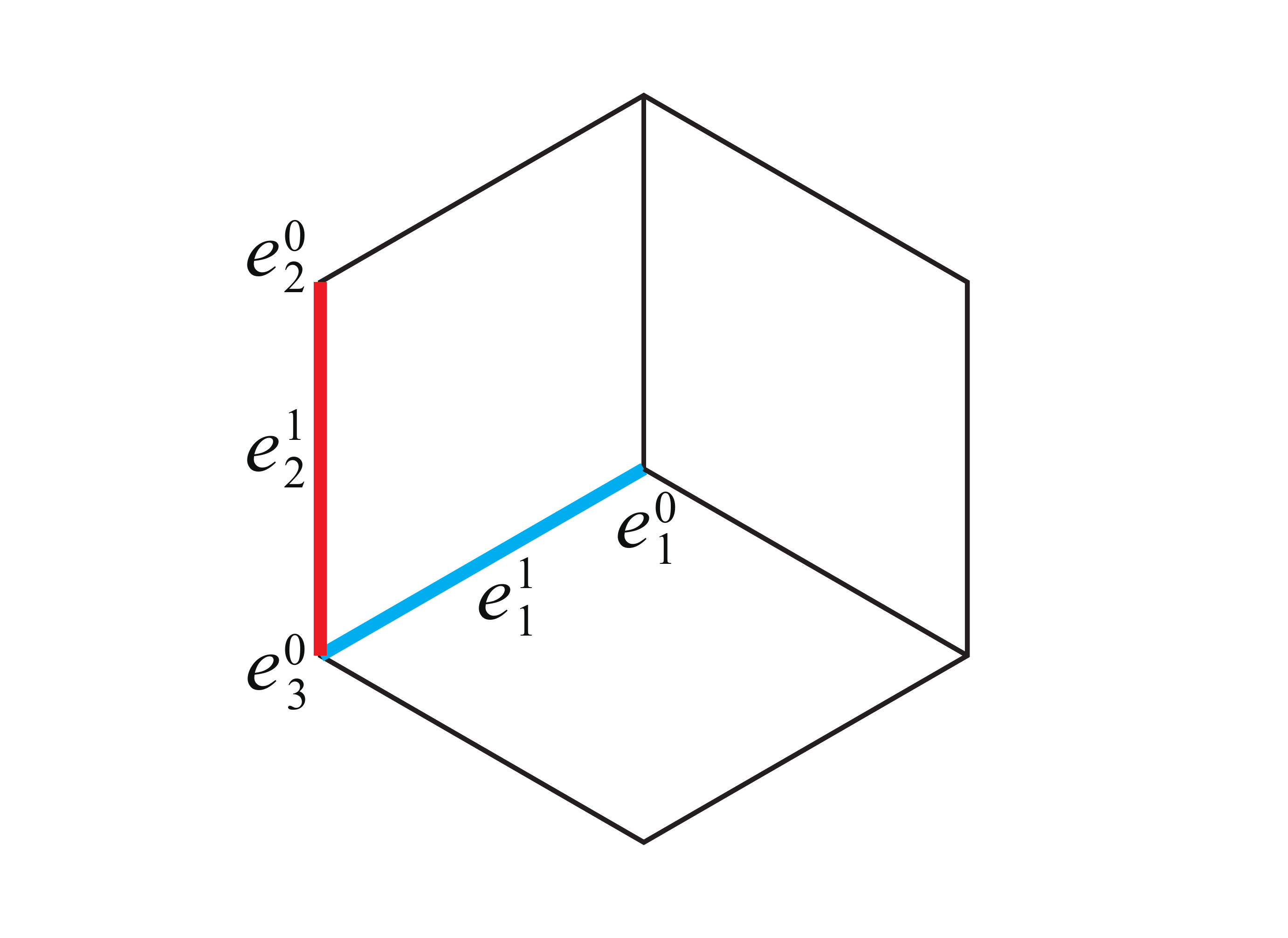}
	    \caption{\label{appendix_cell_complex_p3}WG $p3$}
\end{figure}

\begin{figure}[H]
    \centering
	\includegraphics[width=0.3\textwidth]{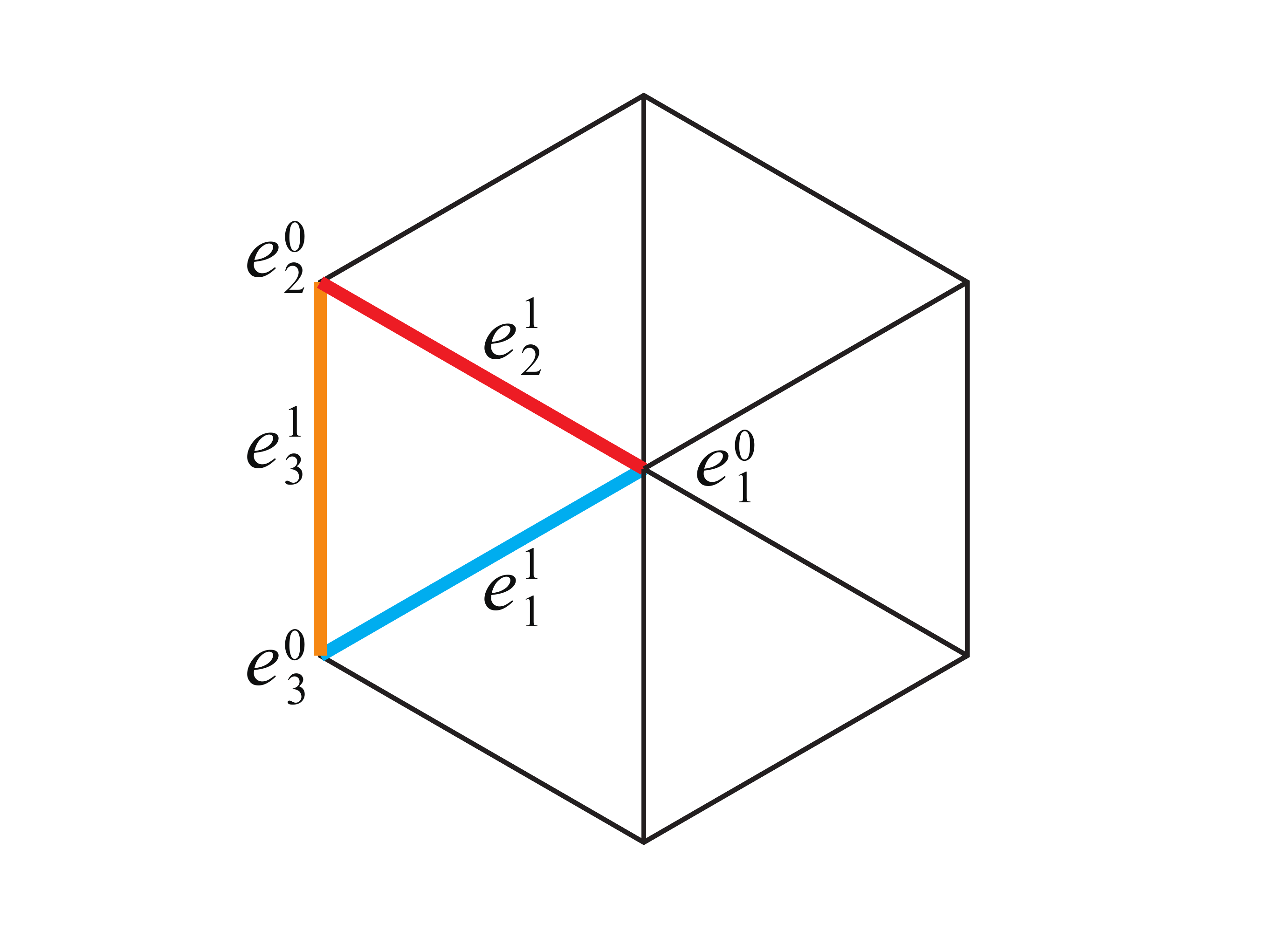}
	 \caption{\label{appendix_cell_complex_p3m1}WG $p3m1$.}
\end{figure}

\begin{figure}[H]
	 \centering
	\includegraphics[width=0.3\textwidth]{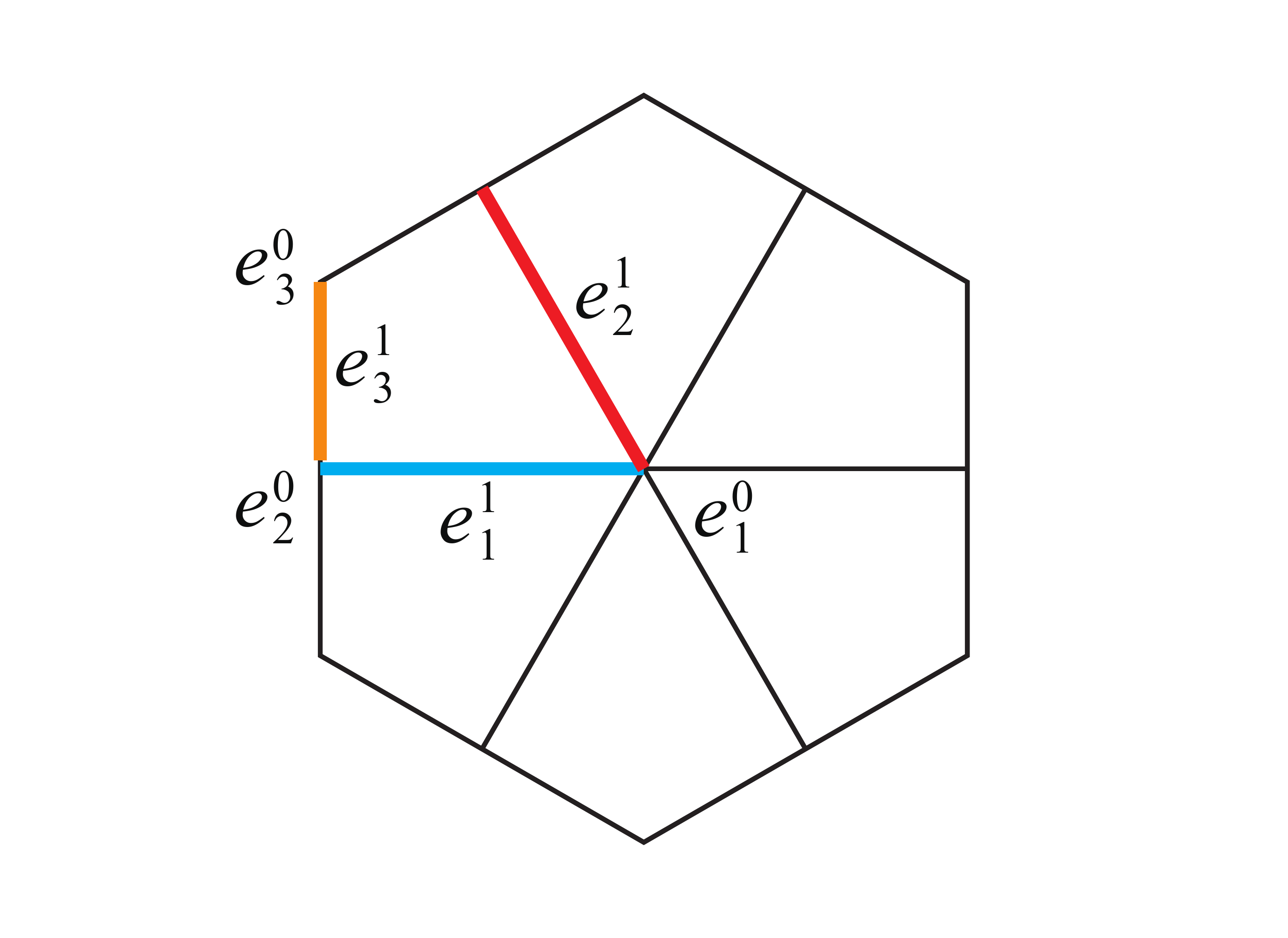}
	\caption{\label{appendix_cell_complex_p31m}WG $p31m$.}
\end{figure}

\begin{figure}[H]
	\centering
	    \includegraphics[width=0.3\textwidth]{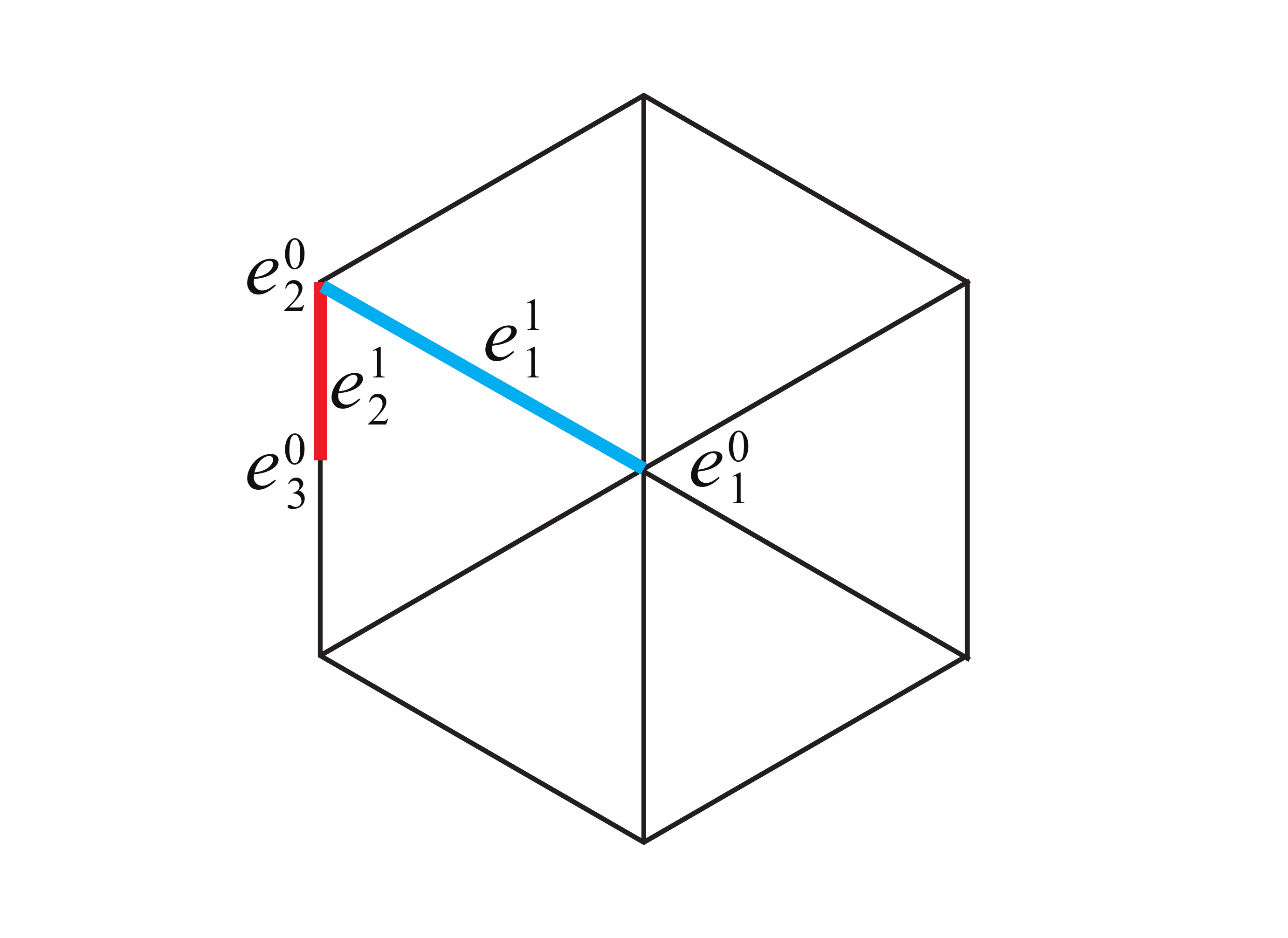}
	    \caption{\label{appendix_cell_complex_p6}WG $p6$}	
\end{figure}

\begin{figure}[H]
	    \centering
	    \includegraphics[width=0.3\textwidth]{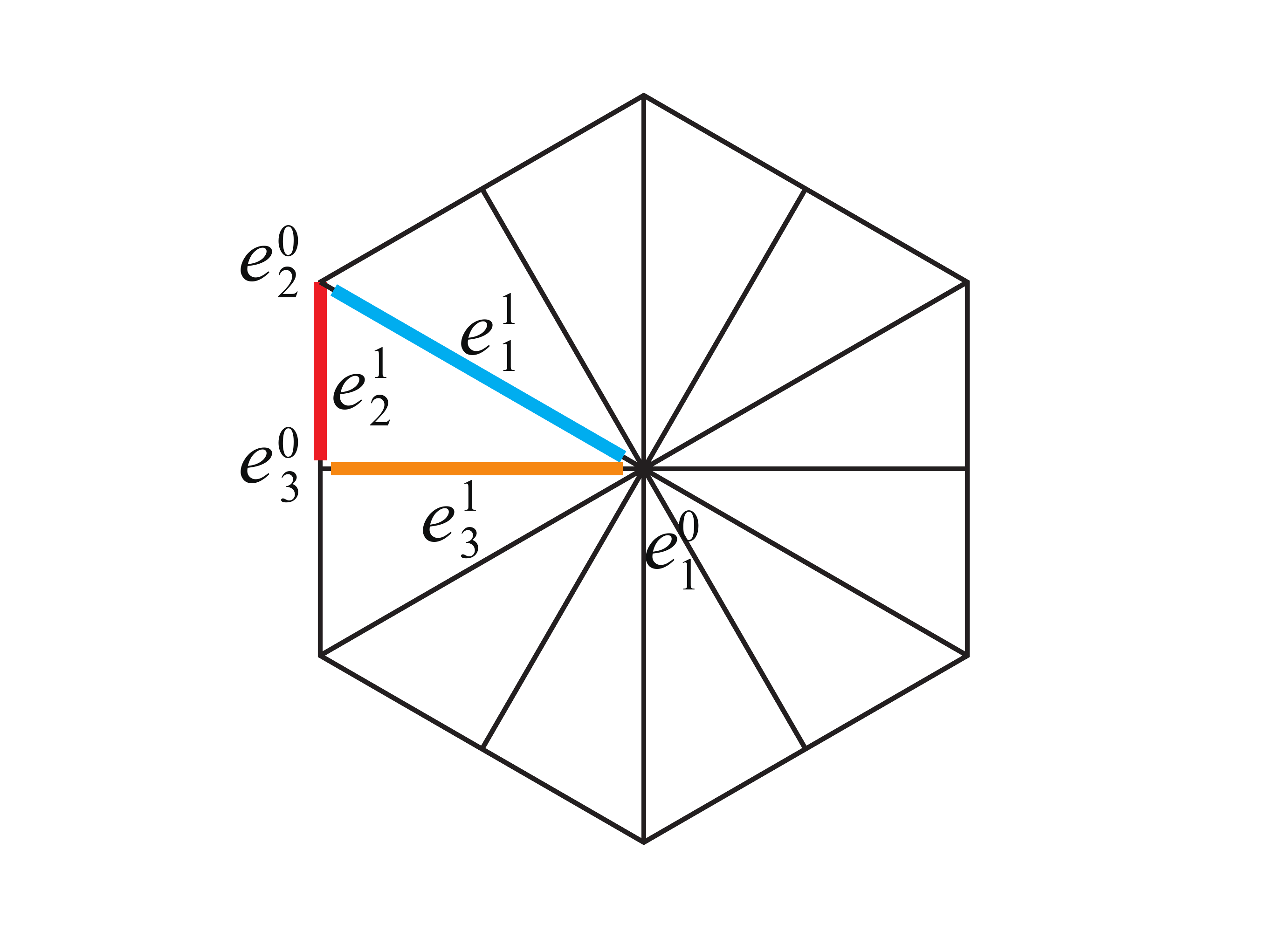}
	    \caption{\label{appendix_cell_complex_p6mm}
	    WG $p6mm$.}
\end{figure}

\section{Cell complex for some LGs}{\label{Appendix cell complex LGs}}

\begin{figure}[H]
	\centering
	\includegraphics[width=0.3\textwidth]{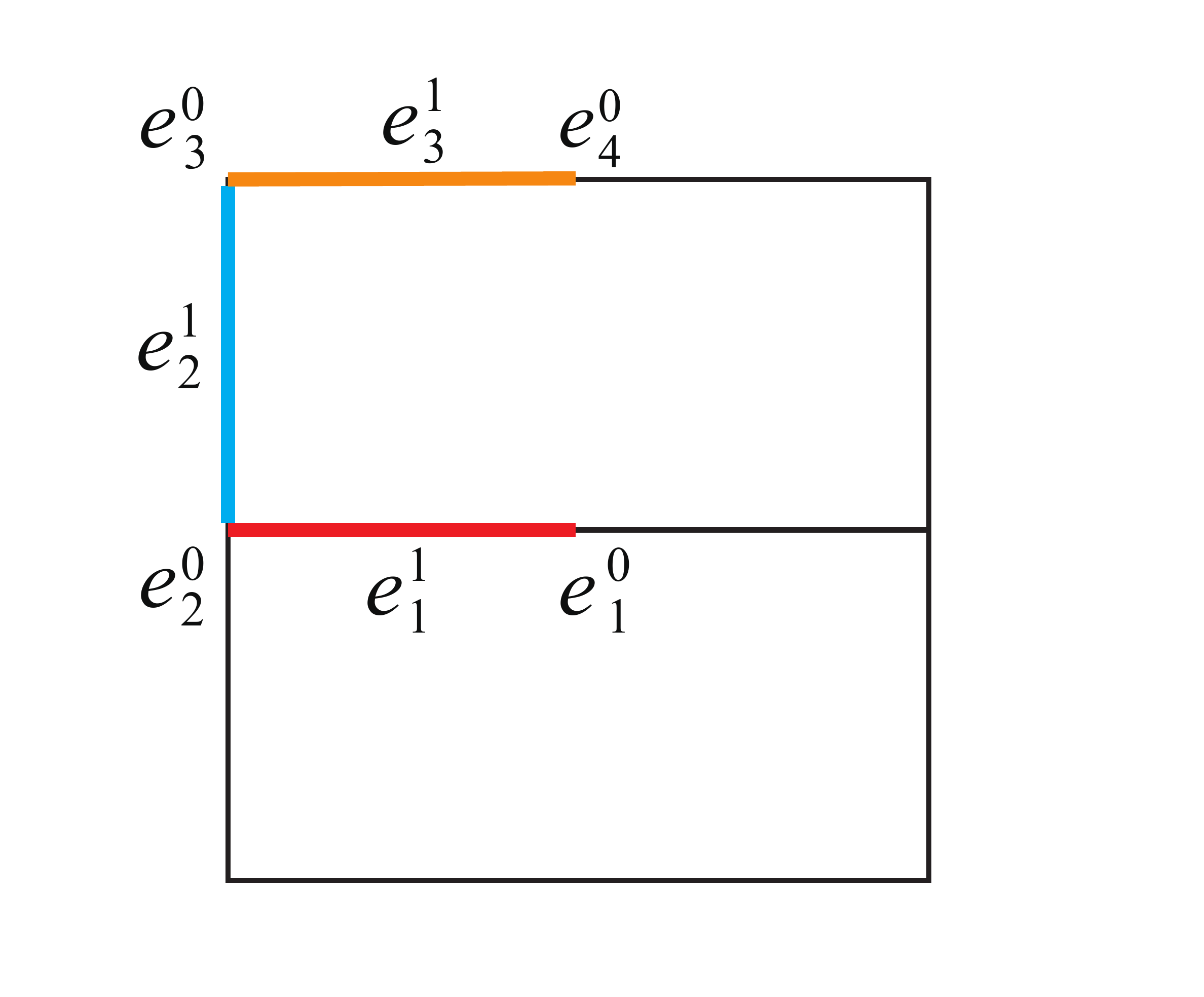}
	\caption{\label{appendix_cell_complex_p-1}
	LG $p\text{-}1$.}
\end{figure}

\begin{figure}[H]
	\centering
	\includegraphics[width=0.3\textwidth]{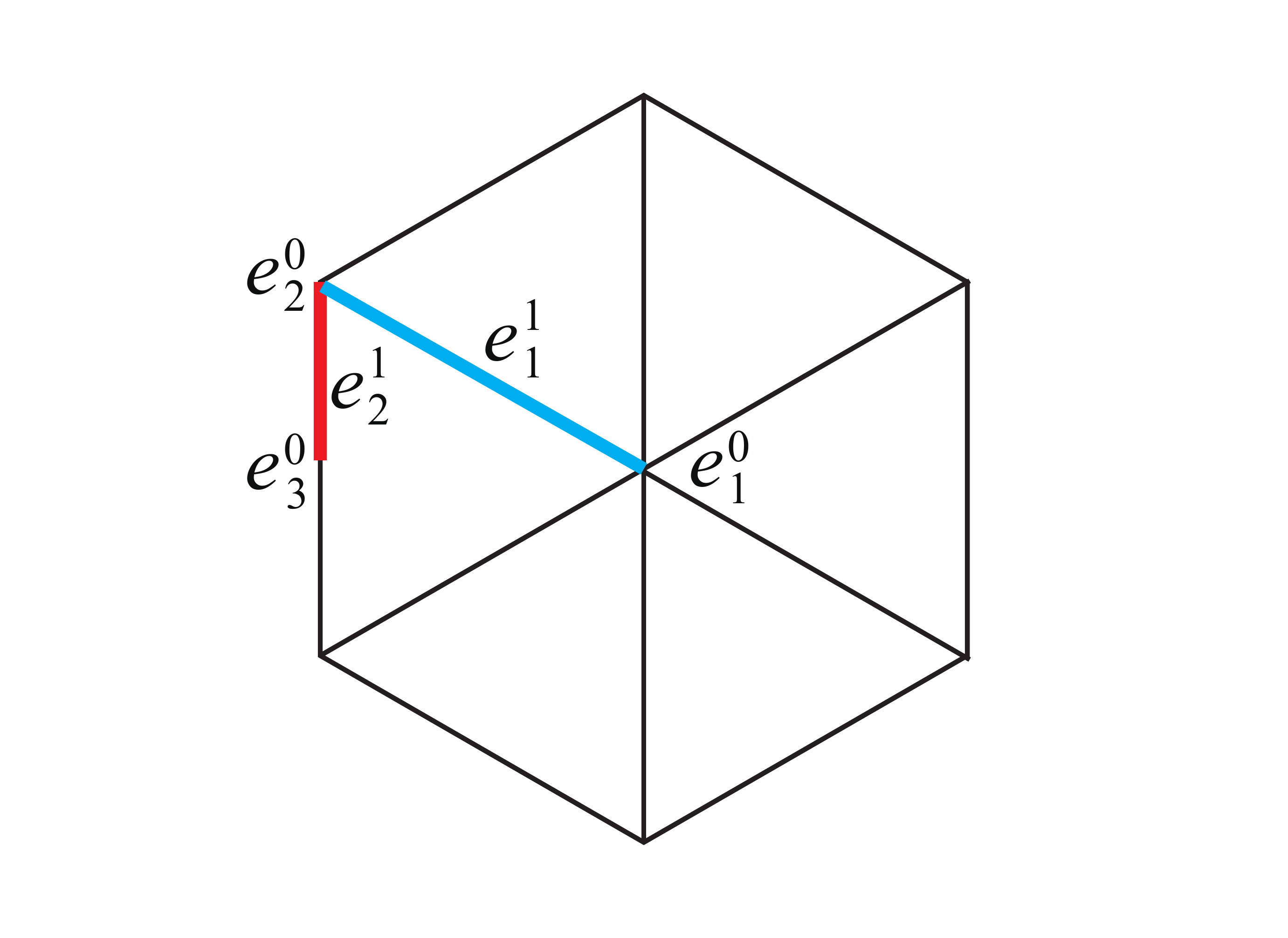}
	\caption{\label{appendix_cell_complex_p-3}
	LG $p\text{-}3$.}
\end{figure}

\begin{figure}[H]
	\centering
	\includegraphics[width=0.3\textwidth]{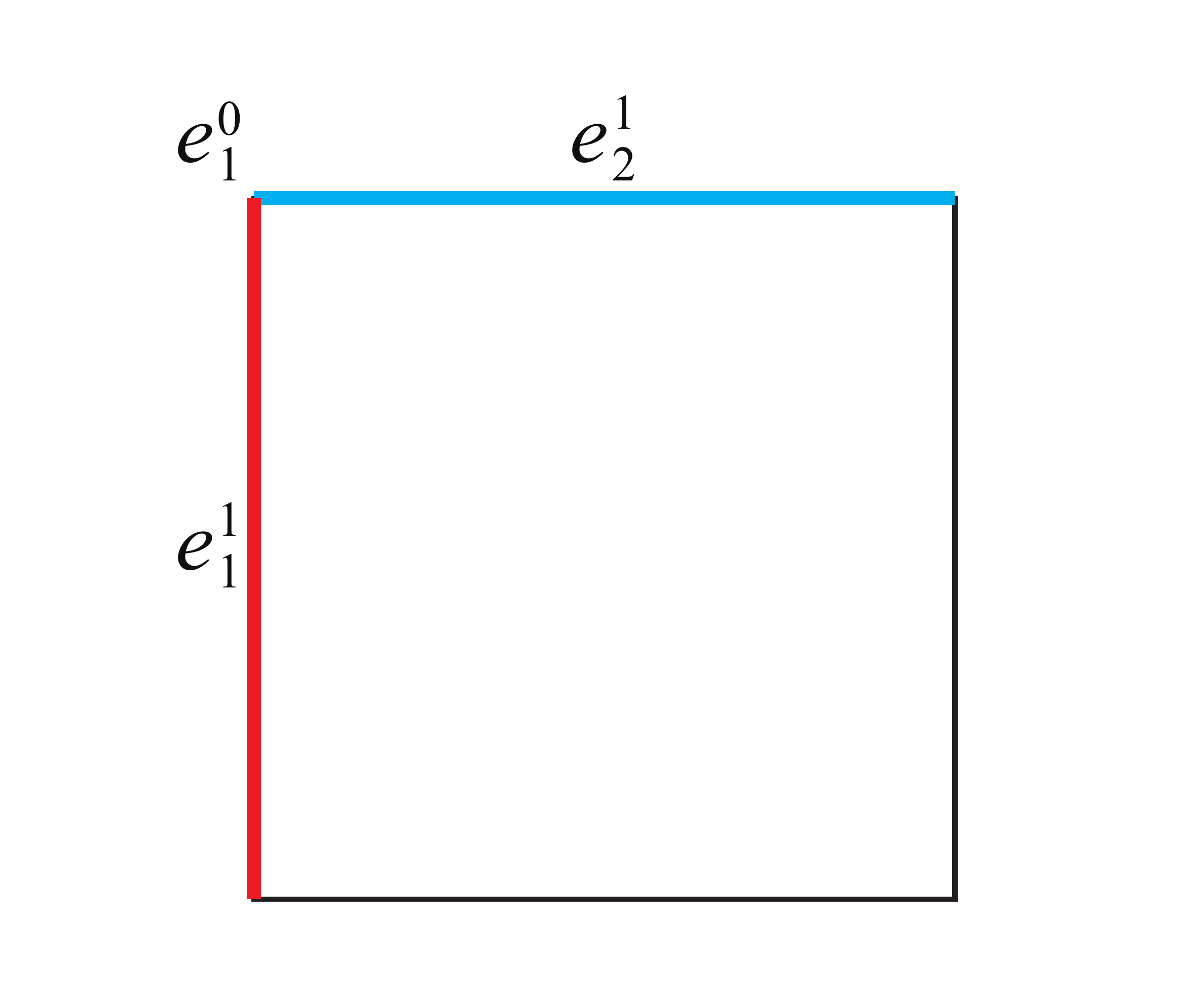}
	\caption{\label{appendix_cell_complex_p11m}
	LG $p11m$.}
\end{figure}

\begin{figure}[H]
	\centering
	\includegraphics[width=0.3\textwidth]{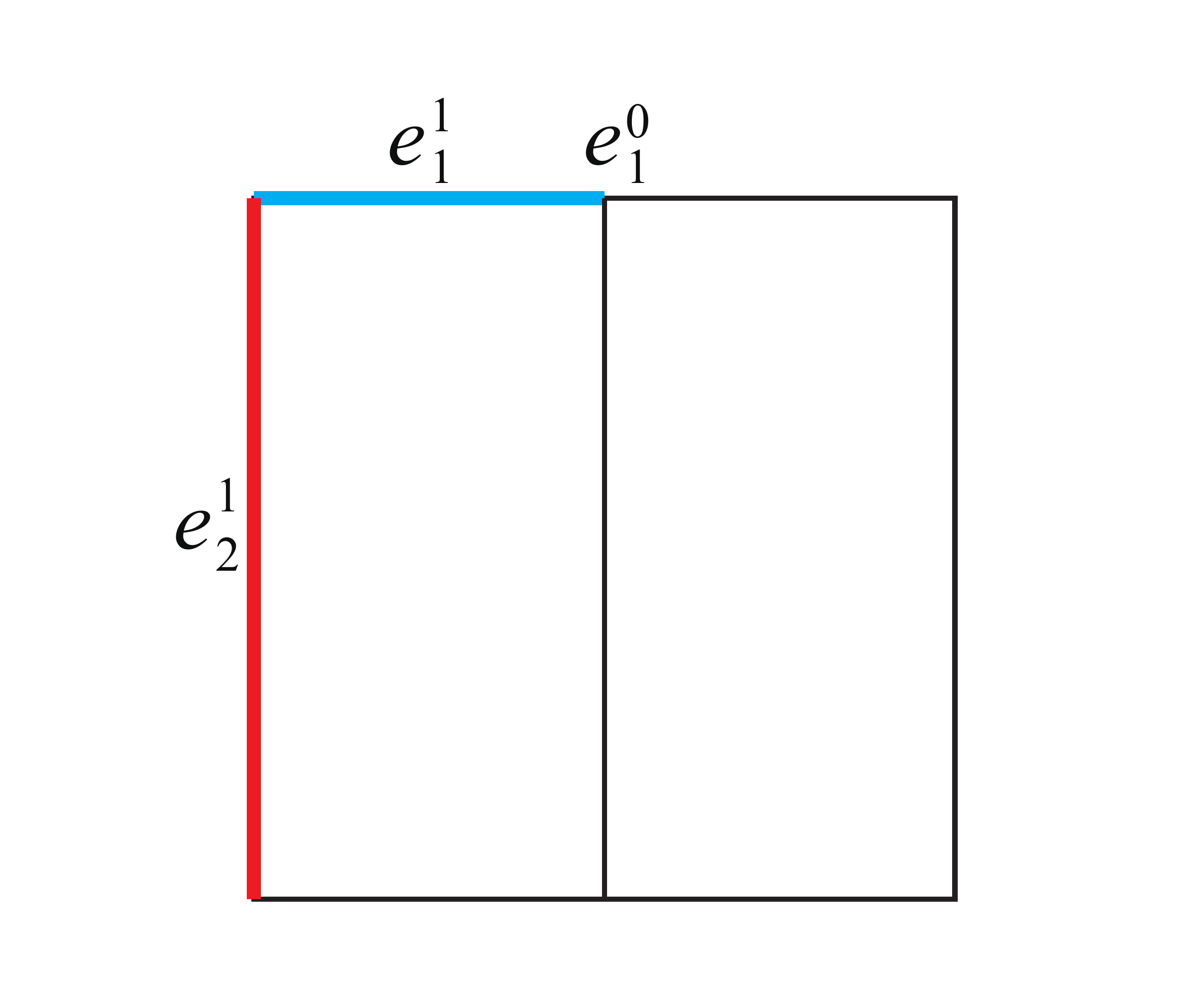}
	\caption{\label{appendix_cell_complex_p11a}
	LG $p11a$.}
\end{figure}

\begin{figure}[H]
	\centering
	\includegraphics[width=0.3\textwidth]{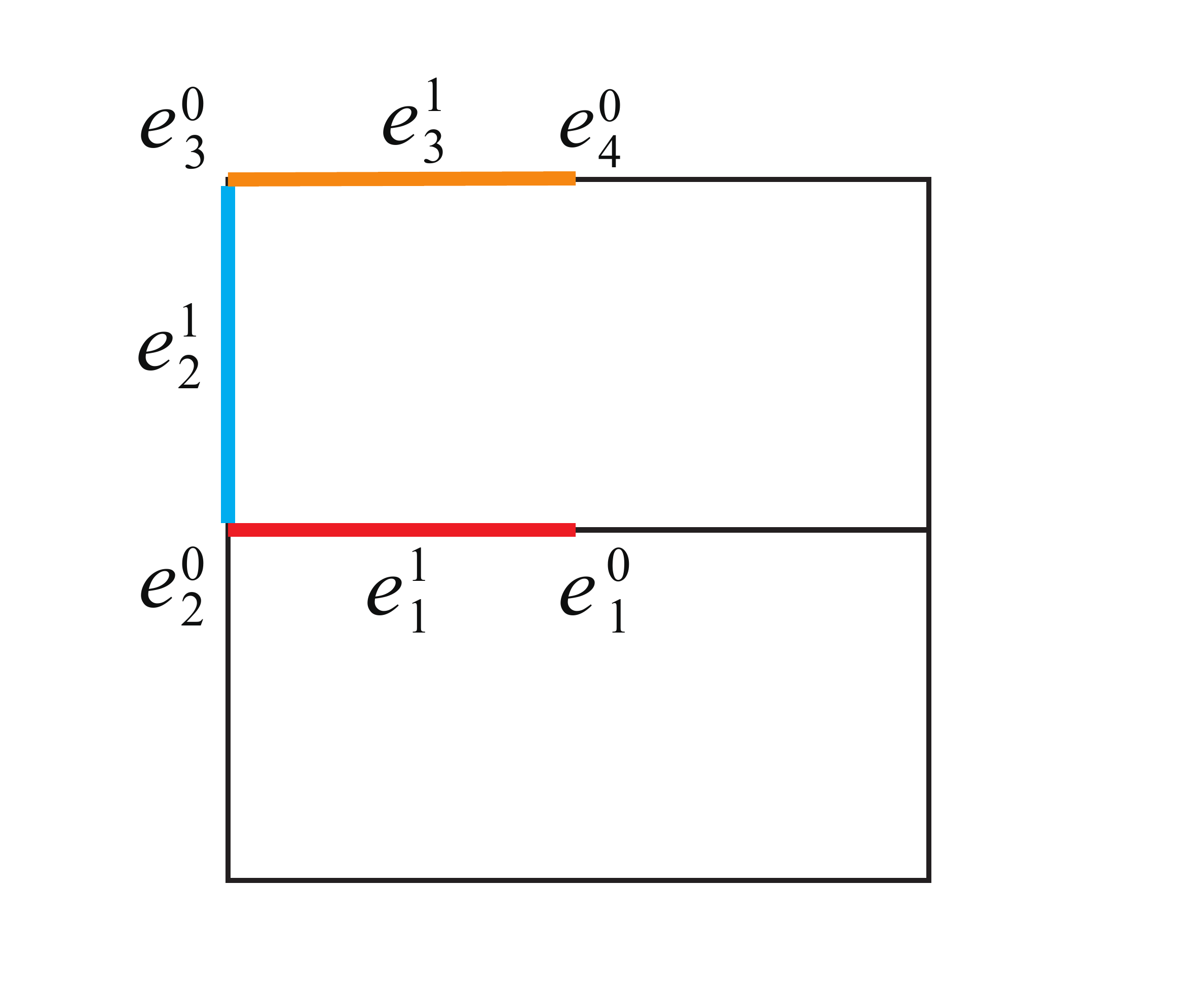}
	\caption{\label{appendix_cell_complex_p112/m}
	LG $p112/m$.}
\end{figure}

\begin{figure}[H]
	\centering
	\includegraphics[width=0.3\textwidth]{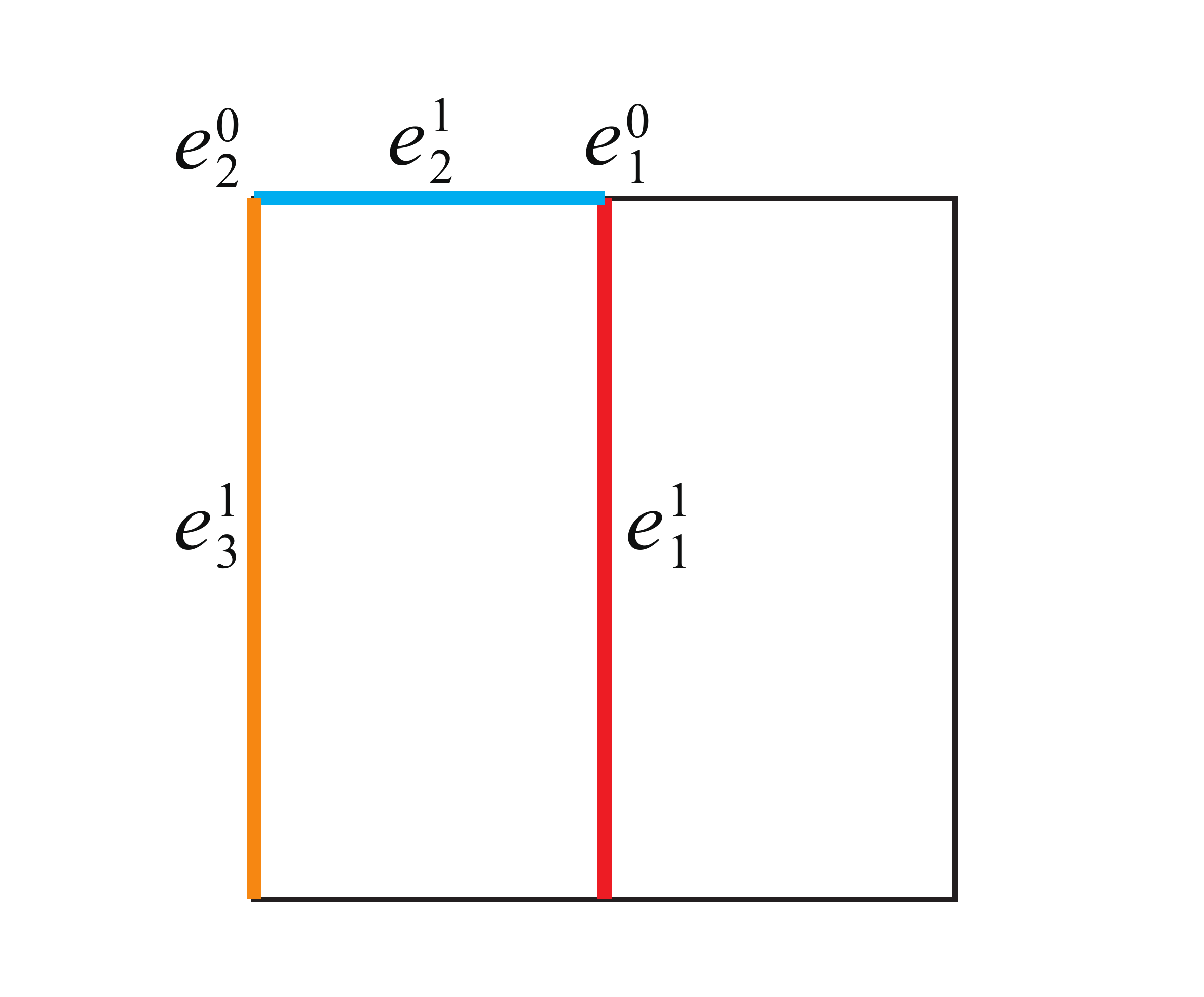}
	\caption{\label{appendix_cell_complex_pm2m}
	LG $pm2m$.}
\end{figure}

\begin{figure}[H]
	\centering
	\includegraphics[width=0.3\textwidth]{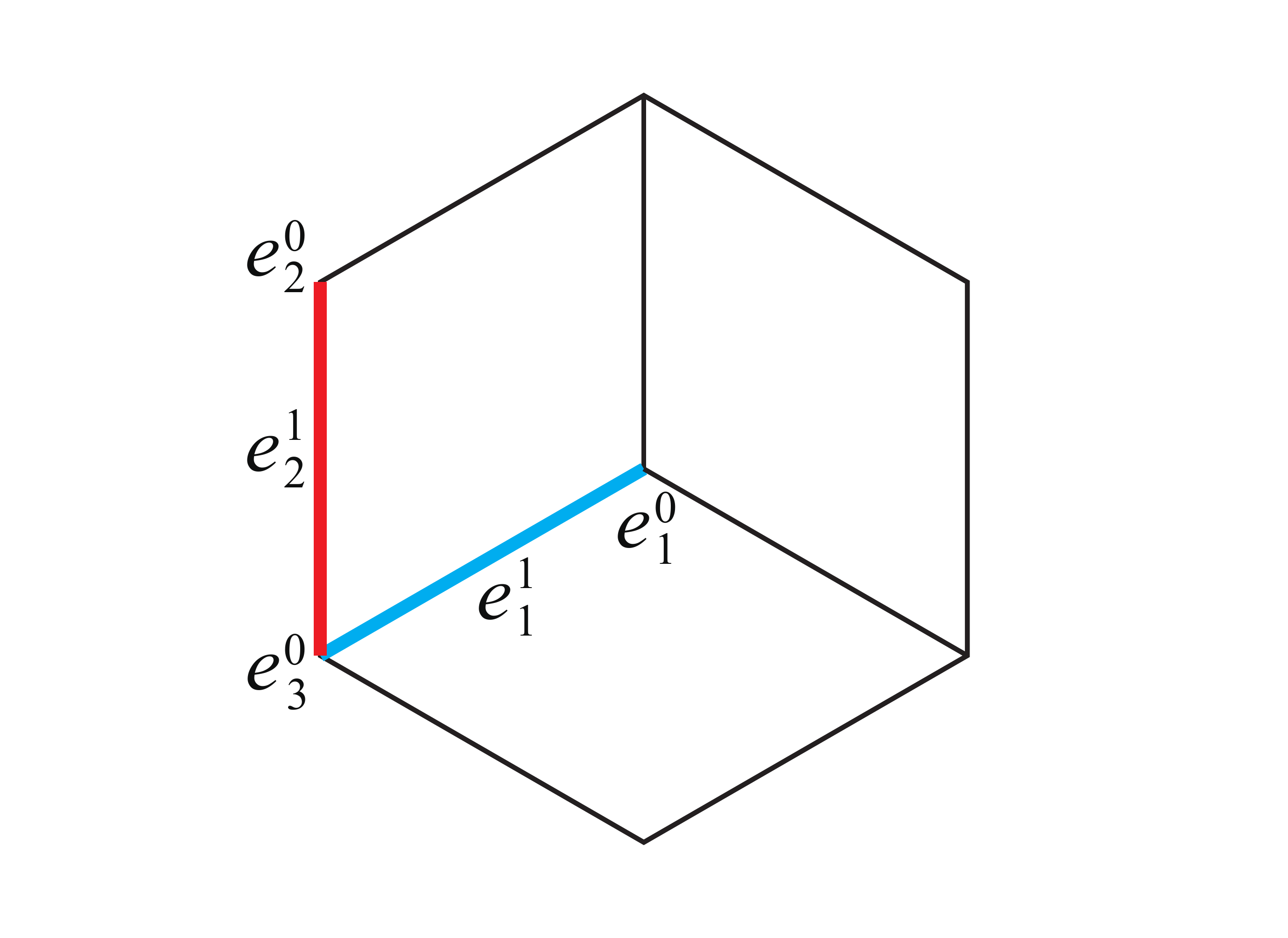}
	\caption{\label{appendix_cell_complex_p3/m}
	LG $p3/m$.}
\end{figure}

\begin{figure}[H]
	\centering
	\includegraphics[width=0.3\textwidth]{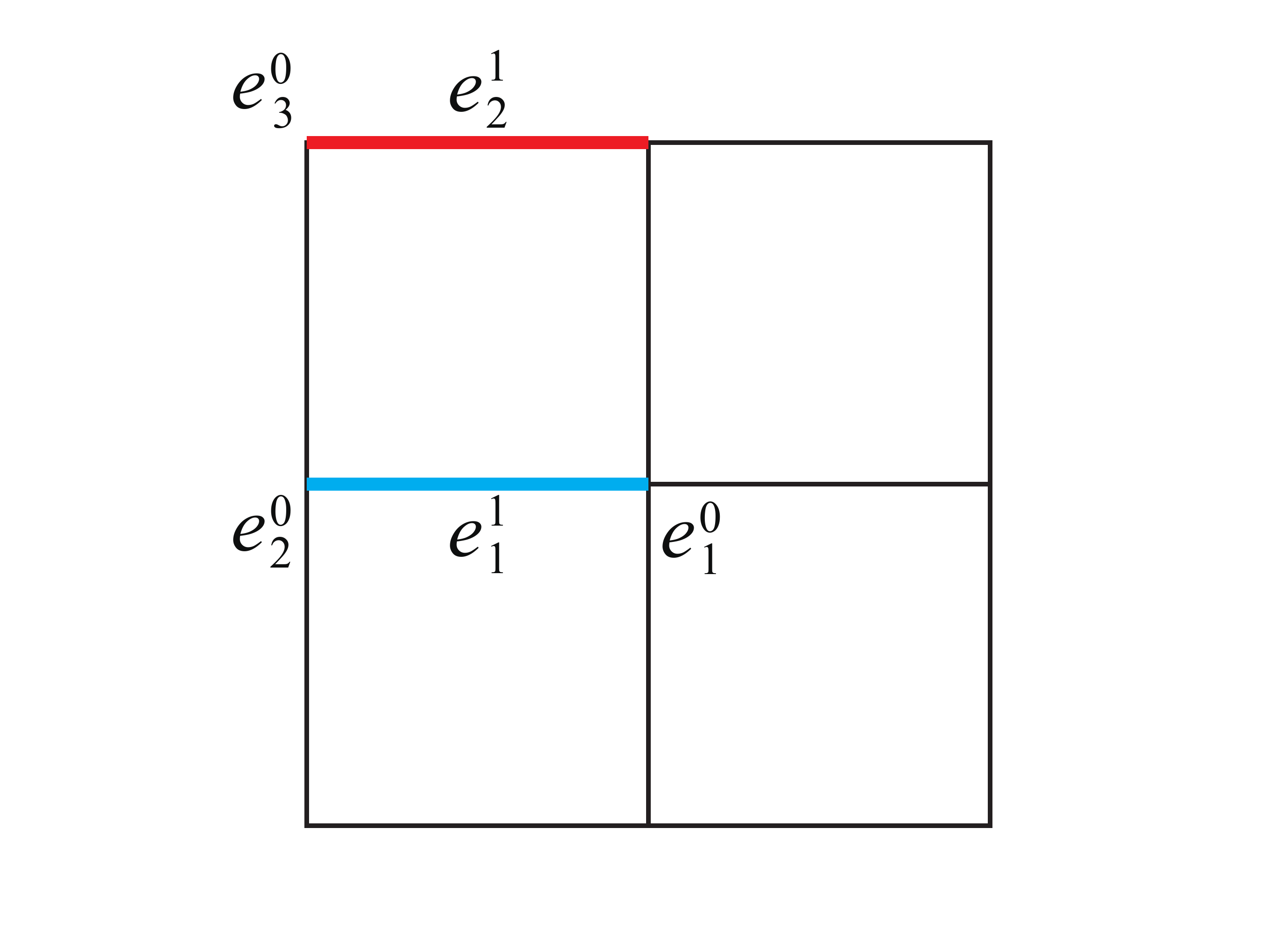}
	\caption{\label{appendix_cell_complex_p4/m}
	LG $p4/m$.}
\end{figure}

\begin{figure}[H]
	\centering
	\includegraphics[width=0.3\textwidth]{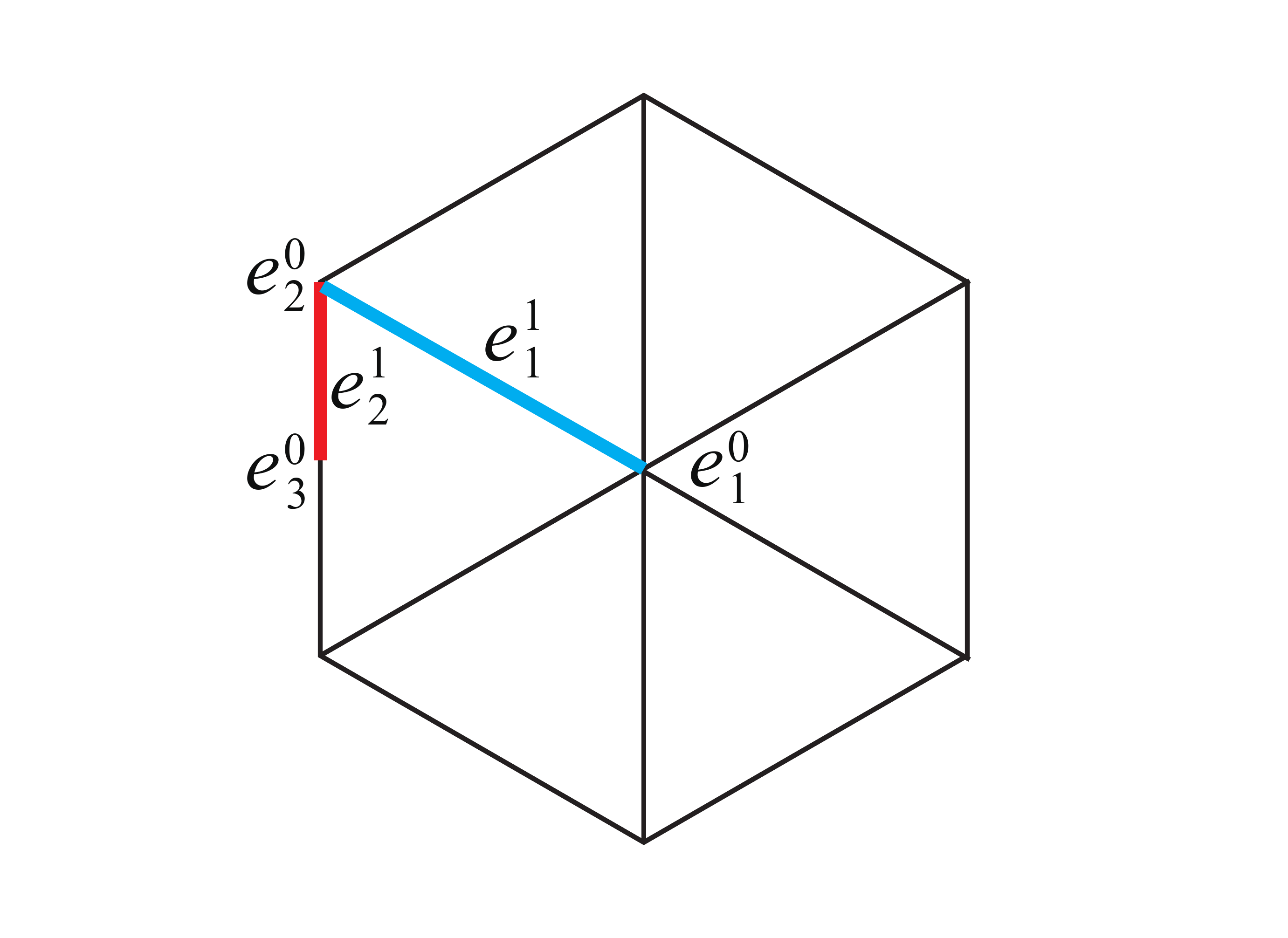}
	\caption{\label{appendix_cell_complex_p6/m}
	LG $p6/m$.}
\end{figure}

\end{document}